\documentclass[reqno,11pt]{amsart}
\usepackage[
  a4paper,
  top=26mm,
  bottom=26mm,
  left=26mm,
  right=26mm
]{geometry}

\usepackage{graphicx}

\usepackage{amsmath,amssymb,amscd}
\usepackage[mathscr]{eucal}
\usepackage{here}
\usepackage{bm}

\usepackage{xcolor}

\usepackage{tikz}
\usetikzlibrary{calc,math,decorations.markings}

\numberwithin{equation}{section}

\newtheorem{thm}{Theorem}[section]
\newtheorem{cor}[thm]{Corollary}
\newtheorem{lem}[thm]{Lemma}
\newtheorem{prop}[thm]{Proposition}

\theoremstyle{definition}
\newtheorem{definition}[thm]{Definition}
\newtheorem{example}[thm]{Example}

\theoremstyle{remark}
\newtheorem{remark}[thm]{Remark}

\def\e{e}
\def\C{{\mathbb C}}
\def\Z{{\mathbb Z}}
\def\ve{{\varepsilon}}

\newcommand\LL{{\mathcal{L}}}
\newcommand\MM{{\mathcal{M}}}

\newcommand\Tr{{\mathrm{Tr}}}

\newcommand{\uu}{{\mathsf u}}
\newcommand{\ww}{{\mathsf w}}
\newcommand{\vv}{{\mathsf v}}

\newcommand{\maru}[1]{\raise0.2ex\hbox{\textcircled{\scriptsize{#1}}}}

\newfont{\bg}{cmr9 scaled\magstep4}
\newcommand{\bigzerol}{\smash{\lower1.0ex\hbox{\bg 0}}}

\tikzset{
  mid arrow/.style={postaction={decorate,decoration={
        markings,
        mark=at position .5 with {\arrow[#1]{latex}}
      }}},
}

\begin{document}

\title[Quantized six-vertex model]
{Quantized six-vertex model at roots of unity:\\
Frobenius property and free-parafermion spectra}

\author[Rei Inoue]{Rei Inoue}
\address{Rei Inoue, Department of Mathematics and Informatics,
   Faculty of Science, Chiba University,
   Chiba 263-8522, Japan.}
\email{reiiy@math.s.chiba-u.ac.jp}

\author[Atsuo Kuniba]{Atsuo Kuniba}
\address{Atsuo Kuniba, Institute of Physics, Graduate School
of Arts and Sciences, University of Tokyo, Komaba, Tokyo, 153-8902, Japan.}
\email{atsuo.s.kuniba@gmail.com}

\author[Yuji Terashima]{Yuji Terashima}
\address{Yuji Terashima, Graduate School of Science, Tohoku University,
6-3, Aoba, Aramaki-aza, Aoba-ku, Sendai, 980-8578, Japan}
\email{yujiterashima@tohoku.ac.jp}

\author[Junya Yagi]{Junya Yagi}
\address{Junya Yagi, Yau Mathematical Sciences Center, Tsinghua University, China}
\email{junyagi@tsinghua.edu.cn}


\begin{abstract}
We study the quantized six-vertex (Q6V) model on arbitrary admissible
diagrams at $q=\ve$, where $\ve$ is a primitive $N$th root of unity
with $N$ odd.  For the commuting layer transfer matrices
$\mathbb{T}(y)$ with mixed boundary conditions, we establish the
quantum Frobenius property
\[
\mathbb{T}(y)\mathbb{T}(\ve y)\cdots
\mathbb{T}(\ve^{N-1}y)
=
\mathscr{P}_N(y^N)\mathbb{I},
\]
where $\mathscr{P}_N$ is a scalar Laurent polynomial obtained directly
from the monomial expansion of $\mathbb{T}(y)$.
Our proof exploits
the underlying three-dimensional integrability, in particular the
tetrahedron equations and their consequences. 
The quantum Frobenius relation essentially determines the joint
eigenvalues of the associated Hamiltonians up to multiplicities and
shows that they take a free-parafermion form, thereby extending such
structures from previously known one-dimensional examples to a broad
class of genuinely two-dimensional quantized six-vertex models.
The construction also
admits independent local gauge choices and recovers the
long-range-interacting $\tau_2$ model, which includes the original 
free-parafermion model as a special case.  
We also apply the same framework to the relativistic quantum Toda
chain at odd roots of unity and obtain the free-parafermion spectra
for the relevant Q6V transfer matrices with fixed boundary states.
\end{abstract}

\maketitle

\setcounter{section}{0}

\section{Introduction and main result}\label{s:intro}

\subsection{Background}
The six-vertex model (cf.~\cite[Chap.~8]{B82}) is a paradigmatic model of
integrable systems in statistical mechanics.
The quantized six-vertex (Q6V) model \cite{IKTY25}
is a natural generalization incorporating both quantum and three-dimensional aspects.
Its local Boltzmann weight, specified in Figure~\ref{fig:6v},
takes values in the $q$-Weyl algebra \eqref{qcom}
generated by $e^{\pm\uu}$ and $e^{\pm\ww}$.
The model is defined on a diagram $G$ on a torus,
subject to a topological condition called admissibility
(Definition~\ref{def:adm}),
with an independent copy of the $q$-Weyl algebra attached to each vertex.

The integrability of the Q6V model is encoded in the transfer matrix
$T_G(x,y)$ \eqref{TGg}, with two spectral parameters associated with
the two homology cycles of the torus.
For admissible $G$, the commutativity
$[T_G(x,y),T_G(x',y')]=0$ was established in \cite[Thm.~3.4]{IKTY25}
through a systematic use of the tetrahedron equations,
the three-dimensional analogues of the Yang--Baxter equation (cf.~\cite{Z80}),
together with inversion relations.
Upon choosing representations of the $q$-Weyl algebras, 
its commuting Laurent coefficients may be regarded as 
Hamiltonians of a two-dimensional quantum integrable system. 
Alternatively, taking matrix elements of the local operators gives scalar Boltzmann weights
for a three-dimensional classical vertex model, with the $q$-Weyl degree of freedom
supplying the third direction, thereby allowing $T_G(x,y)$ to be interpreted 
as a \emph{layer} transfer matrix.
This feature is depicted in \eqref{Tabij} in Example \ref{ex:tabij}.
These complementary viewpoints place the Q6V model at the intersection 
of two-dimensional quantum and three-dimensional classical integrability. 

For generic $q$, the Q6V model includes the quantum relativistic Toda lattice
as a special case and also admits a dimer-model formulation;
see \cite[Chaps.~5, 6]{IKTY25} and the references therein.
Further aspects related to the three-dimensional $R$-matrices have been 
studied in \cite{KMY23} and also in \cite[Sec.7]{IKSTY24},  
the latter including a connection to quantum cluster algebras.

\subsection{Quantum Frobenius property}

In this paper, we establish a quantum Frobenius property of the Q6V model at $q=\ve$, 
where $\ve$ is a primitive $N$th root of unity with $N$ odd.
Let $\mathbb{T}(y)$ denote the transfer matrix \eqref{TTai} at $q=\ve$.
This is a variant of $T_G(x,y)$ with mixed boundary conditions:
periodic boundary conditions are retained in the vertical direction,
while the west and east boundaries are fixed to prescribed states $\mathbf{i}$ and
$\mathbf{a}$, respectively.
Its single spectral parameter $y$ is associated with the periodic direction.
At $q=\ve$, the $N$th powers of the local $q$-Weyl generators become central.
We consider $\mathbb{T}(y)$ on a finite-dimensional quantum space,
with the local $q$-Weyl algebras acting through cyclic representations.
See \eqref{Vcyc} and \eqref{repW} for a typical realization.
The generators $e^{\uu_v}$ and $e^{\ww_v}$ associated with a vertex $v$ in $G$ 
act as a diagonal operator
and a cyclic shift operator, respectively, which may be viewed as the
$N$-state analogues of the local Pauli matrices $\sigma_z$ and $\sigma_x$.
The transfer matrix is a Laurent polynomial in $y$ and satisfies 
the commutativity $[\mathbb{T}(y),\mathbb{T}(y')]=0$.

Our main result, Theorem~\ref{th:main}, is the functional relation
\begin{equation}\label{main0}
\mathbb{T}(y)\mathbb{T}(\ve y)\cdots\mathbb{T}(\ve^{N-1}y)
= \mathscr{P}_N(y^N)\mathbb{I},
\end{equation}
where $\mathbb{I}$ is the identity operator.
The scalar Laurent polynomial $\mathscr{P}_N(y^N)$
is obtained by summing the $N$th powers of the individual terms
in the monomial expansion \eqref{TTy} of $\mathbb{T}(y)$; see \eqref{pdef}.
Thus, the product of transfer matrices over the full $\ve$-orbit of the spectral parameter 
reduces to a scalar operator, with all mixed ``off-diagonal'' contributions cancelling exactly.
This remarkable phenomenon holds for every admissible diagram $G$, 
with the parameters $r_v,s_v,f_v,g_v$ entering the local Boltzmann weights \eqref{L3} 
chosen independently at each vertex $v$.

We call this the \emph{quantum Frobenius property},
in analogy with the quantum Frobenius morphisms for quantum groups
and quantized function algebras at roots of unity,
through which classical structures emerge from suitable
$N$th-power elements \cite{L90,DCL94,G07}.
Heuristically, the condition $\ve^N=1$ implies that taking $N$th powers
eliminates the quantum phases arising from the $\ve$-commutation relations,
leaving behind effectively classical quantities.
The simplest manifestation of this principle is the elementary identity
given in Lemma \ref{le:qfrob}.
In the present setting, the Frobenius phenomenon extends to the transfer matrix,
including its spectral-parameter dependence:
the product over the full $\ve$-orbit
$y,\ve y,\ldots,\ve^{N-1}y$
reduces to a Laurent polynomial obtained directly
from the monomial expansion of $\mathbb{T}(y)$ itself.

The quantum Frobenius property also extends to transfer matrices with
independently chosen local gauges, as established in
Section~\ref{s:t2}.

\subsection{Free-parafermion spectrum}

Suppose that the Laurent polynomial $\mathbb{T}(y)$ has the expansion
$\mathbb{T}(y) = \mathbb{H}_0 y^\kappa  - \mathbb{H}_1 y^{\kappa+1} +
\cdots + (-1)^d \mathbb{H}_d y^{\kappa +d}$,
where $\kappa \in \Z$, $d \in \Z_{\ge 1}$, and
$\mathbb{H}_0,\ldots,\mathbb{H}_d$ are mutually commuting ``Hamiltonians''.
The significance of the quantum Frobenius property \eqref{main0}
is that it essentially determines their joint spectrum up to multiplicities.
For example, one obtains
\begin{align}\label{Hpf0}
\mathrm{Spec}(\mathbb{H}_1) \subseteq
\left\{
\ve^{\ell_1}\mathcal{E}_1+\cdots+\ve^{\ell_d}\mathcal{E}_d
\;\middle|\;
\ell_1,\ldots,\ell_d\in\Z_N
\right\},
\end{align}
where $\mathcal{E}_1,\ldots,\mathcal{E}_d$ are scalars determined directly
from the zeros of $\mathscr{P}_N(y)$.
The derivation is given in Section~\ref{ss:fps}.

A spectrum of the form appearing on the RHS of \eqref{Hpf0}
was first discovered in an $N$-state chain
associated with the superintegrable chiral Potts model \cite{B89}.
The relevant quantum system was later studied further as the
{\em free-parafermion} model in \cite{F14}.
Such a spectrum naturally generalizes the familiar free-fermion spectrum
$\pm\mathcal{E}_1\pm\cdots\pm\mathcal{E}_d$,
which formally corresponds to the case $N=2$.
Free-parafermion spectra and the integrable structures underlying them
have since been investigated in a variety of settings;
see, for example, \cite{AYP14,B14,AP20,BHL23,MEWC25}.
In particular, the free-parafermion model was recognized as a special
case of the long-range-interacting $\tau_2$ model
\cite{B04}.
For a historical account of the early developments, we refer to
\cite[Sec.~1]{AYP14}.

We also apply our results to the relativistic quantum Toda chain at
an odd root of unity, treating Q6V transfer matrices with fixed
boundary states and comparing the resulting formulas with
\cite{PS02}.  The $\tau_2$ model and the relativistic quantum Toda
chain are thereby placed within a unified Q6V framework, with their
free-parafermion spectra arising from the same quantum Frobenius
mechanism.

\subsection{Present work}
In \cite[Sec.~5.3]{IKTY25}, we observed that a suitable specialization
of the diagram $G$ and the local parameters reduces the Q6V model
to the free parafermion model.
This finding provided a principal motivation for the present work.

The results established here show that the quantum Frobenius property
and the consequent free-parafermion spectra
extend to Q6V models on arbitrary admissible diagrams $G$,
which are genuinely two-dimensional.
These Q6V models also accommodate general mixed boundary conditions,
as well as site-dependent local parameters and gauges.
The resulting class thus constitutes a substantial extension
of the specific one-dimensional models from which the subject
originally emerged.

In addition to this broadening of the class of models,
our proof of the main theorem reveals that the quantum Frobenius
property is not an accidental algebraic feature, but is rooted in the
underlying three-dimensional integrability.
Indeed, the crucial steps in the proofs of the key propositions and lemmas
are most transparently represented by three-dimensional diagrams
involving the tetrahedron equations.
See, for example, Figures \ref{fig:mtt}, \ref{fig:emtt}, \ref{fig:rtttt}, and \ref{fig:mzz}.
The use of such three-dimensional structures is therefore not merely
a technical device, but reflects the natural framework of the problem.
Previous studies of free-parafermionic systems have largely proceeded
within one- or two-dimensional formulations.
The three-dimensional viewpoint developed here may be regarded as a natural
resolution of the algebraic complexity that arises in such
lower-dimensional descriptions.
The present work puts this perspective into systematic practice
and thereby provides a new structural understanding of the origin
of the free-parafermion spectrum.

\subsection{Layout of the paper}

In Section \ref{s:gq}, we review the Q6V model at generic $q$ following \cite{IKTY25}.
New results include Proposition \ref{pr:T01} and Lemma \ref{le:TT},
with the key steps in their proofs depicted in Figures \ref{fig:emtt} and \ref{fig:mzz}.

In Section \ref{s:rt}, we turn to the case of odd roots of unity, $q=\ve$.
The main result, Theorem \ref{th:main}, is presented together with a self-contained proof.
We also explain the well-known fact that the quantum Frobenius property
implies a free-parafermion spectrum.

In Section \ref{s:t2}, we introduce an alternative gauge for the 3D
$L$-operator and allow the two gauges to be chosen independently at
each vertex.  We show that the commutativity and quantum Frobenius
properties persist under these local gauge choices; the extension of
Theorem \ref{th:main} is stated in Theorem \ref{th:mixedF}.

In Section \ref{s:tau2}, we use this local gauge freedom to a two-row
Q6V model at a root of unity and identify its reduced transfer matrix
with that of the $\tau_2$ model
(cf.~\cite{BS90,BBP90,B14,AYP14}).
We also recover the $\tau_2$ Hamiltonian and its free-parafermion
specialization.

In Section \ref{s:td}, we apply the quantum Frobenius property to the
relativistic quantum Toda chain and formulate the relevant Q6V
transfer matrices with fixed boundary states.
This realization is related to the classical Lax representation of
\cite{BR89}.
For the homogeneous chain, we express the associated scalar
polynomials in terms of Chebyshev polynomials, obtain the corresponding
free-parafermion spectra, and compare the results with \cite{PS02}.

Section \ref{s:ol} concludes the paper with a summary and outlook.

Appendix \ref{s:ap} presents graphical representations of the
tetrahedron equations and inversion relations collected in Proposition \ref{pr:te4}.
These relations were obtained in \cite{IKTY25} by systematically extending
the earlier results in \cite[(34)]{BS06} and \cite[(3.122)]{K22}.

\section{Quantized six-vertex model at generic $q$}\label{s:gq}

In this section, we develop the Q6V model on admissible diagrams at
generic $q$ and establish the algebraic relations needed for its
root-of-unity specialization.  Following \cite{IKTY25}, we introduce
the 3D $L$-operator and the auxiliary $M$-operators, together with
their tetrahedron and inversion relations and graphical
representations.  The three-dimensional structure encoded in these
local relations, as illustrated by the $MTT$ relation in
Figure~\ref{fig:mtt}, underlies the commutativity of the layer transfer
matrices.  Finally, we extend the results of \cite{IKTY25} to mixed
boundary conditions and derive the commutation and functional
relations used in Section~\ref{s:rt}.

\subsection{The quantized six-vertex model}\label{ss:q6}

Set
\begin{equation}\label{qh}
q=e^\hbar.
\end{equation}
Throughout this section, we assume that the complex parameters $q$ and $\hbar$ are generic.
The algebraic construction below will also be used at roots of unity in
Section~\ref{s:rt}, where $q$ is specialized and a finite-dimensional representation
of the $q$-Weyl algebra is chosen.

Following \cite{IKTY25}, we describe the  local Boltzmann weights.
Let $\uu,\ww$ be canonical variables satisfying $[\uu,\ww]=\hbar$.
We denote by $\mathcal{W}(q)$ the $q$-Weyl algebra generated by
$e^{\pm \uu}$ and $e^{\pm \ww}$ subject to
\begin{align}\label{qcom}
 e^\uu e^\ww=q e^\ww e^\uu.
\end{align}

\begin{figure}[H]
\centering
{\unitlength 0.011in
\begin{picture}(525,75)(-15,30)
\put(6,80){
\put(-11,0){\vector(1,0){23}}\put(0,-10){\vector(0,1){22}}
}
\multiput(81,80.5)(75,0){6}{
\put(-11,0){\vector(1,0){23}}\put(0,-10){\vector(0,1){22}}
}
\put(-74,0){
\put(60.5,77){$i$}\put(77.5,60){$j$}
\put(96,77){$a$}\put(77.5,96.5){$b$}
}
\put(61,77){0}\put(78,60){0}\put(96,77){0}\put(78,96.5){0}
\put(75,0){
\put(61,77){1}\put(78,60){1}\put(96,77){1}\put(78,96.5){1}
}
\put(150,0){
\put(61,77){1}\put(78,60){0}\put(96,77){1}\put(78,96.5){0}
}
\put(225,0){
\put(61,77){0}\put(78,60){1}\put(96,77){0}\put(78,96.5){1}
}
\put(300,0){
\put(61,77){0}\put(78,60){1}\put(96,77){1}\put(78,96.5){0}
}
\put(375,0){
\put(61,77){1}\put(78,60){0}\put(96,77){0}\put(78,96.5){1}
}
\put(78,40){
\put(-77,0){$\mathscr{L}^{ab}_{ij}$}
\put(0,0){$r$} \put(75,0){$s$} \put(144,0){$f e^\uu$}
\put(222,0){$g e^\uu$} \put(295,0){$e^{-\ww}$}
\put(343,0){$rs e^\ww\!+\! fg e^{2\uu+\ww}$}
}
\end{picture}
}
\caption{The operator $\mathscr{L}^{ab}_{ij}(r,s,f,g;q)$.}
\label{fig:6v}
\end{figure}
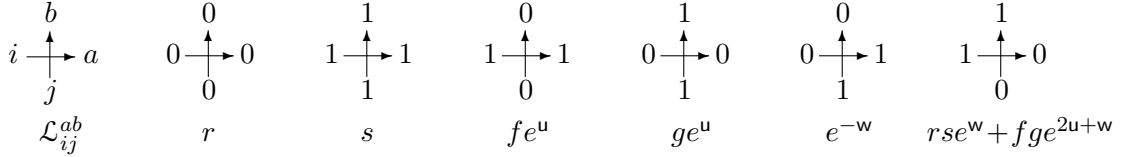

Let $V=\C v_0\oplus \C v_1$.
The local Boltzmann weights of the quantized six-vertex (Q6V) model are encoded in the
three-dimensional (3D) $L$-operator $\LL=\LL(r,s,f,g;q)$ defined by
\begin{subequations}\label{Ldef}
\begin{align}
&\LL(r,s,f,g;q) = \sum_{a,b,i,j=0,1} E_{ai}\otimes E_{bj} \otimes \LL^{ab}_{ij}
\in \mathrm{End}(V \otimes V)  \otimes \mathcal{W}(q),
\label{L1}
\\
&\LL^{ab}_{ij}=0\; \text{unless}\; a+b=i+j,
\label{L2}
\\
&\LL^{00}_{00} = r,\;\;  \LL^{11}_{11} = s,\;\;
\LL^{10}_{10} = f e^\uu,\;\;
\LL^{01}_{01} = g e^\uu,
\;\;  \LL^{10}_{01} = e^{-\ww}, \;\;
\LL^{01}_{10} = rs e^\ww+ fg e^{2\uu+\ww}.
\label{L3}
\end{align}
\end{subequations}
where $r,s,f,g$ are complex parameters and $E_{ij}v_k = \delta_{jk}v_i$.
The $q$-dependence of $\LL$ is through the relation \eqref{qcom}.
A graphical representation is given in Figure~\ref{fig:6v}.

We next describe informally how the local operator $\LL$ gives rise to the
global model; the precise definitions of the wiring diagram $G$, its
admissibility, and the associated transfer matrix $T_G(x,y)$ will be given
in Sections~\ref{ss:ad}--\ref{ss:ct}.

Consider a finite oriented diagram $G$ on a torus whose vertices are all four-valent.
Each edge carries a state $0$ or $1$.
According to the six configurations in Figure~\ref{fig:6v},
we assign to each vertex $v$ the local weight \eqref{L3},
with the parameters $(r,s,f,g)$ and the $q$-Weyl algebra generators
$e^\uu,e^\ww$ replaced by independent copies
$(r_v,s_v,f_v,g_v)$ and $e^{\uu_v},e^{\ww_v}$ satisfying \eqref{qwv}.
Summing products of the local weights over the internal
$\{0,1\}$-valued edge states produces operator-valued partition functions.
For a diagram $G$ with periodic boundary conditions in its two directions,
and with spectral parameters $x$ and $y$, this construction gives the
transfer matrix $T_G(x,y)$.
Formally, $T_G(x,y)$ has the same structure as a two-dimensional six-vertex
partition function, except that its local Boltzmann weights are
noncommutative.

This algebraic object has two complementary interpretations.
Upon choosing a representation of the $q$-Weyl algebra at each vertex,
$T_G(x,y)$ becomes an operator on the tensor product of the corresponding
representation spaces.
For admissible $G$, as defined in Section~\ref{ss:ad},
Theorem~\ref{thm:TT2} shows that the Laurent coefficients of $T_G(x,y)$
in $x$ and $y$ commute, and they may therefore be regarded as Hamiltonians
of a two-dimensional quantum integrable system.

On the other hand, after choosing bases in the local representation spaces,
the matrix elements of the local operators $\LL$ may be regarded as ordinary
scalar Boltzmann weights.
The $q$-Weyl degree of freedom then becomes a third local direction,
represented by the blue arrow in \eqref{LM} below, and $T_G(x,y)$
acquires the interpretation of a layer transfer matrix of a
three-dimensional classical vertex model.
This three-dimensional model is, however, generally highly anisotropic,
with two-state variables in the layer directions and a generally different
$q$-Weyl degree of freedom in the direction perpendicular to the layer.

The operator-valued formulation treats the two-dimensional quantum
and three-dimensional classical viewpoints simultaneously.
These two interpretations apply equally to generic $q$ and to the
root-of-unity specialization considered in Section~\ref{s:rt}.

\subsection{Auxiliary $M$-operators}\label{ss:am}

To establish the integrability of the model, we introduce an auxiliary
$M$-operator.  We first use the operator
$\MM=\MM(r',s',f',g';q)$ defined by
\begin{align}\label{eq:M}
\MM =
 \sum_{a,b,i,j=0,1} E_{ai}\otimes E_{bj} \otimes \MM^{ab}_{ij}
 :=\LL(r',s',f',g';-q)
 \in \mathrm{End}(V\otimes V)\otimes\mathcal{W}(-q).
\end{align}
The elements $\MM^{ab}_{ij}$ are obtained from \eqref{L3} by replacing
$r,s,f,g,e^{\uu},e^{\ww}$ with
$r',s',f',g',e^{\uu'},e^{\ww'}$, respectively, where the generators of
$\mathcal{W}(-q)$ satisfy
$e^{\uu'}e^{\ww'}=-q e^{\ww'}e^{\uu'}$.

The operators $\LL^{ab}_{ij}$ and $\MM^{ab}_{ij}$ are represented graphically as follows:
\begin{align}\label{LM}
\begin{tikzpicture}
\begin{scope}[>=latex]
\draw (-1.2,0.5) node{$\LL_{ij}^{ab}:$};
\draw[->] (0,0.5) node[left]{$i$}--(1,0.5) node[right]{$a$};
\draw[->] (0.5,0) node[below]{$j$}--(0.5,1) node[above]{$b$};
{\color{blue}
\draw[->] (1,0.9) --(0,0.1)[thick];
}
\end{scope}
\begin{scope}[>=latex,xshift=130]
\draw (-1.2,0.5) node{$\MM_{ij}^{ab}:$};
\draw[->] (0,0.5) node[left]{$i$}--(1,0.5) node[right]{$a$};
\draw[->] (0.5,0) node[below]{$j$}--(0.5,1) node[above]{$b$};
{\color{green}
\draw[->] (1,0.9) --(0,0.1)[thick];
}
\end{scope}
\end{tikzpicture}
.
\end{align}
The black lines are oriented wires, independently of their
$\{0,1\}$-valued edge states.  The nonplanar blue (resp.\ green) arrow
completes a right-handed frame with them and carries
$\LL^{ab}_{ij}\in\mathcal{W}(q)$
(resp.\ $\MM^{ab}_{ij}\in\mathcal{W}(-q)$).

Let $h$ be the number operator satisfying
\begin{align}\label{h}
 [h,e^\uu]=0,\quad [h,e^{-\ww}]=e^{-\ww},\quad
 [h,e^{\uu'}]=0,\quad [h,e^{-\ww'}]=e^{-\ww'},\quad
 hv_k=kv_k\,(k=0,1) .
\end{align}
We use the shorthand
$h_1=h\otimes1\otimes1$,
$h_2=1\otimes h\otimes1$, and
$h_3=1\otimes1\otimes h$.
Then both $\LL$ and $\MM$ satisfy the local conservation law
\begin{align}\label{cl}
 [x^{h_1}y^{h_2}(y/x)^{h_3},\LL]=0,\quad
 [x^{h_1}y^{h_2}(y/x)^{h_3},\MM]=0,
\end{align}
where $x$ and $y$ are arbitrary parameters.
According to \eqref{LM}, they are depicted as follows:
\begin{align}\label{LMh}
\begin{tikzpicture}[>=latex,baseline=0.5cm]
\draw (-1.35,0.5) node {$\LL:$};
\draw[->] (0,0.5)--(1,0.5);
\draw[fill=blue!20] (-0.15,0.5) circle[radius=0.15] node[left=2pt] {$x^{h}$};
\draw[->] (0.5,0)--(0.5,1);
\draw[fill=blue!20] (0.5,-0.15) circle[radius=0.15] node[below=2pt] {$y^{h}$};
{\color{blue}\draw[->,thick] (1,0.9)--(0,0.1);}
\draw[fill=blue!20] (1.11,1.01) circle[radius=0.15] node[above=2pt] {$(y/x)^{h}$};
\draw (1.75,0.5) node {$=$};
\begin{scope}[xshift=70]
\draw[->] (0,0.5)--(1,0.5);
\draw[fill=blue!20] (1.15,0.5) circle[radius=0.15] node[right=2pt] {$x^{h}$};
\draw[->] (0.5,0)--(0.5,1);
\draw[fill=blue!20] (0.5,1.15) circle[radius=0.15] node[above=2pt] {$y^{h}$};
{\color{blue}\draw[->,thick] (1,0.9)--(0,0.1);}
\draw[fill=blue!20] (-0.11,-0.01) circle[radius=0.15] node[below=2pt] {$(y/x)^{h}$};
\end{scope}
\end{tikzpicture}
\qquad
\begin{tikzpicture}[>=latex,baseline=0.5cm]
\draw (-1.35,0.5) node {$\MM:$};
\draw[->] (0,0.5)--(1,0.5);
\draw[fill=blue!20] (-0.15,0.5) circle[radius=0.15] node[left=2pt] {$x^{h}$};
\draw[->] (0.5,0)--(0.5,1);
\draw[fill=blue!20] (0.5,-0.15) circle[radius=0.15] node[below=2pt] {$y^{h}$};
{\color{green}\draw[->,thick] (1,0.9)--(0,0.1);}
\draw[fill=blue!20] (1.11,1.01) circle[radius=0.15] node[above=2pt] {$(y/x)^{h}$};
\draw (1.75,0.5) node {$=$};
\begin{scope}[xshift=70]
\draw[->] (0,0.5)--(1,0.5);
\draw[fill=blue!20] (1.15,0.5) circle[radius=0.15] node[right=2pt] {$x^{h}$};
\draw[->] (0.5,0)--(0.5,1);
\draw[fill=blue!20] (0.5,1.15) circle[radius=0.15] node[above=2pt] {$y^{h}$};
{\color{green}\draw[->,thick] (1,0.9)--(0,0.1);}
\draw[fill=blue!20] (-0.11,-0.01) circle[radius=0.15] node[below=2pt] {$(y/x)^{h}$};
\end{scope}
\end{tikzpicture}
\end{align}

For the applications below, we specialize this auxiliary operator to an
oscillator representation.
Set $p=-q^{-1}$, and let $\mathrm{Osc}(p)$ be the $p$-oscillator algebra
generated by $\mathbf{k},\mathbf{a}^{+}$, and $\mathbf{a}^{-}$ subject to
\begin{align}\label{oscr}
\mathbf{k}\,\mathbf{a}^{\pm}
=p^{\pm1}\mathbf{a}^{\pm}\mathbf{k},
\quad
\mathbf{a}^{+}\mathbf{a}^{-}
=1-\mathbf{k}^{2},
\quad
\mathbf{a}^{-}\mathbf{a}^{+}
=1-p^{2}\mathbf{k}^{2}.
\end{align}
We identify these generators with their standard Fock representation on
\begin{equation}\label{Vp}
\mathcal{V}_{+}
= \bigoplus_{m\in\Z_{\geq0}}\C(q)|m\rangle,
\end{equation}
given by
\begin{align}\label{posc}
\mathbf{k}|m\rangle
=p^{m}|m\rangle,
\quad
\mathbf{a}^{+}|m\rangle
=|m+1\rangle,
\quad
\mathbf{a}^{-}|m\rangle
=(1-p^{2m})|m-1\rangle .
\end{align}

\begin{figure}[H]
\centering
{\unitlength 0.011in
\begin{picture}(490,75)(-15,30)
\put(6,80){
\put(-11,0){\vector(1,0){23}}\put(0,-10){\vector(0,1){22}}
}
\multiput(81,80.5)(75,0){6}{
\put(-11,0){\vector(1,0){23}}\put(0,-10){\vector(0,1){22}}
}
\put(-74,0){
\put(60.5,77){$i$}\put(77.5,60){$j$}
\put(96,77){$a$}\put(77.5,96.5){$b$}
}
\put(61,77){0}\put(78,60){0}\put(96,77){0}\put(78,96.5){0}
\put(75,0){
\put(61,77){1}\put(78,60){1}\put(96,77){1}\put(78,96.5){1}
}
\put(150,0){
\put(61,77){1}\put(78,60){0}\put(96,77){1}\put(78,96.5){0}
}
\put(225,0){
\put(61,77){0}\put(78,60){1}\put(96,77){0}\put(78,96.5){1}
}
\put(300,0){
\put(61,77){0}\put(78,60){1}\put(96,77){1}\put(78,96.5){0}
}
\put(375,0){
\put(61,77){1}\put(78,60){0}\put(96,77){0}\put(78,96.5){1}
}
\put(78,40){
\put(-87,0){$M(\alpha)^{ab}_{ij}$}
\put(0,0){$1$} \put(75,0){$1$} \put(144,0){$\alpha \mathbf{k}$}
\put(205,0){$(q\alpha)^{-1}\mathbf{k}$}
\put(298,0){$\mathbf{a}^+$} \put(375,0){$\mathbf{a}^-$}
}
\end{picture}
}
\caption{The operators $M(\alpha)^{ab}_{ij}$.}
\label{fig:6vM}
\end{figure}
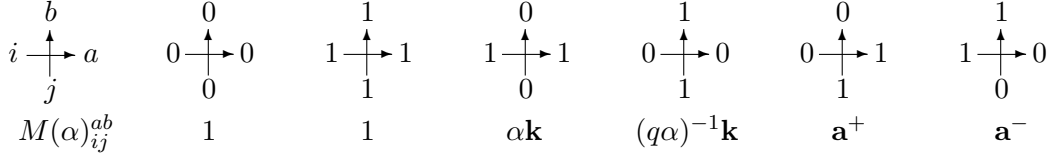

Specializing the operator $\MM$ in \eqref{eq:M} to
$\MM(1,1,\alpha,(q\alpha)^{-1};q)$,
we let it act on $V\otimes V\otimes\mathcal{V}_{+}$ by identifying
$e^{\uu'}$, $e^{-\ww'}$, and $e^{\ww'}(1-e^{2\uu'})$
with $\mathbf{k}$, $\mathbf{a}^+$, and $\mathbf{a}^-$, respectively,
through an embedding
$\mathrm{Osc}(p)\hookrightarrow\mathcal{W}(-q)$.
For the precise construction, see \cite[Sec.~2.2]{IKTY25}.
The resulting operator is denoted by
\begin{align}\label{M}
M=M(\alpha) =\sum_{a,b,i,j=0,1}
E_{ai}\otimes E_{bj}\otimes M(\alpha)^{ab}_{ij}
\in
\mathrm{End}(V\otimes V\otimes\mathcal{V}_{+}).
\end{align}
Its nonzero components $M(\alpha)^{ab}_{ij}$ satisfy
$a+b=i+j$ and are displayed in Figure~\ref{fig:6vM}.

From the definition in Figure \ref{fig:6vM} and \eqref{h} and \eqref{cl}, we have
\begin{align}
M(\alpha)^{ab}_{ij} 
&= \alpha^{a-j}M(1)^{ab}_{ij}= \alpha^{i-b}M(1)^{ab}_{ij},
\label{malp}
\\
y^a z^bM(\alpha)^{ab}_{ij}
&=\bigl(\tfrac{y}{z}\bigr)^h
M(\alpha)^{ab}_{ij}
\bigl(\tfrac{y}{z}\bigr)^{-h}
y^iz^j.
\label{Mh}
\end{align}

\subsection{Tetrahedron and inversion relations}\label{ss:ti}

Set
$ \mathbf{V}:=V^{\otimes4}\otimes\mathcal{W}(q)\otimes\mathcal{W}(-q)$.
For $1\leq i<j\leq4$ and $k=5,6$, let $\LL_{ijk}$ and 
$\MM_{ijk}$ denote the operators $\LL$ and $\MM$, respectively, acting on the
$(i,j,k)$th components of $\mathbf{V}$.
The integrability of the Q6V model is based on the
following local relations, whose consequences will be used throughout.

\begin{prop}[{\cite[Prop.~2.3, Lem.~2.5]{IKTY25}}]\label{pr:te4}
The operators $\LL=\LL(r,s,f,g;q)$ and $\MM=\MM(r',s',f',g';q)$ 
satisfy the following four tetrahedron equations:
\begin{align}
\text{\rm(o)}\quad&
 \MM_{126}\MM_{346}\LL_{135}\LL_{245}
 =
 \LL_{245}\LL_{135}\MM_{346}\MM_{126},
 \label{LM-o}\\
\text{\rm(h)}\quad&
 [\MM_{346}\LL_{315}\LL_{425}\MM_{126}]_{\ast6}
 =
 [\MM_{126}\LL_{425}\LL_{315}\MM_{346}]_{\ast6},
 \label{LM-h}\\
\text{\rm(v)}\quad&
 \MM_{126}\LL_{315}\LL_{425}\MM_{346}
 =
 \MM_{346}\LL_{425}\LL_{315}\MM_{126},
 \label{LM-v}\\
\text{\rm(t)}\quad&
 \LL_{135}\LL_{245}\MM_{126}\MM_{346}
 =
 \MM_{346}\MM_{126}\LL_{245}\LL_{135}.
 \label{LM-t}
\end{align}
Here \eqref{LM-h} requires $r'=s'$ and $g'=qf'$, whereas
\eqref{LM-v} requires $r'=s'$ and $g'=q^{-1}f'$.
The symbol $\ast6$ indicates that the multiplication order in 
$\mathcal{W}(-q)$ acting on the sixth component is reversed.
In addition, the following inversion relations hold:
\begin{align}
\text{\rm(I)}\quad&
\MM(s',r',qg',q^{-1}f';q)\MM(r',s',f',g';q) = r's' \mathrm{Id}, 
 \label{M-I}
 \\
\text{\rm(I')}\quad&
[\MM(r',s',f',g';q)\MM(s',r',q^{-1}g',qf';q)]_{\ast 3} = r's' \mathrm{Id},
 \label{M-I'}
\end{align}
where $\ast3$ indicates that the multiplication order in 
$\mathcal{W}(-q)$ acting on the third component is reversed.
\end{prop}

Recall that $M(\alpha)$ is obtained from
$\MM(1,1,\alpha,(q\alpha)^{-1};q)$.
Thus, \eqref{M-I} and \eqref{M-I'} imply the inversion relations
\begin{subequations}\label{MMinv}
\begin{align}
\text{(I)}&\quad M(\alpha^{-1})M(\alpha) = \mathrm{Id},
\label{MM1}
\\
\text{(I')}& \quad  [M(\alpha) M(q^{-2}\alpha^{-1})]_{\ast 3} = \mathrm{Id},
\label{MM2}
\end{align}
\end{subequations}
where $\ast 3$ means that the multiplication in
$\mathrm{End}(\mathcal{V}_+)$ is reversed.
Here (I) and (I') refer to the corresponding 2D projection diagrams in
\eqref{2D-I}.
The relations \eqref{MM1} and \eqref{MM2} are the special cases of 
\eqref{MMa1} for $(\alpha_1,\alpha_2)=(\alpha^{-1},\alpha)$ and 
\eqref{MMa2} for $(\alpha_1,\alpha_2)=(\alpha, q^{-2}\alpha^{-1})$,
respectively.

\begin{lem}\label{le:MM}
For arbitrary parameters $\alpha_1, \alpha_2$, the operators
$M(\alpha)^{ab}_{ij}$ in Figure \ref{fig:6vM} satisfy  
\begin{subequations}
\begin{align}
&\sum_{k,l=0,1}
(\alpha_1\alpha_2)^{l}
M(\alpha_1)^{ab}_{kl} M(\alpha_2)^{kl}_{ij} = (\alpha_1\alpha_2)^a\delta_{ai}\delta_{bj},
\label{MMa1}
\\
&\sum_{k,l=0,1}(q^2\alpha_1\alpha_2)^{l}
M(\alpha_1)^{kl}_{ij} M(\alpha_2)^{ab}_{kl} = (q^2\alpha_1\alpha_2)^i\delta_{ai}\delta_{bj}.
\label{MMa2}
\end{align}
\end{subequations}
\end{lem}

\begin{proof}
By \eqref{malp}, these relations reduce to the case $\alpha_1=\alpha_2=1$.
The claim simply follows from \eqref{MM1} and \eqref{MM2}.
 We now present the proof of \eqref{MMa2}. 
Setting $\alpha=1$ in \eqref{MM2} and applying \eqref{malp}, we obtain
\[
  {[M(1) M(q^{-2})]_{\ast 3}\,}^{ab}_{ij} 
  = q^{-2i} \sum_{k,l=0,1} q^{2l}  M(1)^{kl}_{ij} M(1)^{ab}_{kl}
= \delta_{ai} \delta_{bj}.
\]
The second equality is equivalent to \eqref{MMa2} 
in the case $\alpha_1 = \alpha_2 = 1$.
The relations may also be verified directly case by case.
For example, the LHS of 
\eqref{MMa2} for $(a,b,i,j)=(1,0,1,0)$ and $(0,1,1,0)$ evaluates to 
\begin{align*}
&M(1)^{10}_{10}M(1)^{10}_{10} + q^2M(1)^{01}_{10}M(1)^{10}_{01}
= \mathbf{k}^2 + q^2\mathbf{a}^- \mathbf{a}^+=q^2,
\\
&M(1)^{10}_{10}M(1)^{01}_{10} + q^2M(1)^{01}_{10}M(1)^{01}_{01}
 =\mathbf{k} \,\mathbf{a}^- + q^2\mathbf{a}^- q^{-1}\mathbf{k} = 0,
 \end{align*} 
 where the final equalities follow from \eqref{oscr} with $p=-q^{-1}$.
\end{proof}

\subsection{2D graphical representations}\label{ss:gr}
By composing the local vertex diagrams in \eqref{LM},
the relations \eqref{LM-o}--\eqref{M-I'} admit three-dimensional
graphical representations, which are given in Appendix~\ref{s:ap}.
Here we present their two-dimensional projections:
\begin{align}\label{2D-LM}
\begin{tikzpicture}
\begin{scope}[>=latex,xshift=0pt]
\draw (0,2.5) node[left] {(o)};
\draw[->] (0,1)--(2,1);
\draw[->] (1,0)--(1,2);
{\color{green}
\draw[->] (0.5,2)--(2,0.5)[thick];
}
\draw (2.5,1) node {$=$}; 
\draw[->] (3,1)--(5,1);
\draw[->] (4,0)--(4,2);
{\color{green}
\draw[->] (3,1.5)--(4.5,0)[thick];
}
\end{scope}
\begin{scope}[>=latex,xshift=200pt]
\draw (0,2.5) node[left] {(h)};
\draw[<-] (0,1)--(2,1);
\draw[->] (1,0)--(1,2);
{\color{green}
\draw[->] (0.5,2)--(2,0.5)[thick];
}
\draw (2.5,1) node {$=$}; 
\draw[<-] (3,1)--(5,1);
\draw[->] (4,0)--(4,2);
{\color{green}
\draw[->] (3,1.5)--(4.5,0)[thick];
}
\end{scope}
\begin{scope}[>=latex,yshift=-100]
\draw (0,2.5) node[left] {(v)};
\draw[->] (0,1)--(2,1);
\draw[<-] (1,0)--(1,2);
{\color{green}
\draw[->] (0.5,2)--(2,0.5)[thick];
}
\draw (2.5,1) node {$=$}; 
\draw[->] (3,1)--(5,1);
\draw[<-] (4,0)--(4,2);
{\color{green}
\draw[->] (3,1.5)--(4.5,0)[thick];
}
\end{scope}
\begin{scope}[>=latex,xshift=200pt,yshift=-100]
\draw (0,2.5) node[left] {(t)};
\draw[<-] (0,1)--(2,1);
\draw[<-] (1,0)--(1,2);
{\color{green}
\draw[->] (0.5,2)--(2,0.5)[thick];
}
\draw (2.5,1) node {$=$}; 
\draw[<-] (3,1)--(5,1);
\draw[<-] (4,0)--(4,2);
{\color{green}
\draw[->] (3,1.5)--(4.5,0)[thick];
}
\end{scope}
\end{tikzpicture}
\end{align}

\begin{align}\label{2D-I}
\begin{tikzpicture}
\begin{scope}[>=latex,xshift=0pt]
\draw (0,2.5) node[left] {(I)};
\draw[->] (0,2)--(2,0);
{\color{green}
\draw[->] (0,1.5)--(0.5,1.5) to [out=0, in=90] (1.5,0.5)--(1.5,0)[thick];
}
\draw (2.5,1) node {$=$}; 
\draw[->] (3,2)--(5,0);
{\color{green}
\draw[->] (2.7,1.7)--(4.7,-0.3)[thick];
}
\end{scope}
\begin{scope}[>=latex,xshift=200pt]
\draw (0,2.5) node[left] {(I')};
\draw[<-] (0,2)--(2,0);
{\color{green}
\draw[->] (0,1.5)--(0.5,1.5) to [out=0, in=90] (1.5,0.5)--(1.5,0)[thick];
}
\draw (2.5,1) node {$=$}; 
\draw[<-] (3,2)--(5,0);
{\color{green}
\draw[->] (2.7,1.7)--(4.7,-0.3)[thick];
}
\end{scope}
\end{tikzpicture}
\end{align}
We also have the inverses (iI) of (I) and the inverse (iI') of (I') depicted as follows.
\begin{align}\label{2D-iI}
\begin{tikzpicture}
\begin{scope}[>=latex,xshift=0pt]
\draw (0,2.5) node[left] {(iI)};
\draw[->] (0,2)--(2,0);
{\color{green}
\draw[->] (0.3,2.3)--(2.3,0.3)[thick];
}
\draw (2.5,1) node {$=$}; 
\draw[->] (3,2)--(5,0);
{\color{green}
\draw[->] (3.5,2)--(3.5,1.5) to [out=-90, in=180] (4.5,0.5)--(5,0.5)[thick];
}
\end{scope}
\begin{scope}[>=latex,xshift=200pt]
\draw (0,2.5) node[left] {(iI')};
\draw[<-] (0,2)--(2,0);
{\color{green}
\draw[->] (0.3,2.3)--(2.3,0.3)[thick];
}
\draw (2.5,1) node {$=$}; 
\draw[<-] (3,2)--(5,0);
{\color{green}
\draw[->] (3.5,2)--(3.5,1.5) to [out=-90, in=180] (4.5,0.5)--(5,0.5)[thick];
}
\end{scope}
\end{tikzpicture}
\end{align}
 
 The eight local moves in \eqref{2D-LM}--\eqref{2D-iI} 
 are grouped  as follows:
\begin{align}\label{types}
\text{type A: {\rm(v)}, {\rm(I)}, {\rm(iI)}},\qquad
\text{type B: {\rm(h)}, {\rm(I')}, {\rm(iI')}},\qquad
\text{type C: {\rm(o)}, {\rm(t)}}.
\end{align}
It follows from Proposition~\ref{pr:te4} that
type A requires $(r',g')=(s',q^{-1}f')$, type B requires
$(r',g')=(s',qf')$, whereas type C imposes no restriction on these parameters.
We will use these relations only after reducing
$\MM(r',s',f',g';q)$ to $M(\alpha)$, which is obtained from
$\MM(1,1,\alpha,(q\alpha)^{-1};q)$.
Consequently, type A requires $\alpha=\pm1$ and
type B requires $\alpha=\pm q^{-1}$, while type C imposes no restriction on
$\alpha$.

\subsection{Admissible diagrams}\label{ss:ad}

Consider a wiring diagram $G$ on a torus consisting of finitely many oriented
closed wires, with exactly two wires meeting at each vertex.
Choose a fundamental domain whose boundary contains no vertices.

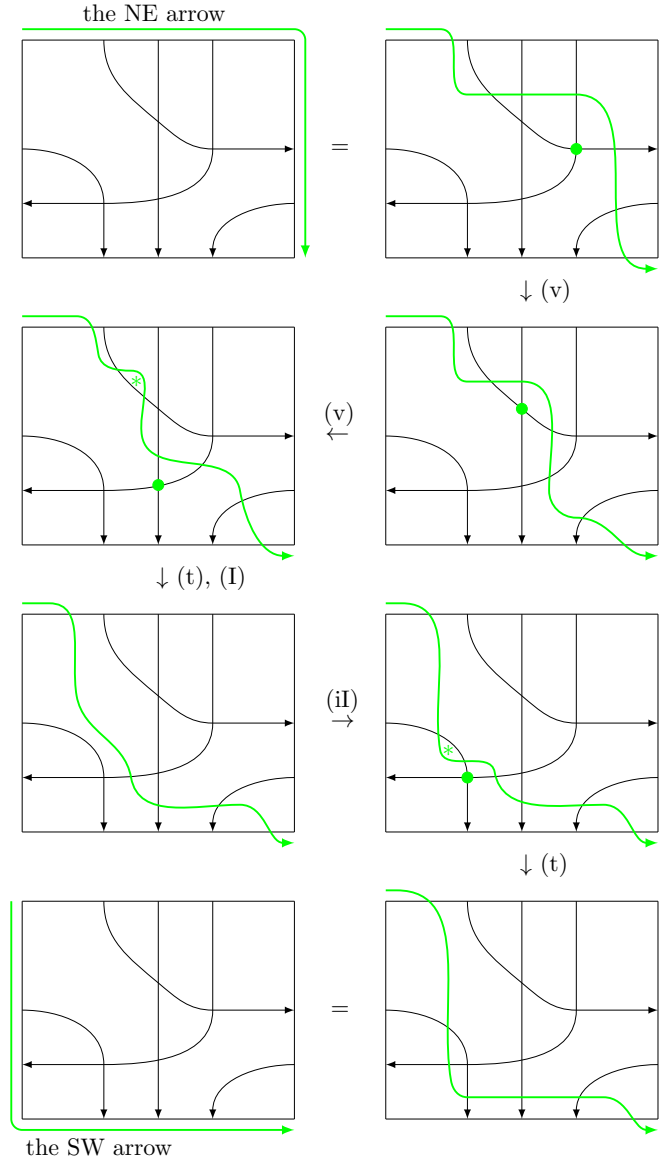
\begin{figure}[H]
\[
\scalebox{0.9}{%
\begin{tikzpicture}[scale=0.8, font=\small]
\begin{scope}[>=latex,xshift=0pt]
\draw [<-] (2.5,0)--(2.5,4);
\draw [<-] (1.5,0)--(1.5,1) to [out=90,in=0](0,2);
\draw [<-] (5,2)--(3.5,2) to [out=180,in=-40](2.5,2.5) to [out=140,in=-90](1.5,4);
\draw [<-] (3.5,0) to [out=90,in=180](5,1);
\draw [<-] (0,1)--(1.5,1) to [out=0,in=-90](3.5,2)--(3.5,4);
{\color{black}
\draw [-] (0,4)--(5,4);
\draw [-] (0,0)--(5,0);
\draw [-] (0,0)--(0,4);
\draw [-] (5,0)--(5,4);
}
{\color{green}
\draw [->] (0,4.2)--(5,4.2) to [out=0,in=90] (5.2,4)--(5.2,0)[thick];
}
\draw (2,4.5) node[right=-24pt] {\small the NE arrow};
\end{scope}
\begin{scope}[>=latex,xshift=190pt]
\draw [<-] (2.5,0)--(2.5,4);
\draw [<-] (1.5,0)--(1.5,1) to [out=90,in=0](0,2);
\draw [<-] (5,2)--(3.5,2) to [out=180,in=-40](2.5,2.5) to [out=140,in=-90](1.5,4);
\draw [<-] (3.5,0) to [out=90,in=180](5,1);
\draw [<-] (0,1)--(1.5,1) to [out=0,in=-90](3.5,2)--(3.5,4);
{\color{black}
\draw [-] (0,4)--(5,4);
\draw [-] (0,0)--(5,0);
\draw [-] (0,0)--(0,4);
\draw [-] (5,0)--(5,4);
}
{\color{green}
\draw [->] (0,4.2)--(1,4.2) to [out=0,in=180] (1.5,3)--(3.5,3) to [out=0,in=180] (5,-0.2)[thick];
\draw[fill] (3.5,2) circle[radius=0.1];
}
\draw (-0.5,2) node[left] {$=$};
\draw (2.3,-0.6) node[right] {$\downarrow$ ~(v)};
\end{scope}
\begin{scope}[>=latex,xshift=190pt, yshift=-150]
\draw [<-] (2.5,0)--(2.5,4);
\draw [<-] (1.5,0)--(1.5,1) to [out=90,in=0](0,2);
\draw [<-] (5,2)--(3.5,2) to [out=180,in=-40](2.5,2.5) to [out=140,in=-90](1.5,4);
\draw [<-] (3.5,0) to [out=90,in=180](5,1);
\draw [<-] (0,1)--(1.5,1) to [out=0,in=-90](3.5,2)--(3.5,4);
{\color{black}
\draw [-] (0,4)--(5,4);
\draw [-] (0,0)--(5,0);
\draw [-] (0,0)--(0,4);
\draw [-] (5,0)--(5,4);
}
{\color{green}
\draw [->] (0,4.2)--(1,4.2) to [out=0,in=180] (1.5,3)--(2.5,3) to [out=0,in=90] (3,1) to [out=-90,in=180] (3.5,0.5) to [out=0,in=180] (5,-0.2)[thick];
\draw[fill] (2.5,2.5) circle[radius=0.1];
}
\draw (-0.5,2) node[left] {$\leftarrow$};
\draw (-0.4,2.4) node[left] {(v)};

\end{scope}
\begin{scope}[>=latex,yshift=-150]
\draw [<-] (2.5,0)--(2.5,4);
\draw [<-] (1.5,0)--(1.5,1) to [out=90,in=0](0,2);
\draw [<-] (5,2)--(3.5,2) to [out=180,in=-40](2.5,2.5) to [out=140,in=-90](1.5,4);
\draw [<-] (3.5,0) to [out=90,in=180](5,1);
\draw [<-] (0,1)--(1.5,1) to [out=0,in=-90](3.5,2)--(3.5,4);
{\color{black}
\draw [-] (0,4)--(5,4);
\draw [-] (0,0)--(5,0);
\draw [-] (0,0)--(0,4);
\draw [-] (5,0)--(5,4);
}
{\color{green}
\draw [->] (0,4.2)--(1,4.2) to [out=0,in=100] (1.4,3.5) to [out=-80,in=180] (2,3.2) to [out=0,in=100] (2.2,2) to [out=-80,in=100] (4,1) to [out=-80,in=180] (5,-0.2)[thick];
\draw[fill] (2.5,1.1) circle[radius=0.1];
\draw (2.1,3) node {$\ast$};
}
\draw (2.3,-0.6) node[right] {$\downarrow$ ~(t),~(I)};
\end{scope}
\begin{scope}[>=latex,yshift=-300]
\draw [<-] (2.5,0)--(2.5,4);
\draw [<-] (1.5,0)--(1.5,1) to [out=90,in=0](0,2);
\draw [<-] (5,2)--(3.5,2) to [out=180,in=-40](2.5,2.5) to [out=140,in=-90](1.5,4);
\draw [<-] (3.5,0) to [out=90,in=180](5,1);
\draw [<-] (0,1)--(1.5,1) to [out=0,in=-90](3.5,2)--(3.5,4);
{\color{black}
\draw [-] (0,4)--(5,4);
\draw [-] (0,0)--(5,0);
\draw [-] (0,0)--(0,4);
\draw [-] (5,0)--(5,4);
}
{\color{green}
\draw [->] (0,4.2)--(0.5,4.2) to [out=0,in=100] (1,2.5) to [out=-80,in=100] (2,1) to [out=-80,in=180] (4,0.5) to [out=0,in=180] (5,-0.2)[thick];
}
\draw (5.5,2) node[right] {$\rightarrow$};
\draw (5.4,2.4) node[right] {(iI)};
\end{scope}
\begin{scope}[>=latex,xshift=190,yshift=-300]
\draw [<-] (2.5,0)--(2.5,4);
\draw [<-] (1.5,0)--(1.5,1) to [out=90,in=0](0,2);
\draw [<-] (5,2)--(3.5,2) to [out=180,in=-40](2.5,2.5) to [out=140,in=-90](1.5,4);
\draw [<-] (3.5,0) to [out=90,in=180](5,1);
\draw [<-] (0,1)--(1.5,1) to [out=0,in=-90](3.5,2)--(3.5,4);
{\color{black}
\draw [-] (0,4)--(5,4);
\draw [-] (0,0)--(5,0);
\draw [-] (0,0)--(0,4);
\draw [-] (5,0)--(5,4);
}
{\color{green}
\draw [->] (0,4.2)--(0.3,4.2) to [out=0,in=100] (1,1.5) to [out=-80,in=100] (2,1.1) to [out=-80,in=180] (4,0.5) to [out=0,in=180] (5,-0.2)[thick];
\draw[fill] (1.5,1) circle[radius=0.1];
\draw (1.15,1.5) node {$\ast$};
}
\draw (2.3,-0.6) node[right] {$\downarrow$ ~(t)};
\end{scope}
\begin{scope}[>=latex,xshift=190,yshift=-450]
\draw [<-] (2.5,0)--(2.5,4);
\draw [<-] (1.5,0)--(1.5,1) to [out=90,in=0](0,2);
\draw [<-] (5,2)--(3.5,2) to [out=180,in=-40](2.5,2.5) to [out=140,in=-90](1.5,4);
\draw [<-] (3.5,0) to [out=90,in=180](5,1);
\draw [<-] (0,1)--(1.5,1) to [out=0,in=-90](3.5,2)--(3.5,4);
{\color{black}
\draw [-] (0,4)--(5,4);
\draw [-] (0,0)--(5,0);
\draw [-] (0,0)--(0,4);
\draw [-] (5,0)--(5,4);
}
{\color{green}
\draw [->] (0,4.2)--(0.2,4.2) to [out=0,in=100] (1.2,0.7) to [out=-80,in=180] (1.5,0.4) -- (4,0.4) to [out=0,in=180] (5,-0.2)[thick];
}
\draw (-0.5,2) node[left] {$=$};
\end{scope}
\begin{scope}[>=latex,yshift=-450pt]
\draw [<-] (2.5,0)--(2.5,4);
\draw [<-] (1.5,0)--(1.5,1) to [out=90,in=0](0,2);
\draw [<-] (5,2)--(3.5,2) to [out=180,in=-40](2.5,2.5) to [out=140,in=-90](1.5,4);
\draw [<-] (3.5,0) to [out=90,in=180](5,1);
\draw [<-] (0,1)--(1.5,1) to [out=0,in=-90](3.5,2)--(3.5,4);
{\color{black}
\draw [-] (0,4)--(5,4);
\draw [-] (0,0)--(5,0);
\draw [-] (0,0)--(0,4);
\draw [-] (5,0)--(5,4);
}
{\color{green}
\draw [->] (-0.2,4)--(-0.2,0) to [out=-90,in=180] (0,-0.2)--(5,-0.2)[thick];
}
\draw (-0.1,-0.5) node[right] {\small the SW arrow};
\end{scope}
\end{tikzpicture}
}
\]
\caption{A wiring diagram illustrating the transformation from the NE arrow to the SW arrow.
The vertices through which the green arrow passes at the next step are indicated by green dots.
The regions marked by green asterisks indicate where inversion
relations are applied.
Only transformations of types A and C are used.}
\label{fig:adG}
\end{figure}

Along the boundary of this domain (the black rectangle in Figure~\ref{fig:adG}),
introduce an auxiliary green arrow running from the NW corner to the SE corner
via the NE corner; we call this the NE arrow.
Similarly, the arrow running from the NW corner to the SE corner via the SW
corner is called the SW arrow.
By successively applying the eight operations in
\eqref{2D-LM}--\eqref{2D-iI}, one can transform the NE arrow into the SW arrow.
See Figure~\ref{fig:adG} for an example.

\begin{definition}\label{def:adm}
The wiring diagram $G$ is \emph{admissible} if the NE arrow can be
transformed into the SW arrow by a finite sequence of the eight operations
listed in \eqref{types}, under the additional condition that moves of
types A and B do not occur together.
We say that $G$ requires type A moves or type B moves according to which of
them is needed in such a transformation.
The remaining possibility is that the transformation can be carried out
using type C moves alone.
\end{definition}

\begin{remark}
Call a face of a wiring diagram \emph{oriented} if its boundary is oriented.
It is known \cite[Prop.~3.9]{IKTY25} that a wiring diagram $G$ on a torus is non-admissible
if it contains an oriented face. 
The converse, which is equivalent to the assertion that acyclicity
implies admissibility, remains conjectural.
\end{remark}

\subsection{Partition functions and commuting transfer matrices}\label{ss:ct}

Suppose that $G$ is admissible, with $m$ and $n$ boundary crossings on the
east and north sides of the fundamental domain, respectively.
Let $K$ be the number of vertices of $G$, and let
$\mathcal{W}_K(q)$ denote the tensor product of $K$ independent copies of
$\mathcal{W}(q)$.

For arrays
$\mathbf{a} = (a_1,\ldots, a_m)$, $\mathbf{i} = (i_1,\ldots, i_m) \in \{0,1\}^m$
and
$\mathbf{b} = (b_1,\ldots, b_n)$, $\mathbf{j} = (j_1,\ldots, j_n) \in \{0,1\}^n$,
we define $T^{\mathbf{a}, \mathbf{b}}_{\mathbf{i},\mathbf{j}}
\in \mathcal{W}_K(q)$
to be the partition function obtained by fixing
$\mathbf{a}$, $\mathbf{b}$, $\mathbf{i}$, and $\mathbf{j}$ on the east, north,
west, and south boundaries of the fundamental domain, respectively, and
summing the products of the local weights over all $\{0,1\}$-valued internal
edge states.  Thus $\mathbf{b},\mathbf{a}$ specify the NE boundary data,
whereas $\mathbf{j},\mathbf{i}$ specify the SW boundary data.
The local Boltzmann weight at a vertex $v$ is
$\LL^{ab}_{ij}(r_v,s_v,f_v,g_v;q) \in \mathcal{W}(q)$,
as defined in \eqref{L3} and Figure~\ref{fig:6v}.
Here $(r_v,s_v,f_v,g_v)$ is a set of parameters that may be chosen
independently at each vertex $v$.
The corresponding copy of $\mathcal{W}(q)$ is generated by
$e^{\pm \uu_v}, e^{\pm \ww_v}$ satisfying
\begin{equation}\label{qwv}
e^{\uu_v}e^{\ww_{v'}}=q^{\delta_{v,v'}}e^{\ww_{v'}}e^{\uu_v}
\quad (v,v'=1,\ldots, K).
\end{equation}
The definition of
$T^{\mathbf{a}, \mathbf{b}}_{\mathbf{i},\mathbf{j}}$ itself does not require
the periodic boundary conditions
$\mathbf{a}=\mathbf{i}$ and $\mathbf{b}=\mathbf{j}$.

We introduce the integers
\begin{align}\label{ztc}
||\mathbf{a}|| = \sum_{k=1}^m s_x(k) a_k,
\quad
||\mathbf{b}|| = \sum_{k=1}^n s_y(k) b_k,
\quad
||\mathbf{i}|| = \sum_{k=1}^m s_x(k) i_k,
\quad
||\mathbf{j}|| = \sum_{k=1}^n s_y(k) j_k,
\end{align}
where $s_x(k)$ and $s_y(k)$ record the wire orientations at the boundary:
\begin{subequations}\label{sdef}
\begin{align}
 s_x(k)&=
 \begin{cases}
  1,&\!\!\text{if the $k$-th east-boundary wire segment is oriented outward},\\
 -1,&\!\!\text{if it is oriented inward},
 \end{cases}
 \quad \!(k=1,\ldots, m),
\label{sdx}
\\
 s_y(k) &=
 \begin{cases}
  1,&\!\!\text{if the $k$-th north-boundary wire segment is oriented outward},\\
 -1,&\!\!\text{if it is oriented inward},
 \end{cases}
 \quad\!\! (k=1,\ldots, n).
\label{sdy}
\end{align}
\end{subequations}
Here the boundary wires are numbered from top to bottom along
the vertical boundaries and from left to right along the horizontal
boundaries.
The conservation law \eqref{L2} implies
\begin{align}\label{tzero}
T_{\mathbf{i},\mathbf{j}}^{\mathbf{a},\mathbf{b}}=0 \;\;\text{unless}\;\;
||\mathbf{a}|| + ||\mathbf{b}|| = ||\mathbf{i}|| + ||\mathbf{j}||.
\end{align}

\begin{example}\label{ex:tabij}
Consider the following admissible diagram, for which $(m,n)=(2,3)$ and $K=4$:
\begin{equation}\label{Tabij}
\begin{tikzpicture}[baseline={(current bounding box.center)},scale=0.82,font=\small]
\begin{scope}[>=latex,xshift=0pt]

\node[xshift=-1.8cm] at (-2.5,1.82)
{$T_{\mathbf{i},\mathbf{j}}^{\mathbf{a},\mathbf{b}}=$};
\node[xshift=-1.0cm] at (-1.1,1.82)
{$\displaystyle\sum$};

\node[xshift=-1.0cm,below=1pt] at (-1.1,1.50)
{$\scriptstyle \text{inner edges}\in\{0,1\}$};

\def\TopY{3.55}
\def\MidY{1.15}

\coordinate (v1) at (1.046,\MidY);
\coordinate (v2) at (2.18,\MidY);
\coordinate (v3) at (3.55,\MidY);
\coordinate (v4) at (3.55,2.12);

\draw[->] (0,\MidY)--(5,\MidY);

\draw[->] (3.55,0)--(3.55,\TopY);

\draw[->]
(1.52,0)
.. controls (1.511,0.339) and (1.312,0.770) .. (v1)
.. controls (0.723,1.610) and (0.301,1.995) .. (0,2.05);

\draw[->]
(2.28,0)
to[out=92,in=-82] (v2)
to[out=102,in=-70] (1.27,\TopY);

\draw[->]
(5,2.08)
to[out=180,in=-6] (v4)
	to[out=174,in=-70] (2.28,\TopY);

\draw[blue,<-,thick]
(0.666,0.84) -- (1.426,1.46); 

\draw[blue,<-,thick]
(1.80,0.84) -- (2.56,1.46);   

\draw[blue,<-,thick]
(3.17,0.84) -- (3.93,1.46);   

\draw[blue,<-,thick]
(3.17,1.81) -- (3.93,2.43);   

{\color{black}
\draw (0,\TopY)--(5,\TopY);
\draw (0,0)--(5,0);
\draw (0,0)--(0,\TopY);
\draw (5,0)--(5,\TopY);
}

\node[above] at (1.27,\TopY) {$b_1$};
\node[above] at (2.28,\TopY) {$b_2$};
\node[above] at (3.55,\TopY) {$b_3$};

\node[below] at (1.52,0) {$j_1$};
\node[below] at (2.28,0) {$j_2$};
\node[below] at (3.55,0) {$j_3$};

\node[left] at (0,2.05) {$i_1$};
\node[left] at (0,\MidY) {$i_2$};

\node[right] at (5,2.08) {$a_1$};
\node[right] at (5,\MidY) {$a_2$};

\node at (1.05,0.75) {$1$};
\node at (2.46,0.89) {$2$};
\node at (3.86,0.89) {$3$};
\node at (3.79,1.83) {$4$};

\end{scope}
\end{tikzpicture}
\end{equation}
For this particular example, the wire orientations \eqref{sdef} are given by
$s_x(1)=-1$, $s_x(2)=1$ and
$s_y(1)=s_y(2)=s_y(3)=1$.
Hence
$||\mathbf{a}||=-a_1+a_2$, 
$||\mathbf{b}|| = b_1+b_2+b_3$,
$||\mathbf{i}||=-i_1+i_2$ and 
$||\mathbf{j}|| = j_1+j_2+j_3$.

For $\mathbf{a}=(1,1)$ and $\mathbf{i}=(0,0)$, nonzero 
$T_{\mathbf{i},\mathbf{j}}^{\mathbf{a},\mathbf{b}}$ are given as follows:
\begin{equation}\label{Tex}
\begin{array}{@{}l@{\quad}l@{}}
T^{11,001}_{00,001}
 =\e^{-\ww_3-\ww_4}r_1r_2,
&
T^{11,001}_{00,010}
 =\e^{\uu_3-\ww_2-\ww_4}f_3r_1,
\\[1mm]
T^{11,001}_{00,100}
 =\e^{\uu_2+\uu_3-\ww_1-\ww_4}f_2f_3,
&
T^{11,010}_{00,001}
 =\e^{\uu_4-\ww_3}g_4r_1r_2,
\\[1mm]
T^{11,010}_{00,010}
 =\e^{\uu_3+\uu_4-\ww_2}f_3g_4r_1,
&
T^{11,010}_{00,100}
 =\e^{\uu_2+\uu_3+\uu_4-\ww_1}f_2f_3g_4,
\\[1mm]
T^{11,011}_{00,011}
 =\e^{-\ww_2}r_1s_3s_4,
&
T^{11,011}_{00,101}
 =\e^{\uu_2-\ww_1}f_2s_3s_4,
\\[1mm]
T^{11,101}_{00,011}
 =\e^{\uu_2-\ww_3-\ww_4}g_2r_1,
&
T^{11,101}_{00,110}
 =\e^{\uu_3-\ww_1-\ww_4}f_3s_2,
\\[1mm]
\multicolumn{2}{@{}l@{}}{
T^{11,101}_{00,101}
 =
 \e^{2\uu_2-\ww_1+\ww_2-\ww_3-\ww_4}f_2g_2
 +\e^{-\ww_1+\ww_2-\ww_3-\ww_4}r_2s_2,
}
\\[1mm]
T^{11,110}_{00,011}
 =\e^{\uu_2+\uu_4-\ww_3}g_2g_4r_1,
&
T^{11,110}_{00,110}
 =\e^{\uu_3+\uu_4-\ww_1}f_3g_4s_2,
\\[1mm]
\multicolumn{2}{@{}l@{}}{
T^{11,110}_{00,101}
 =
 \e^{2\uu_2+\uu_4-\ww_1+\ww_2-\ww_3}f_2g_2g_4
 +\e^{\uu_4-\ww_1+\ww_2-\ww_3}g_4r_2s_2,
}
\\[1mm]
\multicolumn{2}{@{}l@{}}{
T^{11,111}_{00,111}
 =\e^{-\ww_1}s_2s_3s_4.
}
\end{array}
\end{equation}
\end{example}

In \eqref{Tabij}, both the sum over internal edge states and the short
blue arrows were displayed explicitly.  In subsequent graphical formulas,
however, the internal-state sum is always understood and its symbol will often
be omitted.  
Likewise, since the blue arrows indicate the third, 
quantum direction attached to the vertices rather than the planar boundary data, they may be suppressed when only the underlying planar wiring diagram is needed in the argument.

Define the monodromy matrix
$\mathcal{T}(x,y)=\mathcal{T}_G(x,y)
\in\mathrm{End}(V^{\otimes m+n})\otimes\mathcal{W}_K(q)$
with spectral parameters $x$ and $y$ by
\begin{equation}
\begin{split}
\mathcal{T}(x,y) = \sum_{\mathbf{a}, \mathbf{b}, \mathbf{i}, \mathbf{j}}
&x^{s_x(1)h}E_{a_1, i_1} \otimes \cdots \otimes x^{s_x(m)h}E_{a_m, i_m}
\\
\otimes &
\,y^{s_y(1)h}E_{b_1, j_1} \otimes \cdots \otimes y^{s_y(n)h}E_{b_n, j_n}
\otimes T^{\mathbf{a}, \mathbf{b}}_{\mathbf{i}, \mathbf{j}},
\end{split}
\end{equation}
where $h$ is defined in \eqref{h}, 
and the sum runs over $\mathbf{a}, \mathbf{i} \in \{0,1\}^m$ and
$\mathbf{b}, \mathbf{j} \in \{0,1\}^n$.
This is a natural generalization of $\LL$ in \eqref{L1},
which corresponds to the simplest case $m=n=1$ and $x=y=1$.
By definition, it maps a SW boundary configuration to an NE one,
assigning to each transition a ``Boltzmann weight'' in
$\mathcal{W}_K(q)[x^{\pm1},y^{\pm1}]$, according to
\begin{equation}
\begin{split}
&\mathcal{T}(x,y)
\bigl(
v_{i_1}\otimes\cdots\otimes v_{i_m}
\otimes
v_{j_1}\otimes\cdots\otimes v_{j_n}
\otimes w
\bigr)
\\
&\qquad =
\sum_{\mathbf a,\mathbf b}
x^{\|\mathbf a\|}
y^{\|\mathbf b\|}
v_{a_1}\otimes\cdots\otimes v_{a_m}
\otimes
v_{b_1}\otimes\cdots\otimes v_{b_n}
\otimes
T^{\mathbf a,\mathbf b}_{\mathbf i,\mathbf j}\,w,
\end{split}
\end{equation}
where \(w\in\mathcal W_K(q)\).

We regard $V^{\otimes m+n}$ as the auxiliary space of $\mathcal{T}(x,y)$.
The transfer matrix $T(x,y)=T_G(x,y)$ is defined by taking the trace
of the monodromy matrix over the auxiliary space:
\begin{equation}\label{TGg}
 T(x,y)=\Tr_{V^{\otimes m+n}}\bigl(\mathcal{T}(x,y)\bigr)=
 \sum_{\substack{\mathbf{i}\in\{0,1\}^m\\
                  \mathbf{j}\in\{0,1\}^n}}
 T^{\mathbf{i},\mathbf{j}}_{\mathbf{i},\mathbf{j}}\,
 x^{\|\mathbf{i}\|}y^{\|\mathbf{j}\|}
 \in\mathcal{W}_K(q).
\end{equation}
This realizes explicitly the two interpretations described in
Section~\ref{ss:q6}.  After choosing a representation of each local copy
of $\mathcal{W}(q)$ on $\mathcal{H}_v$, $T(x,y)$ acts on the quantum space
$\mathcal{H}_1\otimes\cdots\otimes\mathcal{H}_K$.
From the three-dimensional classical viewpoint, the same operator is the
transfer matrix of the layer represented by $G$, acting in the direction
perpendicular to the layer, along the blue arrows (cf. \eqref{Tabij}).
The spectral parameters $x$ and $y$ play the role of ``boundary magnetic fields''
associated with the two fundamental directions of the torus.
The main result of \cite{IKTY25} is the following:
\begin{thm}[{\cite[Thm.~3.4]{IKTY25}}]\label{thm:TT2}
For any admissible diagram $G$,
the associated transfer matrices $T(x,y)=T_G(x,y)$
form a two-parameter commuting family:
\begin{align}
 [T(x,y),T(x',y')]=0.
\end{align}
\end{thm}
Thus, in the two-dimensional quantum interpretation,
$T(x,y)$ is a generating Laurent polynomial for mutually commuting
``Hamiltonians''.

\subsection{The $MTT$ relation}\label{ss:mt}

\begin{figure}[H]
\centering
\includegraphics[page=1,trim=65bp 320bp 65bp 315bp,clip,width=\textwidth]{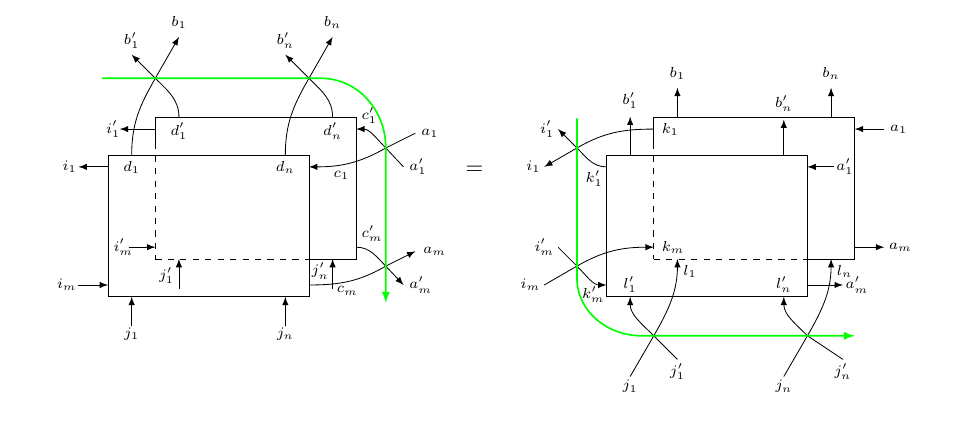}
\caption{Graphical representation of the $MTT$ relation \eqref{mtt}.
The green arrow is carried by the auxiliary $M$-operator \eqref{M}.  
Each rectangle represents a fixed-boundary layer amplitude of the quantized
six-vertex model; the blue arrows in the quantum direction, which are
concatenated when these layer amplitudes are multiplied, are suppressed
in the figure.
The displayed boundary-wire orientations correspond to the concrete case
$s_x(1)=-1$, $s_x(m)=1$ and $s_y(1)=s_y(n)=1$ of \eqref{sdef};
the omitted boundary orientations are read by the same convention.}
\label{fig:mtt}
\end{figure}

Theorem \ref{thm:TT2} follows from the ``$MTT$ relation'' depicted in Figure \ref{fig:mtt}
\begin{equation}\label{mtt}
\begin{split}
&\sum_{\mathbf{c}, \mathbf{c}',\mathbf{d}, \mathbf{d}'}
\tilde{M}(\alpha_\ast)^{a_m, a'_m}_{c_m,c'_m}\cdots 
\tilde{M}(\alpha_\ast)^{a_1, a'_1}_{c_1,c'_1}
\tilde{M}(\alpha_\ast)^{b_n, b'_n}_{d_n,d'_n}\cdots 
\tilde{M}(\alpha_\ast)^{b_1, b'_1}_{d_1,d'_1}
\otimes 
T^{\mathbf{c}, \mathbf{d}}_{\mathbf{i},\mathbf{j}}
T^{\mathbf{c}', \mathbf{d}'}_{\mathbf{i}',\mathbf{j}'}
\\
&=\sum_{\mathbf{k}, \mathbf{k'},  \mathbf{l}, \mathbf{l'}}
\tilde{M}(\alpha_\ast)^{l_n, l'_n}_{j_n,j'_n}\cdots 
\tilde{M}(\alpha_\ast)^{l_1, l'_1}_{j_1,j'_1}
\tilde{M}(\alpha_\ast)^{k_m, k'_m}_{i_m, i'_m}\cdots 
\tilde{M}(\alpha_\ast)^{k_1, k'_1}_{i_1, i'_1}
\otimes 
T^{\mathbf{a}', \mathbf{b}'}_{\mathbf{k}',\mathbf{l}'}
T^{\mathbf{a}, \mathbf{b}}_{\mathbf{k},\mathbf{l}},
\end{split}
\end{equation}
where 
$\mathbf{a}, \mathbf{a}', \mathbf{i}, \mathbf{i'} \in \{0,1\}^m$ and 
$\mathbf{b}, \mathbf{b}', \mathbf{j}, \mathbf{j'} \in \{0,1\}^n$ are fixed.
The sums run over 
$\mathbf{c}, \mathbf{c}', \mathbf{k}, \mathbf{k'} \in \{0,1\}^m$ and 
$\mathbf{d}, \mathbf{d}', \mathbf{l}, \mathbf{l'} \in \{0,1\}^n$.
Here we set 
\begin{subequations}\label{mta}
\begin{align}
&(\tilde{M}(\alpha)^{a_t, a'_t}_{c_t,c'_t},\tilde{M}(\alpha)^{k_t, k'_t}_{i_t,i'_t} )
= \begin{cases} 
(M(\alpha)^{a_t,a'_t}_{c_t, c'_t}, M(\alpha)^{k_t, k'_t}_{i_t,i'_t}) & s_x(t) = 1, \\
(M(\alpha)_{a_t,a'_t}^{c_t, c'_t}, M(\alpha)_{k_t, k'_t}^{i_t,i'_t})  & s_x(t) = -1,
\end{cases}\quad (t=1,\ldots, m),
\label{acki}\\
&(\tilde{M}(\alpha)^{b_t, b'_t}_{d_t,d'_t},\tilde{M}(\alpha)^{l_t, l'_t}_{j_t,j'_t} )
= \begin{cases} 
(M(\alpha)^{b_t,b'_t}_{d_t, d'_t}, M(\alpha)^{l_t, l'_t}_{j_t,j'_t}) & s_y(t) = 1, \\
(M(\alpha)_{b_t,b'_t}^{d_t, d'_t}, M(\alpha)_{l_t, l'_t}^{j_t,j'_t})  & s_y(t) = -1,
\end{cases}\quad (t=1,\ldots, n).
\label{bdlj}
 \end{align}
 \end{subequations}
in terms of $M(\alpha)^{ab}_{ij}$ defined in Figure \ref{fig:6vM} for general $\alpha$.
Thus, depending on the orientation of the corresponding boundary wire segments,
$\tilde{M}$ is either $M$ in \eqref{M} or its transpose with respect to the
$\mathrm{End}(V\otimes V)$ component.
The value $\alpha_\ast$ of the parameter $\alpha$ in \eqref{mtt}
is determined according to Definition \ref{def:adm} as follows:
\begin{equation}\label{alp}
\alpha_\ast = \begin{cases} 
\pm 1 & \text{if $G$ requires type A moves}, \\
\pm q^{-1} & \text{if $G$ requires type B moves},\\
\text{arbitrary} & \text{otherwise}.
\end{cases}
\end{equation}
The last possibility corresponds to the special case mentioned at the end of
Definition \ref{def:adm}, in which type C moves alone suffice to transform
the NE arrow into the SW arrow.

The $MTT$ relation \eqref{mtt} is an identity in
$\mathrm{End}(\mathcal{V}_+)\otimes \mathcal{W}_K(q)$ and is obtained by an
appropriate combination of the tetrahedron and inversion relations in
Proposition \ref{pr:te4}.

Later, we will use the property
\begin{equation}\label{Mth}
\begin{split}
y^{s_y(t)b_t}z^{s_y(t)b'_t}\tilde{M}(\alpha)^{b_t,b'_t}_{d_t,d'_t}
&=\bigl(\tfrac{y}{z}\bigr)^h \tilde{M}(\alpha)^{b_t,b'_t}_{d_t,d'_t}\bigl(\tfrac{y}{z}\bigr)^{-h} 
y^{s_y(t)d_t}z^{s_y(t)d'_t},
\\
y^{s_y(t)l_t}z^{s_y(t)l'_t}\tilde{M}(\alpha)^{l_t,l'_t}_{j_t,j'_t}
&=\bigl(\tfrac{y}{z}\bigr)^h \tilde{M}(\alpha)^{l_t,l'_t}_{j_t,j'_t}\bigl(\tfrac{y}{z}\bigr)^{-h} 
y^{s_y(t)j_t}z^{s_y(t)j'_t}
\end{split}
\end{equation}
for $t=1,\ldots, n$,
which follows from \eqref{bdlj} and \eqref{Mh} for general $\alpha$.

Using $\alpha_\ast$ in \eqref{alp}, define $\alpha'_\ast$ by
\begin{equation}\label{alal}
\alpha_\ast'\alpha_\ast = \begin{cases}
q^{-2} & \text{if $s_y(n)=1$},
\\
1 & \text{if $s_y(n)=-1$}.
\end{cases}
\end{equation}
The resulting choices of $(\alpha_\ast, \alpha'_\ast)$
are summarized in Table \ref{tab:alp}, together with the corresponding
inversion relations \eqref{MMinv}.

\begin{table}[ht]
\centering
{\renewcommand{\arraystretch}{1.20}
$\begin{array}{c|c|c|c}
 & \text{type A} & \text{type B} & \text{type C}
\\
\hline
s_y(n)=1
&
\begin{gathered}
\rule{0pt}{2.8ex}
[M(\alpha_\ast)M(\alpha'_\ast)]_{\ast 3}=\mathrm{Id}
\\
(\alpha_\ast,\alpha'_\ast)=(\pm1,\pm q^{-2})
\end{gathered}
&
\begin{gathered}
\rule{0pt}{2.8ex}
[M(\alpha_\ast)M(\alpha'_\ast)]_{\ast 3}=\mathrm{Id}
\\
(\alpha_\ast,\alpha'_\ast)=(\pm q^{-1},\pm q^{-1})
\end{gathered}
&
\begin{gathered}
\rule{0pt}{2.8ex}
[M(\alpha_\ast)M(\alpha'_\ast)]_{\ast 3}=\mathrm{Id}
\\
(\alpha_\ast,\alpha'_\ast)=(\alpha,(q^2\alpha)^{-1})
\end{gathered}
\\
\hline
s_y(n)=-1
&
\begin{gathered}
\rule{0pt}{2.8ex}
M(\alpha'_\ast)M(\alpha_\ast)=\mathrm{Id}
\\
(\alpha_\ast,\alpha'_\ast)=(\pm1,\pm 1)
\end{gathered}
&
\begin{gathered}
\rule{0pt}{2.8ex}
M(\alpha'_\ast)M(\alpha_\ast)=\mathrm{Id}
\\
(\alpha_\ast,\alpha'_\ast)=(\pm q^{-1},\pm q)
\end{gathered}
&
\begin{gathered}
\rule{0pt}{2.8ex}
M(\alpha'_\ast)M(\alpha_\ast)=\mathrm{Id}
\\
(\alpha_\ast,\alpha'_\ast)=(\alpha, \alpha^{-1})
\end{gathered}
\end{array}$}

\vspace{3mm}
\caption{Choices of
$(\alpha_\ast,\alpha'_\ast)$ implied by \eqref{alp} and \eqref{alal},
together with the relevant inversion relations.
For $s_y(n)=1$ and $s_y(n)=-1$, these are \eqref{MM2} and \eqref{MM1},
respectively, with $\alpha=\alpha_\ast$.
Here types A and B mean that $G$ requires type A and type B moves,
respectively, in the sense of Definition \ref{def:adm}.
Type C refers to the remaining case described there, in which
$\alpha_\ast=\alpha$ may be chosen arbitrarily.}
\label{tab:alp}
\end{table}

\begin{lem}\label{le:minv}
The operator $\tilde{M}$ in \eqref{bdlj} with $t=n$
satisfies the inversion relations
\begin{subequations}
\begin{align}
&\sum_{l,l'=0,1}\tilde{M}(\alpha'_\ast)^{l,l'}_{j,j'}\tilde{M}(\alpha_\ast)^{o,o'}_{l,l'}
= \delta_{o, j} \delta_{o', j'} \quad (j,j',o,o' \in \{0,1\}),
\label{invu}
\\
&\sum_{j,j'=0,1}q^{-2s_y(n)j'}\tilde{M}(\alpha'_\ast)^{l,l'}_{j,j'}
\tilde{M}(\alpha_\ast)^{j,j'}_{d,d'} = q^{-2s_y(n)l}\delta_{l,d}\delta_{l',d'}
\quad (l,l',d,d' \in \{0,1\}).
\label{qmm}  
\end{align}
\end{subequations}
\end{lem}

\begin{proof}
For $s_y(n)=1$ and $s_y(n)=-1$, 
the relation \eqref{invu} follows from 
\eqref{MMa2} and \eqref{MMa1} 
by setting $(\alpha_1,\alpha_2)=(\alpha'_\ast,\alpha_\ast)$, respectively.
Similarly, for $s_y(n)=1$ and $s_y(n)=-1$, 
the relation \eqref{qmm} follows from 
\eqref{MMa1} and \eqref{MMa2}
by setting $(\alpha_1,\alpha_2)=(\alpha'_\ast,\alpha_\ast)$, respectively.
\end{proof}

The last important ingredient involving $\tilde{M}$ is the quantum $R$-matrix
$R(\zeta)\in \mathrm{End}(V^{\otimes n-1} \otimes V^{\otimes n-1})$ 
constructed in matrix product form as 
\begin{subequations}\label{Rmat}
\begin{align}
&R(\zeta)= \sum_{\hat{\mathbf{b}}, \hat{\mathbf{b}}', 
\hat{\mathbf{j}}', \hat{\mathbf{j}}\in \{0,1\}^{n-1}}
R(\zeta)^{\hat{\mathbf{b}}, \hat{\mathbf{b}}'}_{\hat{\mathbf{j}}, \hat{\mathbf{j}}'}
E_{b_1,j_1}\otimes \cdots \otimes E_{b_{n-1},j_{n-1}} \otimes 
E_{b'_1,j'_1}\otimes \cdots \otimes E_{b'_{n-1},j'_{n-1}},
\label{Rmat1}
\\
&R(\zeta)^{\hat{\mathbf{b}}, \hat{\mathbf{b}}'}_{\hat{\mathbf{j}}, \hat{\mathbf{j}}'}
= \mathrm{Tr}_{\mathcal{V}_+}\Bigl(\zeta^h
\tilde{M}(\alpha_\ast)^{b_{n-1}, b'_{n-1}}_{j_{n-1}, j'_{n-1}}
\cdots \tilde{M}(\alpha_\ast)^{b_1, b'_1}_{j_1, j'_1}\Bigr),
\label{Rmat2}
\end{align} 
\end{subequations}
where $\mathcal{V}_+$ is defined in \eqref{Vp}.
It follows from \cite[Lem.~2.2]{IKTY25} and the 
last paragraph of the proof of \cite[Thm.~3.4]{IKTY25}
that $R(\zeta)$ is invertible for generic values of the spectral parameter $\zeta$.
For further details, see \cite[Chap.~11]{K22}.

\subsection{Monodromy and transfer matrices with mixed boundary condition}\label{ss:mb}
Let us introduce the monodromy matrix 
$\mathcal{T}(\mathbf{i}, \mathbf{a}|y)=\mathcal{T}_G(\mathbf{i}, \mathbf{a}|y)$,
which is labeled by $\mathbf{a}, \mathbf{i} \in \{0,1\}^m$ 
and depends on a single spectral parameter:
\begin{align}
\mathcal{T}({\mathbf{i}}, {\mathbf{a}}|y)
= \sum_{\mathbf{b}, \mathbf{j} \in \{0,1\}^n}
y^{s_y(1)h}E_{b_1, j_1} \otimes \cdots \otimes y^{s_y(n)h}E_{b_n, j_n}
\otimes T^{\mathbf{a}, \mathbf{b}}_{\mathbf{i}, \mathbf{j}}
\in \mathrm{End}(V^{\otimes n}) \otimes \mathcal{W}_K(q),
\label{ctai}
\end{align}
An example with $(m,n)=(2,3)$, $\mathbf{a}=(a_1,a_2)$, 
$\mathbf{i}=(i_1,i_2)$, $(s_x(1),s_x(2)) = (-1,-1)$, and 
$(s_y(1), s_y(2), s_y(3))=(1,1,-1)$ is shown below:
\begin{equation}
\begin{tikzpicture}[baseline={(current bounding box.center)},scale=0.82,font=\small]
\begin{scope}[>=latex,xshift=0pt]

\node[xshift=-2.2cm] at (-1,1.82)
{$\displaystyle
\mathcal{T}(\mathbf{i},\mathbf{a}\mid y)
=
\sum\limits_{\text{inner edges}\in\{0,1\}}
$};

\def\TopY{3.55}
\def\MidY{1.15}

\coordinate (v1) at (1.046,\MidY);
\coordinate (v2) at (2.18,\MidY);
\coordinate (v3) at (3.55,\MidY);
\coordinate (v4) at (3.55,2.12);

\draw[<-] (0,\MidY)--(5,\MidY);

\draw[<-] (3.55,-0.5)--(3.55,\TopY);

\draw[->]
(1.52,-0.5)
.. controls (1.511,0.339) and (1.312,0.770) .. (v1)
.. controls (0.723,1.610) and (0.301,1.995) .. (0,2.05);

\draw[->]
(2.28,-0.5)
to[out=92,in=-82] (v2)
to[out=102,in=-90] (1.27,\TopY);

\draw[->]
(5,2.08)
to[out=180,in=-6] (v4)
	to[out=174,in=-90] (2.28,\TopY);

\draw[<-] (1.27,4.7) -- (1.27,\TopY);
\draw[fill=blue!15] (1.27,4.1) circle[radius=0.15] node[left=1pt] {$y^{h}$}; 
\draw[<-] (2.28,4.7)-- (2.28,\TopY);
\draw[fill=blue!15] (2.28,4.1) circle[radius=0.15] node[left=1pt] {$y^{h}$}; 
\draw[->] (3.55,4.7)-- (3.55,\TopY);
\draw[fill=blue!15] (3.55,4.1) circle[radius=0.15] node[left=1pt] {$y^{-h}$};

\draw[blue,<-,thick]
(0.666,0.84) -- (1.426,1.46); 

\draw[blue,<-,thick]
(1.80,0.84) -- (2.56,1.46);   

\draw[blue,<-,thick]
(3.17,0.84) -- (3.93,1.46);   

\draw[blue,<-,thick]
(3.17,1.81) -- (3.93,2.43);   

{\color{black}
\draw (0,\TopY)--(5,\TopY);
\draw (0,0)--(5,0);
\draw (0,0)--(0,\TopY);
\draw (5,0)--(5,\TopY);
}



\node[left] at (0,2.05) {$i_1$};
\node[left] at (0,\MidY) {$i_2$};

\node[right] at (5,2.08) {$a_1$};
\node[right] at (5,\MidY) {$a_2$};

\node at (1.05,0.75) {$1$};
\node at (2.46,0.89) {$2$};
\node at (3.86,0.89) {$3$};
\node at (3.79,1.83) {$4$};

\end{scope}
\end{tikzpicture}
\label{iiaa}
\end{equation}

Using the monodromy matrix, we define the transfer matrix 
$T({\mathbf{i}}, {\mathbf{a}}|y)=T_G({\mathbf{i}}, {\mathbf{a}}|y)$ by
\begin{equation}
T({\mathbf{i}}, {\mathbf{a}}|y)
= \mathrm{Tr}_{V^{\otimes n}}(\mathcal{T}({\mathbf{i}}, {\mathbf{a}}|y) )
=
\sum_{\mathbf{b} \in \{0,1\}^n}
T^{\mathbf{a}, \mathbf{b}}_{\mathbf{i}, \mathbf{b}}\, y^{||\mathbf{b}||}
\in \mathcal{W}_K(q).
\label{tai}
\end{equation}
This corresponds to a mixed boundary condition:
periodic in the north-south direction,
while the east and west boundaries are fixed to 
$\mathbf{a}=(a_1,\ldots, a_m)$ and 
$\mathbf{i}=(i_1,\ldots, i_m) \in \{0,1\}^m$, respectively.
In view of \eqref{tzero}, we have 
$T({\mathbf{i}}, {\mathbf{a}}|y)=0$ unless $||\mathbf{a}||=||\mathbf{i}||$.
It is related to the full transfer matrix $T(x,y)$ in \eqref{TGg} by
\begin{align}\label{taa}
T(x,y) = \sum_{\mathbf{a} \in \{0,1\}^m}
x^{||\mathbf{a}||}T({\mathbf{a}}, {\mathbf{a}}|y).
\end{align}

\begin{prop}\label{pr:tai}
For any $\mathbf{a}, \mathbf{i}$ such that $||\mathbf{a}|| = ||\mathbf{i}||$,
the transfer matrices \eqref{tai} form a commuting family:
\begin{equation}\label{ttz}
[T({\mathbf{i}}, {\mathbf{a}}|y), 
T({\mathbf{i}}, {\mathbf{a}}|z)]=0.
\end{equation}
\end{prop}
This is proved by the same method as in \cite[Thm.~3.4]{IKTY25},
using the $MTT$ relation \eqref{mtt} with $\mathbf{a}'=\mathbf{a}$ and 
$\mathbf{i}'=\mathbf{i}$, together with the identities 
$M^{a,a}_{i,j}=\delta_{i,a}\delta_{j,a}$ 
and $M^{a,b}_{i,i}=\delta_{a,i}\delta_{b,i}$.

\begin{example}\label{ex:tabij2}
For Example \ref{ex:tabij}, $T(00,11|y)$ is given by 
\begin{equation*}
T(00,11|y)=y(
T^{11,001}_{00,001}
+T^{11,010}_{00,010})
+y^2(
T^{11,011}_{00,011}
+T^{11,101}_{00,101}
+T^{11,110}_{00,110}
)+y^3T^{11,111}_{00,111}.
\end{equation*}
\end{example}

\subsection{Refined transfer matrices and a functional relation}\label{ss:rf}

From now on, we write simply
\begin{equation}\label{ryak}
\mathcal{T}(y) = \mathcal{T}(\mathbf{i},\mathbf{a}|y),\qquad 
T(y) = T(\mathbf{i},\mathbf{a}|y)
\end{equation}
with fixed $\mathbf{a}, \mathbf{i} \in \{0,1\}^m$ satisfying 
$||\mathbf{a}|| = ||\mathbf{i}||$.
For $\beta, \gamma \in \{0,1\}$, 
we introduce a further refinement of the monodromy and transfer matrices by
\begin{align}
\mathcal{T}^\beta_\gamma(y) 
&= \!\!\sum_{\hat{\mathbf{b}}, \hat{\mathbf{j}} \in \{0,1\}^{n-1}}
y^{s_y(1)h}E_{b_1, j_1} \otimes \cdots \otimes y^{s_y(n-1)h}E_{b_{n-1}, j_{n-1}}
\otimes T^{\mathbf{a}, \hat{\mathbf{b}}\beta}_{\mathbf{i}, \hat{\mathbf{j}}\gamma}
\in \mathrm{End}(V^{\otimes n-1}) \otimes \mathcal{W}_K(q),
\label{tbg}
\\
T_\beta(y) &= 
\mathrm{Tr}_{V^{\otimes n-1}}(\mathcal{T}^\beta_\beta(y))
=
\sum_{\hat{\mathbf{b}}\in \{0,1\}^{n-1}}
T^{\mathbf{a}, \hat{\mathbf{b}}\beta}_{\mathbf{i},\, \hat{\mathbf{b}}\beta}
\, y^{||\hat{\mathbf{b}}||},
\label{T01}
\end{align}
where $\hat{\mathbf{b}}\beta=(b_1,\ldots, b_{n-1},\beta)$ for 
 $\hat{\mathbf{b}}=(b_1,\ldots, b_{n-1})  \in \{0,1\}^{n-1}$ 
 and $\beta \in \{0,1\}$. 
The array $\hat{\mathbf{j}}\gamma$ is defined similarly.
The transfer matrix 
$T_\beta(y)$ corresponds to imposing an additional fixed boundary condition
on the rightmost wire segments crossing the north and south boundaries of $G$.
By isolating the $n$th component in \eqref{ztc}, we have
$||\mathbf{b}|| = ||\hat{\mathbf{b}}|| + s_y(n)b_n$
for $\mathbf{b}=(b_1,\ldots, b_{n-1},b_n)$.
Thus, these definitions lead to the decomposition
\begin{align}
T(y) &= T_0(y) + y^{s_y(n)}T_1(y).
\label{Tdec}
\end{align}
By the same argument as in Proposition \ref{pr:tai}, one can show that
\begin{equation}\label{t01c}
[T_0(y), T_0(z)] =0,
\quad  [T_1(y), T_1(z)]=0.
\end{equation} 
Combining these relations with \eqref{Tdec} and 
$[T(y), T(z)]=0$ from Proposition \ref{pr:tai},
we obtain
\begin{equation}\label{t011}
z^{s_y(n)}T_0(y)T_1(z) + 
y^{s_y(n)}T_1(y)T_0(z)  =
y^{s_y(n)}T_0(z)T_1(y) + 
z^{s_y(n)}T_1(z)T_0(y),
\end{equation}
where $s_y(n)$ is defined in \eqref{sdy}.

\begin{example}\label{ex:tabij3}
For Examples \ref{ex:tabij} and \ref{ex:tabij2}, we have
\begin{subequations}\label{T01e}
\begin{align}
T_0(y)
&=yT^{11,010}_{00,010} +y^2\,T^{11,110}_{00,110}
\nonumber\\
&=y\,\e^{\uu_3+\uu_4-\ww_2}f_3g_4r_1
+y^2\,\e^{\uu_3+\uu_4-\ww_1}f_3g_4s_2,
\\
T_1(y)&=
T^{11,001}_{00,001}
+y\,T^{11,011}_{00,011}
+y\,T^{11,101}_{00,101}
+y^2\,T^{11,111}_{00,111}
\nonumber\\
&=\e^{-\ww_3-\ww_4}r_1r_2
+y\,\e^{-\ww_2}r_1s_3s_4
+y\,\e^{2\uu_2-\ww_1+\ww_2-\ww_3-\ww_4}f_2g_2
\nonumber\\
&+y\,\e^{-\ww_1+\ww_2-\ww_3-\ww_4}r_2s_2
+y^2\,\e^{-\ww_1}s_2s_3s_4.
\end{align}
\end{subequations}
They satisfy \eqref{t011} with $s_y(n=3)=1$.
\end{example}

We now establish a key functional relation that will be used in the proof of our main result.
\begin{prop}\label{pr:T01}
The transfer matrices $T_0(y)$ and $T_1(y)$ 
in \eqref{T01} satisfy the 
functional relation
\begin{equation}\label{kore1}
(q^2y-z)(T_0(y)T_1(z)-T_1(y)T_0(z))
= (y \leftrightarrow z),
\end{equation}
where the RHS denotes the expression obtained from the LHS by interchanging $y$ and $z$.
\end{prop}

The remainder of this subsection is devoted to the proof of Proposition \ref{pr:T01}.
The following lemma gives a cornered version of the
$MTT$ relation \eqref{mtt}.
\begin{lem}[Cornered $MTT$ relation]\label{le:mtt2}
Given 
$\mathbf{a}, \mathbf{a}', \mathbf{i}, \mathbf{i'} \in \{0,1\}^m$ and 
$\mathbf{b}, \mathbf{b}'$, 
$\mathbf{j}=(j_1,\ldots, j_n)$, 
$\mathbf{j'} =(j'_1,\ldots, j'_n) \in \{0,1\}^n$, 
set 
$\hat{\mathbf{j}}=(j_1,\ldots, j_{n-1})$ and 
$\hat{\mathbf{j}}'=(j'_1,\ldots, j'_{n-1})$.
Choose $\alpha_\ast$ and $\alpha'_\ast$ as specified in Table \ref{tab:alp}.
Then the following identity holds.
\begin{equation}\label{mtt2}
\begin{split}
&\sum_{\mathbf{c}, \mathbf{c}', \mathbf{d}, \mathbf{d}',l_n,l'_n}
\tilde{M}(\alpha_\ast')^{l_n,l'_n}_{j_n,j'_n}
\tilde{M}(\alpha_\ast)^{a_m, a'_m}_{c_m, c'_m}
\cdots \tilde{M}(\alpha_\ast)^{a_1, a'_1}_{c_1, c'_1}
\tilde{M}(\alpha_\ast)^{b_n, b'_n}_{d_n, d'_n}
\cdots \tilde{M}(\alpha_\ast)^{b_1, b'_1}_{d_1, d'_1}
\otimes
T^{\mathbf{c}, \mathbf{d}}_{\mathbf{i},\, \hat{\mathbf{j}}l_n}
T^{\mathbf{c}', \mathbf{d}'}_{\mathbf{i}',\, \hat{\mathbf{j}}'l'_n}
\\
&=
\sum_{\mathbf{k}, \mathbf{k}', \hat{\mathbf{l}}, \hat{\mathbf{l}}'}
\tilde{M}(\alpha_\ast)^{l_{n-1}, l'_{n-1}}_{j_{n-1}, j'_{n-1}}
\cdots \tilde{M}(\alpha_\ast)^{l_1, l'_1}_{j_1, j'_1}
\tilde{M}(\alpha_\ast)^{k_m, k'_m}_{i_m, i'_m}
\cdots \tilde{M}(\alpha_\ast)^{k_1, k'_1}_{i_1, i'_1}
\otimes
T^{\mathbf{a}', \mathbf{b}'}_{\mathbf{k}', \,\hat{\mathbf{l}}'j'_n}
T^{\mathbf{a}, \mathbf{b}}_{\mathbf{k}, \,\hat{\mathbf{l}}j_n}.
\end{split}
\end{equation}
The sums on the LHS run over 
$\mathbf{c}, \mathbf{c}'\in \{0,1\}^m$, $\mathbf{d}, \mathbf{d}'\in \{0,1\}^n$
and $l_n, l'_n\in \{0,1\}$.
The sums on the RHS run over 
$\mathbf{k}, \mathbf{k}' \in \{0,1\}^m$, 
$\hat{\mathbf{l}}=(l_1,\ldots, l_{n-1}), 
\hat{\mathbf{l}}'=(l'_1,\ldots, l'_{n-1})  \in \{0,1\}^{n-1}$.
We have set 
$\hat{\mathbf{j}}l_n=(j_1,\ldots, j_{n-1}, l_n)$,
$\hat{\mathbf{j}}'l'_n=(j'_1,\ldots, j'_{n-1}, l'_n)$,
$\hat{\mathbf{l}}j_n= (l_1,\ldots, l_{n-1}, j_n)$ and 
$\hat{\mathbf{l}}'j'_n =(l'_1,\ldots, l'_{n-1},j'_n)$.
\end{lem}

\begin{figure}[H]
\centering
\includegraphics[page=1,trim=65bp 75bp 90bp 565bp,clip,width=.96\textwidth]{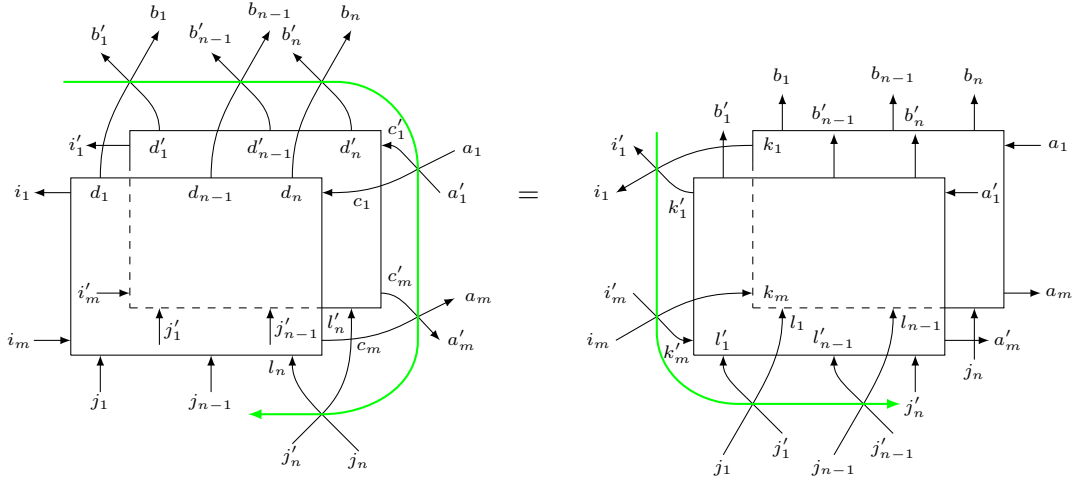}
\caption{Graphical representation of the cornered $MTT$ relation
in \eqref{mtt2}.  The rightmost pair of
boundary indices is left untraced.  The displayed boundary-wire orientations
correspond to $s_x(1)=-1$, $s_x(m)=1$ and $s_y(1)=s_y(n)=1$ in \eqref{sdef},
with the omitted signs determined by the diagram.}
\label{fig:emtt}
\end{figure}

The identity in Lemma \ref{le:mtt2} is represented graphically in
Figure \ref{fig:emtt}.

\begin{proof}
The LHS has the same form as that in \eqref{mtt},
apart from the leftmost factor $\tilde{M}(\alpha_\ast')^{l_n,l'_n}_{j_n,j'_n}$.
Thus the sums over
$\mathbf{c},  \mathbf{c}',  \mathbf{d},  \mathbf{d}'$ can be performed, giving
 \begin{equation}
 \begin{split}
 \sum_{l_n,l'_n}\tilde{M}(\alpha_\ast')^{l_n,l'_n}_{j_n,j'_n}
 \sum_{\mathbf{k}, \mathbf{k}', \hat{\mathbf{l}}, \hat{\mathbf{l}}',o,o'}
& \tilde{M}(\alpha_\ast)^{o,o'}_{l_n,l'_n}
\tilde{M}(\alpha_\ast)^{l_{n-1}, l'_{n-1}}_{j_{n-1}, j'_{n-1}}
\cdots \tilde{M}(\alpha_\ast)^{l_1, l'_1}_{j_1, j'_1}
\\
&
\quad \times \tilde{M}(\alpha_\ast)^{k_m, k'_m}_{i_m, i'_m}
\cdots \tilde{M}(\alpha_\ast)^{k_1, k'_1}_{i_1, i'_1}
\otimes
T^{\mathbf{a}', \mathbf{b}'}_{\mathbf{k}', \,\hat{\mathbf{l}}'o'}
T^{\mathbf{a}, \mathbf{b}}_{\mathbf{k}, \,\hat{\mathbf{l}}o},
\end{split}
\end{equation}
where the sums over $\mathbf{l}, \mathbf{l}'$ on the RHS of \eqref{mtt} 
are written here as sums over $\hat{\mathbf{l}}o, \hat{\mathbf{l}}' o'$.
We can now apply the inversion relation \eqref{invu} to perform the sums over $l_n, l'_n$,
which reduces the leftmost two $\tilde{M}$ factors to Kronecker deltas.
The remaining sums over $o,o'$ then give the RHS of \eqref{mtt2}.
\end{proof}

\begin{prop}\label{pr:rtt}
The $R$ matrix in \eqref{Rmat} and the monodromy matrix in \eqref{tbg}
satisfy the following relation in $\mathrm{End}(V^{\otimes n-1} \otimes V^{\otimes n-1}) 
\otimes \mathcal{W}_K(q)$:
\begin{equation}\label{rtt}
R\bigl(\tfrac{y}{z}\bigr) (\mathcal{T}^0_0(y)\mathcal{T}^1_1(z)
+q^{-2s_y(n)}\mathcal{T}^1_1(y)\mathcal{T}^0_0(z))
=
(\mathcal{T}^0_0(z)\mathcal{T}^1_1(y)
+q^{-2s_y(n)}\mathcal{T}^1_1(z)\mathcal{T}^0_0(y))R\bigl(\tfrac{y}{z}\bigr),
\end{equation}
where $s_y(n)$ is defined in \eqref{sdy}.
\end{prop}

\begin{proof}
Consider the special case 
$\mathbf{a}'=\mathbf{a}$ and $\mathbf{i}'=\mathbf{i}$ of the 
cornered $MTT$ relation \eqref{mtt2}.
Using the simplification mentioned after Proposition \ref{pr:tai},
it takes the form
\begin{equation}\label{mtt3}
\begin{split}
&\sum_{\mathbf{d}, \mathbf{d}',l_n,l'_n}
\tilde{M}(\alpha_\ast')^{l_n,l'_n}_{j_n,j'_n}
\tilde{M}(\alpha_\ast)^{b_n, b'_n}_{d_n, d'_n}
\cdots \tilde{M}(\alpha_\ast)^{b_1, b'_1}_{d_1, d'_1}
\otimes
T^{\mathbf{a}, \mathbf{d}}_{\mathbf{i},\, \hat{\mathbf{j}}l_n}
T^{\mathbf{a}, \mathbf{d}'}_{\mathbf{i},\, \hat{\mathbf{j}}'l'_n}
\\
&=
\sum_{\hat{\mathbf{l}}, \hat{\mathbf{l}}'}
\tilde{M}(\alpha_\ast)^{l_{n-1}, l'_{n-1}}_{j_{n-1}, j'_{n-1}}
\cdots \tilde{M}(\alpha_\ast)^{l_1, l'_1}_{j_1, j'_1}
\otimes
T^{\mathbf{a}, \mathbf{b}'}_{\mathbf{i}, \,\hat{\mathbf{l}}'j'_n}
T^{\mathbf{a}, \mathbf{b}}_{\mathbf{i}, \,\hat{\mathbf{l}} j_n}.
\end{split}
\end{equation}
Multiply both sides by $y^{||\mathbf{b}||}z^{||\mathbf{b}'||}$, where 
$||\mathbf{b}||$ is defined in \eqref{ztc} and, similarly, 
$||\mathbf{b}'|| = \sum_{t=1}^ns_y(t)b'_t$.
Applying \eqref{Mth} and then multiplying both sides from the right by $\bigl(\tfrac{y}{z}\bigr)^h$,
we obtain
\begin{equation}\label{mtt4}
\begin{split}
&\sum_{\hat{\mathbf{d}}, \hat{\mathbf{d}}',d_n,d'_n,l_n,l'_n}
\bigl(\tfrac{y}{z}\bigr)^h \tilde{M}(\alpha_\ast')^{l_n,l'_n}_{j_n,j'_n}
\tilde{M}(\alpha_\ast)^{b_n, b'_n}_{d_n, d'_n}
\tilde{M}(\alpha_\ast)^{b_{n-1}, b'_{n-1}}_{d_{n-1}, d'_{n-1}}
\cdots \tilde{M}(\alpha_\ast)^{b_1, b'_1}_{d_1, d'_1}
\\
& \qquad\qquad\qquad \quad \otimes 
y^{||\hat{\mathbf{d}}||+s_y(n)(d_n+j_n-l_n)}
T^{\mathbf{a}, \hat{\mathbf{d}}d_n}_{\mathbf{i},\, \hat{\mathbf{j}}l_n}
z^{||\hat{\mathbf{d}}'||+s_y(n)(d'_n+j'_n-l'_n)}
T^{\mathbf{a}, \hat{\mathbf{d}'}d'_n}_{\mathbf{i},\, \hat{\mathbf{j}}'l'_n}
\\
&=
\sum_{\hat{\mathbf{l}}, \hat{\mathbf{l}}'}
\tilde{M}(\alpha_\ast)^{l_{n-1}, l'_{n-1}}_{j_{n-1}, j'_{n-1}}
\cdots \tilde{M}(\alpha_\ast)^{l_1, l'_1}_{j_1, j'_1}\bigl(\tfrac{y}{z}\bigr)^h
\otimes
z^{||\hat{\mathbf{b}}'||+s_y(n)b'_n}
T^{\mathbf{a}, \hat{\mathbf{b}}'b'_n}_{\mathbf{i}, \,\hat{\mathbf{l}}'j'_n}
y^{||\mathbf{b}||+s_y(n)b_n}
T^{\mathbf{a}, \hat{\mathbf{b}}b_n}_{\mathbf{i}, \,\hat{\mathbf{l}}j_n},
\end{split}
\end{equation}
where we have set 
$\mathbf{d}=\hat{\mathbf{d}}d_n$, $\mathbf{d}'=\hat{\mathbf{d}'}d'_n$, 
$\hat{\mathbf{d}}=(d_1,\ldots,d_{n-1})$, 
$\hat{\mathbf{d}}'=(d'_1,\ldots,d'_{n-1})$,
 on the LHS of \eqref{mtt3}, 
and 
$\mathbf{b}=\hat{\mathbf{b}}b_n$,  $\mathbf{b}'=\hat{\mathbf{b}'}b'_n$, 
$\hat{\mathbf{b}}=(b_1,\ldots,b_{n-1})$, 
$\hat{\mathbf{b}}'=(b'_1,\ldots,b'_{n-1})$
on the RHS.

Set $b_n=j_n$, $b'_n=j'_n$, multiply by 
$q^{-2s_y(n)j'_n}y^{-s_y(n)j_n}z^{-s_y(n)j'_n}$, and 
sum over $j_n,j'_n=0,1$ subject to the condition $j_n+j'_n=1$.
The inversion relation \eqref{qmm} reduces the leftmost two $\tilde{M}$s 
on the LHS of \eqref{mtt4}
to Kronecker deltas.
Therefore, the subsequent sum over $d_n,d'_n$ amounts to setting $d_n=l_n$ and $d'_n=l'_n$.
The result is
\begin{equation}\label{mtt5}
\begin{split}
&\sum_{\substack{\hat{\mathbf{d}}, \hat{\mathbf{d}}',l_n,l'_n \\ l_n+l'_n=1}}
q^{-2s_y(n)l_n}
\bigl(\tfrac{y}{z}\bigr)^h 
\tilde{M}(\alpha_\ast)^{b_{n-1}, b'_{n-1}}_{d_{n-1}, d'_{n-1}}
\cdots \tilde{M}(\alpha_\ast)^{b_1, b'_1}_{d_1, d'_1}
\otimes 
y^{||\hat{\mathbf{d}}||}
T^{\mathbf{a}, \hat{\mathbf{d}}l_n}_{\mathbf{i},\, \hat{\mathbf{j}}l_n}
z^{||\hat{\mathbf{d}}'||}
T^{\mathbf{a}, \hat{\mathbf{d}'}l'_n}_{\mathbf{i},\, \hat{\mathbf{j}}'l'_n}
\\
&=
\sum_{\substack{\hat{\mathbf{l}}, \hat{\mathbf{l}}', j_n,j'_n \\ j_n+j'_n=1}}
q^{-2s_y(n)j'_n}
\tilde{M}(\alpha_\ast)^{l_{n-1}, l'_{n-1}}_{j_{n-1}, j'_{n-1}}
\cdots \tilde{M}(\alpha_\ast)^{l_1, l'_1}_{j_1, j'_1}\bigl(\tfrac{y}{z}\bigr)^h
\otimes
z^{||\hat{\mathbf{b}}'||}
T^{\mathbf{a}, \hat{\mathbf{b}}'j'_n}_{\mathbf{i}, \,\hat{\mathbf{l}}'j'_n}
y^{||\mathbf{b}||}
T^{\mathbf{a}, \hat{\mathbf{b}}j_n}_{\mathbf{i}, \,\hat{\mathbf{l}}j_n},
\end{split}
\end{equation}
This is an identity in $\mathrm{End}(\mathcal{V}_+) \otimes \mathcal{W}_K(q)$.
Taking the trace over $\mathcal{V}_+$
and applying the definitions \eqref{Rmat2} and \eqref{tbg}, we obtain
\begin{equation}\label{rtttt}
\begin{split}
&\sum_{\hat{\mathbf{d}}, \hat{\mathbf{d}}'}
R\bigl(\tfrac{y}{z}\bigr)^{\hat{\mathbf{b}}, \hat{\mathbf{b}}'}_{\hat{\mathbf{d}}, \hat{\mathbf{d}}'}
\bigl(\mathcal{T}^0_0(y)\mathcal{T}^1_1(z) 
+ q^{-2s_y(n)}\mathcal{T}^1_1(y)\mathcal{T}^0_0(z)
\big)^{\hat{\mathbf{d}}, \hat{\mathbf{d}}'}_{\hat{\mathbf{j}}, \hat{\mathbf{j}}'}
\\
&=
\sum_{\hat{\mathbf{l}}, \hat{\mathbf{l}}'}
\bigl(\mathcal{T}^0_0(z)\mathcal{T}^1_1(y) 
+ q^{-2s_y(n)}\mathcal{T}^1_1(z)\mathcal{T}^0_0(y)
\big)^{\hat{\mathbf{b}}, \hat{\mathbf{b}}'}_{\hat{\mathbf{l}}, \hat{\mathbf{l}}'}
R\bigl(\tfrac{y}{z}\bigr)^{\hat{\mathbf{l}}, \hat{\mathbf{l}}'}_{\hat{\mathbf{j}}, \hat{\mathbf{j}}'}.
\end{split}
\end{equation}

\begin{figure}[H]
\centering
\includegraphics[page=2,trim=100bp 420bp 85bp 55bp,clip,width=\textwidth]{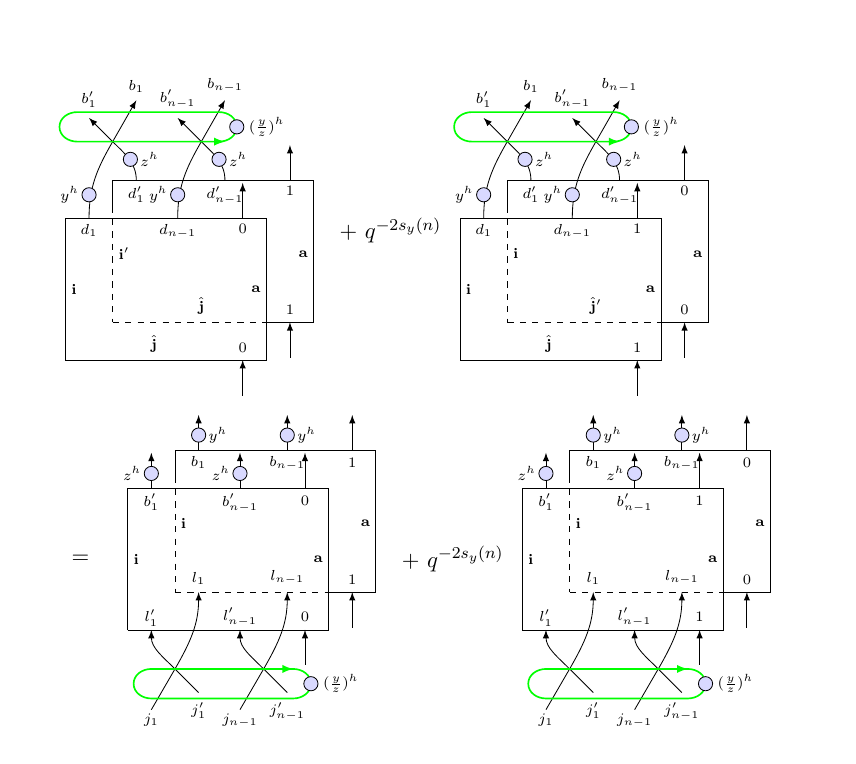}
\caption{Graphical representation of the matrix-element identity
\eqref{rtttt}.  The north-boundary wire segments shown here are all oriented outward,
so this is the concrete case $s_y(1)=\cdots=s_y(n)=1$ of \eqref{sdef}.
The LHS is understood to include the sum over the intermediate states
$\hat{\mathbf{d}},\hat{\mathbf{d}}'$, and similarly the RHS is understood
to include the sum over $\hat{\mathbf{l}},\hat{\mathbf{l}}'$.}
\label{fig:rtttt}
\end{figure}
This is precisely the equality of the matrix elements of the two sides of \eqref{rtt}
corresponding to the transition
$v_{j_1}\otimes \cdots v_{j_{n-1}}\otimes 
v_{j'_1}\otimes \cdots v_{j'_{n-1}} \rightarrow 
v_{b_1}\otimes \cdots v_{b_{n-1}}\otimes 
v_{b'_1}\otimes \cdots v_{b'_{n-1}}$ 
in $V^{\otimes n-1}\otimes V^{\otimes n-1}$.
\end{proof}

\begin{prop}\label{pr:t01}
The transfer matrices $T_0(y)$ and $T_1(y)$ in \eqref{T01}
satisfy the functional relation
\begin{equation}\label{tq01}
T_0(y)T_1(z)+q^{-2s_y(n)}T_1(y)T_0(z)
= T_0(z)T_1(y)+q^{-2s_y(n)}T_1(z)T_0(y).
\end{equation}
\end{prop}
\begin{proof}
As remarked after \eqref{Rmat}, the matrix $R\bigl(\zeta)$ is invertible
for generic $\zeta$.
Thus, multiplying \eqref{rtt} from the right by $R\bigl(\tfrac{y}{z}\bigr)^{-1}$
and taking the trace over $V^{\otimes n-1}\otimes V^{\otimes n-1}$ eliminates
$R\bigl(\tfrac{y}{z}\bigr)$.
In view of the definition \eqref{T01}, the resulting identity in $\mathcal{W}_K(q)$ 
yields \eqref{tq01}.
\end{proof}

We now combine \eqref{t011} with Proposition \ref{pr:t01}
to prove Proposition \ref{pr:T01}.
\begin{proof}[Proof of Proposition \ref{pr:T01}]
For $s_y(n)=1$, take the linear combination
$(1+q^2)\times \eqref{t011}+q^2(y+z)\times \eqref{tq01}$.
For  $s_y(n)=-1$, take the linear combination
$yz(1+q^2)\times \eqref{t011}+(y+z)\times \eqref{tq01}$.
\end{proof}

At $z=q^2y$, Proposition \ref{pr:T01} reduces to
\begin{cor}\label{co:t01}
The transfer matrices satisfy the functional relation
\begin{equation}\label{s01}
T_0(q^2y)T_1(y) = T_1(q^2y)T_0(y).
\end{equation}
\end{cor}

More generally, Proposition \ref{pr:T01}
implies that 
$(T_0(y)T_1(z)-T_0(z)T_1(y))$
is a Laurent polynomial in $y,z$ with coefficients in 
$\mathcal{W}_K(q)$, which is divisible 
by $(q^2z-y)$ with the quotient symmetric in $y$ and $z$.

\begin{example}
We continue with Example \ref{ex:tabij3}, which corresponds to $s_y(n)=1$.
We use the temporary shorthand 
$\mathsf{e}_1=y+z$, $\mathsf{e}_2=yz$, $[2]=1+q^2$.

Proposition \ref{pr:t01} asserts that the following expression is symmetric in
$y$ and $z$.
This can be checked explicitly as
\begin{equation*}
\begin{split}
&q^2T_0(y)T_1(z)+T_1(y)T_0(z)
\\
&=[2]\mathsf{e}_2\,\e^{2\uu_2+\uu_3+\uu_4-\ww_1-\ww_3-\ww_4}f_2f_3g_2g_4r_1
 +[2]\mathsf{e}_2\,\e^{\uu_3+\uu_4-2\ww_2}f_3g_4r_1^2s_3s_4
\\
&\quad
 +[2]\mathsf{e}_1\mathsf{e}_2\,\e^{\uu_3+\uu_4-\ww_1-\ww_2}f_3g_4r_1s_2s_3s_4
 +[2]\mathsf{e}_2^2\,\e^{\uu_3+\uu_4-2\ww_1}f_3g_4s_2^2s_3s_4
\\
&\quad
 +q\mathsf{e}_1\,\e^{\uu_3+\uu_4-\ww_2-\ww_3-\ww_4}f_3g_4r_1^2r_2
 +q\mathsf{e}_1^2\,\e^{\uu_3+\uu_4-\ww_1-\ww_3-\ww_4}f_3g_4r_1r_2s_2
\\
&\quad
 +q\mathsf{e}_1\mathsf{e}_2\,\e^{2\uu_2+\uu_3+\uu_4-2\ww_1+\ww_2-\ww_3-\ww_4}f_2f_3g_2g_4s_2
 +q\mathsf{e}_1\mathsf{e}_2\,\e^{\uu_3+\uu_4-2\ww_1+\ww_2-\ww_3-\ww_4}f_3g_4r_2s_2^2.
\end{split}
\end{equation*}
For Proposition \ref{pr:T01}, we similarly have 
\begin{equation*}
\begin{split}
&q(y-q^2z)^{-1}
(T_0(y)T_1(z)-T_0(z)T_1(y))
\\
&=
 \e^{\uu_3+\uu_4-\ww_2-\ww_3-\ww_4}f_3g_4r_1^2r_2
 +\e^{2\uu_2+\uu_3+\uu_4-2\ww_1+\ww_2-\ww_3-\ww_4}
 \mathsf{e}_2 f_2f_3g_2g_4s_2
\\
&
 +\e^{\uu_3+\uu_4-\ww_1-\ww_3-\ww_4} \mathsf{e}_1f_3g_4r_1r_2s_2
 +\e^{\uu_3+\uu_4-2\ww_1+\ww_2-\ww_3-\ww_4}
\mathsf{e}_2 f_3g_4r_2s_2^2.
\end{split}
\end{equation*}
\end{example}

\subsection{Fixed-boundary amplitudes and commutation relations}\label{ss:ta}

We first recall, in a form adapted to the present purpose, the tangle description
introduced in \cite[Sec.~3.3]{IKTY25}.
Fix an admissible diagram $G$ and a particular sequence of local moves
transforming the NE arrow into the SW arrow as in Definition \ref{def:adm}.
The eight elementary tangle pieces corresponding to the moves
${\rm(o)}, {\rm(h)}, {\rm(v)}, {\rm(t)}, {\rm(I)}, \text{\rm(I')},
{\rm(iI)}, \text{\rm(iI')}$ in \eqref{2D-LM}--\eqref{2D-iI}
are depicted as follows:

\begin{align}\label{tan}
\begin{tikzpicture}
\begin{scope}[>=latex,xshift=0pt]
\draw (-2,0.75) node[right] {type A:};
{\color{green}
\draw [->,dashed] (0,1.5)--(2,1.5);
\draw [->,dashed] (0,0)--(2,0);
}
\draw [->] (0.5,0)--(1.5,1.5);
\draw [->] (0.5,1.5)--(1.5,0);
\draw (1,-0.5) node {(v)};
{\color{green}
\draw [->,dashed] (3,1.5)--(5,1.5);
\draw [->,dashed] (3,0)--(5,0);
}
\draw [->] (3.5,0) to [out=90,in=180] (4,0.5) to [out=0,in=90] (4.5,0);
\draw (4,-0.5) node {(iI)};
{\color{green}
\draw [->,dashed] (6,1.5)--(8,1.5);
\draw [->,dashed] (6,0)--(8,0);
}
\draw [->] (6.5,1.5) to [out=-90,in=180] (7,1) to [out=0,in=-90] (7.5,1.5);
\draw (7,-0.5) node {(I)};
\end{scope}
\begin{scope}[>=latex,yshift=-80pt]
\draw (-2,0.75) node[right] {type B:};
{\color{green}
\draw [->,dashed] (0,1.5)--(2,1.5);
\draw [->,dashed] (0,0)--(2,0);
}
\draw [<-] (0.5,0)--(1.5,1.5);
\draw [<-] (0.5,1.5)--(1.5,0);
\draw (1,-0.5) node {(h)};
{\color{green}
\draw [->,dashed] (3,1.5)--(5,1.5);
\draw [->,dashed] (3,0)--(5,0);
}
\draw [<-] (3.5,0) to [out=90,in=180] (4,0.5) to [out=0,in=90] (4.5,0);
\draw (4,-0.5) node {(iI')};
{\color{green}
\draw [->,dashed] (6,1.5)--(8,1.5);
\draw [->,dashed] (6,0)--(8,0);
}
\draw [<-] (6.5,1.5) to [out=-90,in=180] (7,1) to [out=0,in=-90] (7.5,1.5);
\draw (7,-0.5) node {(I')};
\end{scope}
\begin{scope}[>=latex,yshift=-160pt]
\draw (-2,0.75) node[right] {type C:};
{\color{green}
\draw [->,dashed] (0,1.5)--(2,1.5);
\draw [->,dashed] (0,0)--(2,0);
}
\draw [->] (0.5,0)--(1.5,1.5);
\draw [<-] (0.5,1.5)--(1.5,0);
\draw (1,-0.5) node {(o)};
{\color{green}
\draw [->,dashed] (3,1.5)--(5,1.5);
\draw [->,dashed] (3,0)--(5,0);
}
\draw [<-] (3.5,0)--(4.5,1.5);
\draw [->] (3.5,1.5)--(4.5,0);
\draw (4,-0.5) node {(t)};
\end{scope}
\end{tikzpicture}
\end{align}

Here the green dashed arrows correspond to the auxiliary green arrow in
\eqref{2D-LM}--\eqref{2D-iI}, and its successive positions in the
transformation from the NE arrow to the SW arrow are read from top to bottom.
A tangle diagram associated with the chosen admissible transformation is
obtained by stacking these elementary pieces accordingly.  It is not unique
in general.

The local equivalences in \cite[(3.18)]{IKTY25} may be summarized symbolically as
\begin{subequations}\label{tang}
\begin{align}
{\rm(iI)}-{\rm(t)}-{\rm(I)}
&\sim {\rm(v)} \sim {\rm(I)}-{\rm(o)}-{\rm(iI)},
\label{tan-v}\\
\text{\rm(iI')}-{\rm(o)}-\text{\rm(I')}
&\sim {\rm(h)} \sim \text{\rm(I')}-{\rm(t)}-\text{\rm(iI')},
\label{tan-h}\\
{\rm(iI)}-{\rm(I)}
&\sim {\rm Id} \sim \text{\rm(I')}-\text{\rm(iI')},
\label{tid}\\
\text{\rm(iI')}-\text{\rm(I')}
&\sim {\rm Id} \sim {\rm(I)}-{\rm(iI)}.
\label{tidp}
\end{align}
\end{subequations}

The relations in \eqref{tang} are represented graphically as follows.

\begin{figure}[H]
\centering
\begin{tikzpicture}[scale=.82,every node/.style={transform shape}]
\begin{scope}[>=latex,xshift=0pt]
{\color{green}
\draw [->,dashed] (0,1.5)--(4,1.5);
\draw [->,dashed] (0,0)--(4,0);
}
\draw [->] (0.5,0)--(0.5,1.5);
\draw [->] (0.5,1.5) to [out=90,in=180] (1,2) to [out=0,in=90] (1.5,1.5);
\draw [->] (1.5,1.5)--(2.5,0);
\draw [->] (2.5,1.5)--(1.5,0);
\draw [->] (2.5,0) to [out=-90,in=180] (3,-0.5) to [out=0,in=-90] (3.5,0);
\draw [->] (3.5,0)--(3.5,1.5);
\draw (4.5,0.75) node {$=$};
{\color{green}
\draw [->,dashed] (5,1.5)--(7,1.5);
\draw [->,dashed] (5,0)--(7,0);
}
\draw [->] (5.5,0)--(6.5,1.5);
\draw [->] (5.5,1.5)--(6.5,0);
\draw (7.5,0.75) node {$=$};
{\color{green}
\draw [->,dashed] (8,1.5)--(12,1.5);
\draw [->,dashed] (8,0)--(12,0);
}
\draw [->] (8.5,0) to [out=-90,in=180] (9,-0.5) to [out=0,in=-90] (9.5,0);
\draw [->] (8.5,1.5)--(8.5,0);
\draw [->] (9.5,0)--(10.5,1.5);
\draw [->] (10.5,0)--(9.5,1.5);
\draw [->] (11.5,1.5)--(11.5,0);
\draw [->] (10.5,1.5) to [out=90,in=180] (11,2) to [out=0,in=90] (11.5,1.5);
\draw (-0.5,2.5) node[right] {${\rm(iI)}-{\rm(t)}-{\rm(I)}={\rm(v)}={\rm(I)}-{\rm(o)}-{\rm(iI)}$};
\end{scope}

\begin{scope}[>=latex,yshift=-100pt]
{\color{green}
\draw [->,dashed] (0,1.5)--(4,1.5);
\draw [->,dashed] (0,0)--(4,0);
}
\draw [<-] (0.5,0)--(0.5,1.5);
\draw [<-] (0.5,1.5) to [out=90,in=180] (1,2) to [out=0,in=90] (1.5,1.5);
\draw [<-] (1.5,1.5)--(2.5,0);
\draw [<-] (2.5,1.5)--(1.5,0);
\draw [<-] (2.5,0) to [out=-90,in=180] (3,-0.5) to [out=0,in=-90] (3.5,0);
\draw [<-] (3.5,0)--(3.5,1.5);
\draw (4.5,0.75) node {$=$};
{\color{green}
\draw [->,dashed] (5,1.5)--(7,1.5);
\draw [->,dashed] (5,0)--(7,0);
}
\draw [<-] (5.5,0)--(6.5,1.5);
\draw [<-] (5.5,1.5)--(6.5,0);
\draw (7.5,0.75) node {$=$};
{\color{green}
\draw [->,dashed] (8,1.5)--(12,1.5);
\draw [->,dashed] (8,0)--(12,0);
}
\draw [<-] (8.5,0) to [out=-90,in=180] (9,-0.5) to [out=0,in=-90] (9.5,0);
\draw [<-] (8.5,1.5)--(8.5,0);
\draw [<-] (9.5,0)--(10.5,1.5);
\draw [<-] (10.5,0)--(9.5,1.5);
\draw [<-] (11.5,1.5)--(11.5,0);
\draw [<-] (10.5,1.5) to [out=90,in=180] (11,2) to [out=0,in=90] (11.5,1.5);
\draw (-0.5,2.5) node[right] {$\text{\rm(iI')}-{\rm(o)}-\text{\rm(I')}={\rm(h)}=\text{\rm(I')}-{\rm(t)}-\text{\rm(iI')}$};
\end{scope}

\begin{scope}[>=latex,yshift=-160pt]
{\color{green}
\draw [->,dashed] (0,0)--(3,0);
}
\draw [->] (0.5,-0.5) to [out=90,in=180] (1,0.5) to [out=0,in=90] (1.5,0);
\draw [->] (1.5,0) to [out=-90,in=180] (2,-0.5) to [out=0,in=-90] (2.5,0.5);
\draw (3.5,0) node {$=$};
{\color{green}
\draw [->,dashed] (4,0)--(5,0);
}
\draw [<-] (4.5,0.5)--(4.5,-0.5);
\draw (5.5,0) node {$=$};
{\color{green}
\draw [->,dashed] (6,0)--(9,0);
}
\draw [<-] (6.5,0.5) to [out=-90,in=180] (7,-0.5) to [out=0,in=-90] (7.5,0);
\draw [<-] (7.5,0) to [out=90,in=180] (8,0.5) to [out=0,in=90] (8.5,-0.5);
\draw (-0.5,1) node[right] {${\rm(iI)}-{\rm(I)}={\rm Id}=\text{\rm(I')}-\text{\rm(iI')}$};
\end{scope}

\begin{scope}[>=latex,yshift=-230pt]
{\color{green}
\draw [->,dashed] (0,0)--(3,0);
}
\draw [<-] (0.5,-0.5) to [out=90,in=180] (1,0.5) to [out=0,in=90] (1.5,0);
\draw [<-] (1.5,0) to [out=-90,in=180] (2,-0.5) to [out=0,in=-90] (2.5,0.5);
\draw (3.5,0) node {$=$};
{\color{green}
\draw [->,dashed] (4,0)--(5,0);
}
\draw [->] (4.5,0.5)--(4.5,-0.5);
\draw (5.5,0) node {$=$};
{\color{green}
\draw [->,dashed] (6,0)--(9,0);
}
\draw [->] (6.5,0.5) to [out=-90,in=180] (7,-0.5) to [out=0,in=-90] (7.5,0);
\draw [->] (7.5,0) to [out=90,in=180] (8,0.5) to [out=0,in=90] (8.5,-0.5);
\draw (-0.5,1) node[right] {$\text{\rm(iI')}-\text{\rm(I')}={\rm Id}={\rm(I)}-{\rm(iI)}$};
\end{scope}
\end{tikzpicture}
\caption{Graphical form of the local tangle equivalences in \eqref{tang}.
The elementary pieces are stacked geometrically from bottom to top, whereas
the corresponding hyphenated expression is read from right to left, so that
its rightmost factor is the bottom piece.  The dashed green arrows record the
successive positions of the auxiliary arrow.  Reproduced from
\cite[(3.18)]{IKTY25}.}
\label{fig:equiv}
\end{figure}
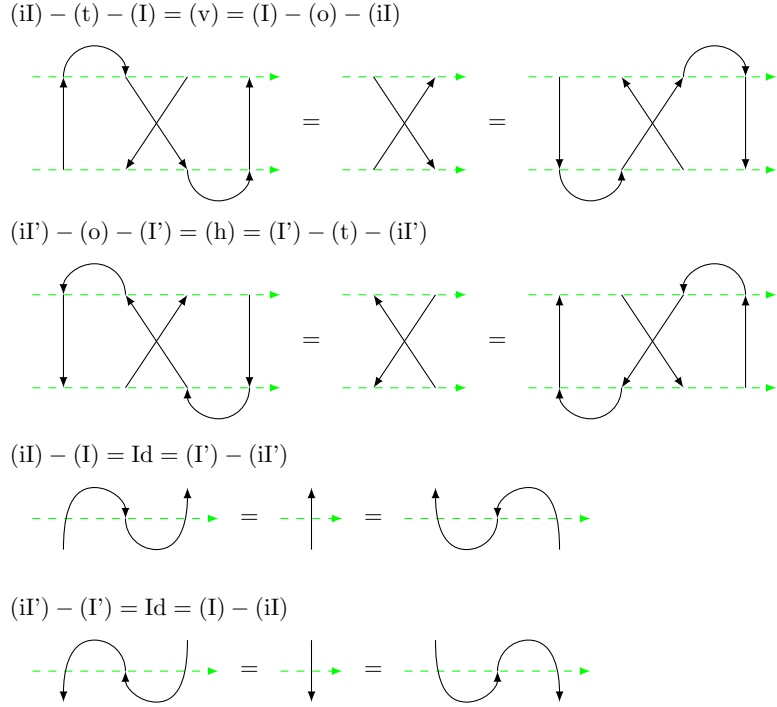

The factors in \eqref{tang} run from top to bottom when read from left
to right, and $\sim$ denotes equivalence of tangle diagrams.  In particular,
the first two relations allow every occurrence of
${\rm(v)}$ or ${\rm(h)}$ to be eliminated.

\begin{definition}\label{def:stan}
A tangle diagram associated with an admissible transformation of an
admissible diagram $G$ is called \emph{simple} if it contains neither an
${\rm(h)}$ move nor a ${\rm(v)}$ move.
Thus, by admissibility, a simple tangle is obtained by concatenating elementary
pieces chosen entirely from one of the following two sets:
\begin{equation}\label{tcls}
\{{\rm(o)},{\rm(t)},{\rm(I)},{\rm(iI)}\},
\qquad
\{{\rm(o)},{\rm(t)},\text{\rm(I')},\text{\rm(iI')}\}.
\end{equation}
\end{definition}

By \eqref{tang}, every tangle diagram is equivalent to a simple
one.  Hence, without loss of generality, we restrict ourselves to simple
tangles in what follows.

\begin{example}\label{ex:st}
\leavevmode

\begin{figure}[H]
\centering
\includegraphics[width=.95\textwidth]{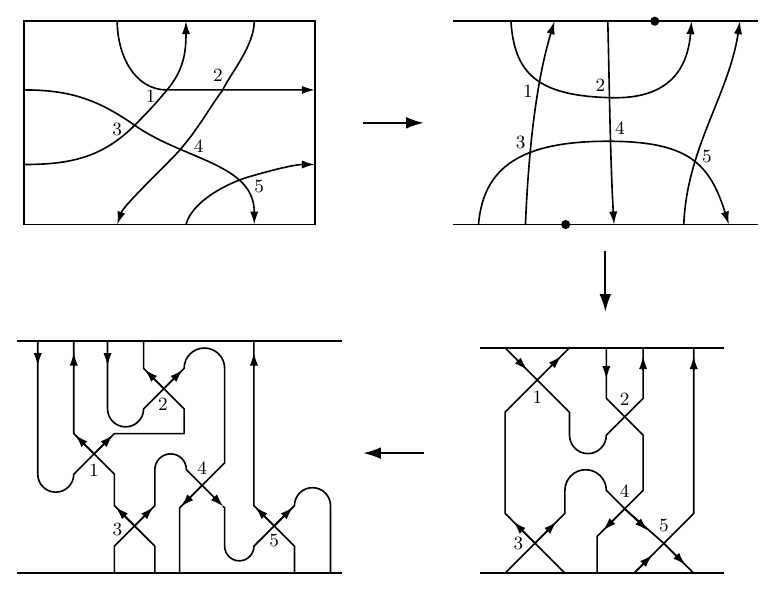}
\caption{An example of a tangle representation. Top left: an admissible diagram $G$.
Top right: the corresponding tangle diagram, with its NE sides placed at the
top and its SW sides at the bottom; the two bullets mark the positions
corresponding to the NE and SW corners of $G$. Bottom right: the decomposition
of this tangle into the elementary pieces in \eqref{tan}. Bottom left: the
simple tangle obtained from the bottom-right decomposition by replacing the
${\rm(v)}$ pieces at vertices $1$, $2$, and $5$ according to \eqref{tan-v};
read from top to bottom, equivalently from left to right in the symbolic
expression, the three replacement pieces are
${\rm(iI)}-{\rm(o)}-{\rm(I)}$.
The resulting simple tangle involves only the four operations listed
on the left in \eqref{tcls}.}
\label{fig:st}
\end{figure}

\end{example}

Let $G$ be an admissible diagram, and 
let $\mathscr{T}$ be the corresponding simple tangle 
consisting of $L$ oriented wires whose orientation is 
specified by the signs $s(1),\ldots, s(L) =\pm1$.

We denote by $A^{\bm{\mu}}_{\bm{\nu}} \in \mathcal{W}_K(q)$  
the fixed-boundary amplitude associated with $\mathscr{T}$, namely the
partition function of the Q6V model on $G$ with boundary
condition fixed to
$\bm{\mu}=(\mu_1,\ldots, \mu_L) \in \{0,1\}^L$ on the top and
$\bm{\nu} =(\nu_1,\ldots, \nu_L) \in \{0,1\}^L$ on the bottom of
$\mathscr{T}$.

In this simple-tangle formulation, the fixed-boundary partition function
$T^{\mathbf{a},\mathbf{b}}_{\mathbf{i},\mathbf{j}}$ introduced at the beginning of
Section \ref{ss:ct} is identified with the amplitude $A^{\bm{\mu}}_{\bm{\nu}}$ as
\begin{subequations}\label{TZ}
\begin{align}
T^{\mathbf{a},\mathbf{b}}_{\mathbf{i},\mathbf{j}}
&= A^{\bm{\mu}}_{\bm{\nu}},
\quad 
\bm{\mu}
= (b_1,\ldots,b_n,a_1,\ldots,a_m),
\quad
\bm{\nu}
= (i_1,\ldots, i_m, j_1,\ldots,j_n),
\label{TZab}
\\
s(k)
&=
\begin{cases}
s_y(k), & 1\le k\le n,\\
s_x(k-n), & n<k\le L.
\end{cases}
\label{TZs}
\end{align}
\end{subequations}
where $L=n+m$, and the signs $s_x(k)$ and $s_y(k)$ are defined in \eqref{sdef}.
The weight conservation condition \eqref{tzero}
then takes the form
\begin{equation}\label{Zw}
 A^{\bm{\mu}}_{\bm{\nu}} = 0 \;\;\text{unless}\;\;
 \sum_{k=1}^{L}s(k)\mu_k =  \sum_{k=1}^{L}s(k)\nu_k.
\end{equation}

Suppose that $\bm{\mu}, \bm{\nu} \in \{0,1\}^L$ satisfy the condition in
\eqref{Zw}.
If $s(k)=s(k+1)$, then the arrays obtained by interchanging
$\mu_k \leftrightarrow \mu_{k+1}$ or
$\nu_k \leftrightarrow \nu_{k+1}$ also satisfy the same condition.
If $(s(k),s(k+1))=(1,-1)$, the condition is likewise preserved by
replacing $(\mu_k,\mu_{k+1})=(0,0)$ with $(1,1)$, or conversely,
and similarly for $(\nu_k,\nu_{k+1})$.
The following lemma gives simple commutation relations among the corresponding
amplitudes $A^{\bm{\mu}}_{\bm{\nu}}$.

\begin{lem}\label{le:TT}
The following commutation relations hold under variations of the
$k$th and $(k+1)$st components of the boundary arrays:
\begin{subequations}\label{ZZ}
\begin{align}
&(s(k),s(k+1))=(\pm1, \pm1): 
\nonumber \\
&\qquad A^{\ldots 01\ldots}_{\bm{\nu}} A^{\ldots10\ldots}_{\bm{\nu}} 
= q^{\pm1} A^{\ldots 10\ldots}_{\bm{\nu}} A^{\ldots 01\ldots}_{\bm{\nu}},
\quad\;
A_{\ldots 01\ldots}^{\bm{\mu}} A_{\ldots 10\ldots}^{\bm{\mu}} 
= q^{\pm 1} A_{\ldots 10\ldots}^{\bm{\mu}} A_{\ldots 01\ldots}^{\bm{\mu}},
\label{ZZ1}\\
&(s(k),s(k+1))=(1,-1): 
\nonumber \\
&\qquad A^{\ldots 00\ldots}_{\bm{\nu}} A^{\ldots 11\ldots}_{\bm{\nu}} 
= q^2 A^{\ldots 11\ldots}_{\bm{\nu}} A^{\ldots 00\ldots}_{\bm{\nu}},
\qquad
A_{\ldots 00\ldots}^{\bm{\mu}} A_{\ldots 11\ldots}^{\bm{\mu}} 
= q^2 A_{\ldots 11\ldots}^{\bm{\mu}} A_{\ldots 00\ldots}^{\bm{\mu}},
\label{ZZ2}
\end{align}
where the entries represented by $\ldots$ are identical on both sides.
\end{subequations}
\end{lem}

\begin{proof}
Set 
\begin{align*}
T^{\mathbf{a}',\mathbf{b}'}_{\mathbf{i}',\mathbf{j}'}
&= A^{\bm{\mu}'}_{\bm{\nu}'},
\quad 
\bm{\mu}'
= (b'_1,\ldots,b'_n,a'_1,\ldots,a'_m),
\quad
\bm{\nu}'
= (i'_1,\ldots, i'_m, j'_1,\ldots,j'_n).
\end{align*}
Let us first consider the case $s(k)=s(k+1)=\pm1$.
Suppose that the arrays $\bm{\mu}', \bm{\nu}'$ 
and those in \eqref{TZab} satisfy 
$\bm{\mu}=\bm{\mu}'$ and $\bm{\nu}=\bm{\nu}'$, except for
$(\mu_k,\mu_{k+1})=(0,1)$ and $(\mu'_k,\mu'_{k+1})=(1,0)$.
Specializing the $MTT$ relation \eqref{mtt} to these boundary conditions, we obtain
\begin{equation}\label{mzz}
\sum_{\rho,\rho',\sigma,\sigma'=0,1}
\tilde{M}(\alpha_\ast)^{01}_{\sigma\sigma'}
\tilde{M}(\alpha_\ast)^{10}_{\rho \rho'}\otimes 
A^{\ldots \rho\sigma \ldots}_{\bm{\nu}}
A^{\ldots \rho'\sigma' \ldots}_{\bm{\nu}}
=A^{\ldots 01 \ldots}_{\bm{\nu}}A^{\ldots 10 \ldots}_{\bm{\nu}}.
\end{equation}

\begin{figure}[H]
\centering
\includegraphics[page=2,trim=100bp 250bp 95bp 430bp,clip,width=.90\textwidth]{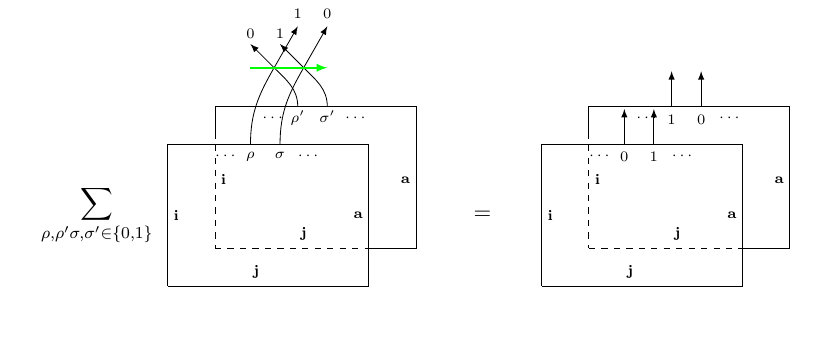}
\caption{Graphical representation of the local summation identity
\eqref{mzz}.}
\label{fig:mzz}
\end{figure}
The condition \eqref{Zw} restricts the sum to
$\rho+\sigma=\rho'+\sigma'=1$.
Subtracting the RHS of \eqref{mzz} from the LHS, we obtain
\begin{equation*}
\begin{split}
0=&(\tilde{M}(\alpha_\ast)^{01}_{10}
\tilde{M}(\alpha_\ast)^{10}_{01}-1)\otimes 
A^{\ldots 01 \ldots}_{\bm{\nu}}
A^{\ldots 10 \ldots}_{\bm{\nu}}
+
\tilde{M}(\alpha_\ast)^{01}_{01}
\tilde{M}(\alpha_\ast)^{10}_{10}\otimes 
A^{\ldots 10 \ldots}_{\bm{\nu}}
A^{\ldots 01 \ldots}_{\bm{\nu}}.
\end{split}
\end{equation*}
Using \eqref{mta} and the matrix elements $M^{ab}_{ij}$ in Figure \ref{fig:6vM},
the expression above is evaluated, for $s(k)=s(k+1)=\pm1$, as
\begin{equation*}
\begin{split}
&\begin{pmatrix}
(M(\alpha_\ast)^{01}_{10}
M(\alpha_\ast)^{10}_{01}-1)\otimes 
A^{\ldots 01 \ldots}_{\bm{\nu}}
A^{\ldots 10 \ldots}_{\bm{\nu}}
+
M(\alpha_\ast)^{01}_{01}
M(\alpha_\ast)^{10}_{10}\otimes 
A^{\ldots 10 \ldots}_{\bm{\nu}}
A^{\ldots 01 \ldots}_{\bm{\nu}}
\\
(M(\alpha_\ast)_{01}^{10}
M(\alpha_\ast)_{10}^{01}-1)\otimes 
A^{\ldots 01 \ldots}_{\bm{\nu}}
A^{\ldots 10 \ldots}_{\bm{\nu}}
+
M(\alpha_\ast)_{01}^{01}
M(\alpha_\ast)_{10}^{10}\otimes 
A^{\ldots 10 \ldots}_{\bm{\nu}}
A^{\ldots 01 \ldots}_{\bm{\nu}}
\end{pmatrix}
\\
&=
\begin{pmatrix}
(\mathbf{a}^-\mathbf{a}^+ -1)\otimes 
A^{\ldots 01 \ldots}_{\bm{\nu}}
A^{\ldots 10 \ldots}_{\bm{\nu}}
+
q^{-1}\mathbf{k}^2 \otimes
A^{\ldots 10 \ldots}_{\bm{\nu}}
A^{\ldots 01 \ldots}_{\bm{\nu}}
\\
(\mathbf{a}^+\mathbf{a}^--1) \otimes 
A^{\ldots 01 \ldots}_{\bm{\nu}}
A^{\ldots 10 \ldots}_{\bm{\nu}}
+
q^{-1}\mathbf{k}^2 \otimes
A^{\ldots 10 \ldots}_{\bm{\nu}}
A^{\ldots 01 \ldots}_{\bm{\nu}}
\end{pmatrix}
\\
&=
\begin{pmatrix}
q^{-2}\mathbf{k}^2\otimes 
(-A^{\ldots 01 \ldots}_{\bm{\nu}}
A^{\ldots 10 \ldots}_{\bm{\nu}}
+
qA^{\ldots 10 \ldots}_{\bm{\nu}}
A^{\ldots 01 \ldots}_{\bm{\nu}})
\\
\mathbf{k}^2\otimes 
(-A^{\ldots 01 \ldots}_{\bm{\nu}}
A^{\ldots 10 \ldots}_{\bm{\nu}}
+
q^{-1}
A^{\ldots 10 \ldots}_{\bm{\nu}}
A^{\ldots 01 \ldots}_{\bm{\nu}})
\end{pmatrix}.
\end{split}
\end{equation*}
Here the last equality follows from \eqref{oscr} with $p=-q^{-1}$.
Since $\mathbf{k}$ is invertible, the second tensor factors in both
components must vanish, which proves the first relation in \eqref{ZZ1}.
The remaining relations follow similarly.
For \eqref{ZZ2}, one also uses \eqref{alp}.
\end{proof}

\section{Quantized six-vertex model at odd roots of unity}\label{s:rt}

In this section, we specialize the Q6V model to an odd root of
unity and establish the quantum Frobenius property for its transfer
matrices.  After introducing the cyclic representation, we state the
main functional relation and prove it in two steps, based respectively
on the expansions \eqref{TTai} and \eqref{TTgv}.  We then show how
this relation determines the transfer-matrix spectrum up to
multiplicities and yields free-parafermion spectra for the associated
Hamiltonians.

\subsection{Specialization of $q$ to odd roots of unity}\label{ss:sp}

Set
\begin{equation}\label{qwn}
\ve = \text{a primitive $N$th root of unity},\quad 
\omega = \ve^2, \quad N \in 2\Z_{\ge 1}+1.
\end{equation}
By definition, $\omega$ is also a primitive $N$th root of unity.
We henceforth consider the Q6V model at $q=\ve$.
For $N=1$, our main result, 
Theorem~\ref{th:main} follows tautologically from the definitions, so we restrict throughout to $N\ge3$.

Set 
\begin{equation}\label{Vcyc}
\mathscr{V}= \bigoplus_{m_1,\ldots, m_K \in \Z_N}
\C|m_1,\ldots, m_K\rangle,
\end{equation}
where $\Z_N = \Z/(N\Z)$.
We write this simply as
$\mathscr{V}= \bigoplus_{\mathbf{m} \in \Z^K_N}|\mathbf{m}\rangle$
in terms of the array $\mathbf{m} = (m_1,\ldots, m_K)$.
At $q=\ve$, the $N$th powers of the generators 
$e^{\pm \uu_v}, e^{\pm \ww_v}$ $(v=1,\ldots, K)$ 
of $\mathcal{W}_K(q)$ become central.
 (See \eqref{qwv}.) 
 The specialized algebra admits the natural representation 
$\varrho: \mathcal{W}_K(q)|_{q=\ve} \rightarrow \mathrm{End}(\mathscr{V})$ given by
\begin{align}\label{repW}
e^{\uu_v} |\mathbf{m}\rangle = \ve^{m_v}  |\mathbf{m} \rangle, \quad 
e^{\ww_v} |\mathbf{m}\rangle =  |\mathbf{m}+\mathbf{e}_v \rangle
\quad (v=1,\ldots, K),
\end{align}
where $\mathbf{e}_v=(\delta_{1,v},\ldots, \delta_{K,v})$.
Hereafter, we denote
$\varrho(e^{\uu_v}|_{q=\ve})$ and $\varrho(e^{\ww_v}|_{q=\ve})$
simply by $e^{\uu_v}$ and $e^{\ww_v}$, respectively.
In this representation, $e^{N\uu_v}=e^{N\ww_v}=1$ holds, and 
$\dim \mathscr{V}=N^K$.
We denote the identity operator in $\mathrm{End}(\mathscr{V})$ by 
$\mathbb{I}$.

As in the previous section, 
let $G$ be an admissible diagram with $m$ boundary crossings on each of the
east and west sides and $n$ on each of the north and south sides, as described
in Section \ref{ss:ct}.
Recall the commuting transfer matrix 
$T(y) = T(\mathbf{i},\mathbf{a}|y)$ defined in \eqref{tai},
which corresponds to the mixed boundary condition specified by 
$\mathbf{a}, \mathbf{i} \in \{0,1\}^m$ satisfying 
$||\mathbf{a}||=||\mathbf{i}||$ in \eqref{ztc}.
Suppose that $T^{\mathbf{a}, \mathbf{b}}_{\mathbf{i}, \mathbf{b}}$ appearing 
therein admits the following expansion into monomials in the generators of
$\mathcal{W}_K(q)$:
\begin{align}
T^{\mathbf{a}, \mathbf{b}}_{\mathbf{i}, \mathbf{b}}
&= \sum_\gamma \Gamma_\gamma(\mathbf{a}, \mathbf{b}, \mathbf{i}) e^{\vv_\gamma}.
\label{tgv}
\end{align}
Here $\vv_\gamma$ ranges over distinct nonzero $\Z$-linear
combinations of the canonical variables $\uu_v$ and $\ww_v$.
By construction, each coefficient
$\Gamma_\gamma(\mathbf{a},\mathbf{b},\mathbf{i})$ is a {\em monomial}
in the parameters $r_v,s_v,f_v,g_v$ $(v=1,\ldots,K)$.
In particular, $\Gamma_\gamma(\mathbf{a},\mathbf{b},\mathbf{i})$
has no explicit dependence on $q$; all the $q$-dependence is carried
by the $q$-Weyl generators and their commutation relations.
Accordingly, the same coefficient appears unchanged after the
specialization $q=\ve$.
Examples are given in \eqref{Tex}.
Combining \eqref{tgv} with \eqref{tai}, we obtain the expansion
\begin{equation}\label{ty}
T(y) = T(\mathbf{i},\mathbf{a}|y) 
= \sum_{\mathbf{b} \in \{0,1\}^n, \gamma}
\Gamma_\gamma(\mathbf{a}, \mathbf{b}, \mathbf{i}) y^{||\mathbf{b}||}e^{\vv_\gamma}
\in \mathcal{W}_K(q).
\end{equation}

The counterparts of these objects at $q=\ve$ are elements of $\mathrm{End}(\mathscr{V})$,
which we denote as follows:
\begin{align}
&\mathbb{T}^{\mathbf{a}, \mathbf{b}}_{\mathbf{i}, \mathbf{b}}
= \varrho(T^{\mathbf{a}, \mathbf{b}}_{\mathbf{i}, \mathbf{b}}|_{q=\ve})
=  \sum_\gamma \Gamma_\gamma(\mathbf{a}, \mathbf{b}, \mathbf{i}) e^{\vv_\gamma},
\label{TTgv}
\\
&\mathbb{T}(y) = \varrho(T(y)|_{q=\ve})
= \sum_{\mathbf{b} \in \{0,1\}^n}
\mathbb{T}^{\mathbf{a}, \mathbf{b}}_{\mathbf{i}, \mathbf{b}}\, y^{||\mathbf{b}||},
\label{TTai} 
\\
&\mathbb{T}_\beta(y) =\varrho(T_\beta(y)|_{q=\ve})
=
\sum_{\hat{\mathbf{b}}\in \{0,1\}^{n-1}}
\mathbb{T}^{\mathbf{a}, \hat{\mathbf{b}}\beta}_{\mathbf{i},\, \hat{\mathbf{b}}\beta}
\, y^{||\hat{\mathbf{b}}||}
\quad (\beta=0,1),
\label{TTs}
\end{align}
where the last line gives the counterparts of the refined transfer matrices 
$T_0(y)$ and $T_1(y)$ in \eqref{T01}.
Substituting \eqref{TTgv} into \eqref{TTai} leads to
\begin{equation}\label{TTy}
\mathbb{T}(y) = \sum_{\mathbf{b} \in \{0,1\}^n, \gamma}
\Gamma_\gamma(\mathbf{a}, \mathbf{b}, \mathbf{i}) y^{||\mathbf{b}||}e^{\vv_\gamma}.
\end{equation}
The RHS expressions in the formulas \eqref{TTgv} and \eqref{TTy}
are formally identical to \eqref{tgv} and \eqref{ty},
respectively, in view of the convention stated after \eqref{repW}.
We note that \eqref{TTai}--\eqref{TTy} also depend on the boundary
choices $\mathbf{i}$ and $\mathbf{a}$, although this dependence is suppressed
in the notation.

The relations \eqref{ttz}, \eqref{Tdec}, \eqref{t01c} and \eqref{s01} 
then specialize naturally to $q=\ve$:
\begin{align}
&[\mathbb{T}(y), \mathbb{T}(z)]=0, \quad
[\mathbb{T}_0(y), \mathbb{T}_0(z)]=0,\quad
[\mathbb{T}_1(y), \mathbb{T}_1(z)]=0, 
\label{TTc}\\
&\mathbb{T}(y) = \mathbb{T}_0(y) + y^{s_y(n)}\mathbb{T}_1(y),
\label{TTd}
\\
&\mathbb{T}_0(\omega y)\mathbb{T}_1(y)
=\mathbb{T}_1(\omega y)\mathbb{T}_0(y).
\label{ss01}
\end{align}

\subsection{Main result}\label{ss:mr}

Referring to the expansion \eqref{TTy}, 
we introduce a Laurent polynomial $\mathscr{P}_N(y)$ in $y$ by
\begin{equation}\label{pdef}
\mathscr{P}_N(y) = 
\sum_{\mathbf{b} \in \{0,1\}^n, \gamma}
\Gamma_\gamma(\mathbf{a}, \mathbf{b}, \mathbf{i})^N y^{||\mathbf{b}||}.
\end{equation}

The main result of this paper is the following:

\begin{thm}[Quantum Frobenius property]\label{th:main}
The transfer matrix $\mathbb{T}(y)$ at $q=\ve$ satisfies the functional relation
\begin{equation}\label{main}
\mathbb{T}(y)\mathbb{T}(\ve y) \cdots \mathbb{T}(\ve^{N-1}y)
= \mathscr{P}_N(y^N)\mathbb{I}.
\end{equation}
\end{thm}
 
By virtue of the commutativity \eqref{TTc},
the LHS is invariant under the change $y \rightarrow \ve y$,
and hence must be a Laurent polynomial depending on $y$ only through $y^N$ as claimed.
The real content of the theorem lies in identifying this operator explicitly:
it is a {\em scalar} operator obtained from the $N$th power of the entire
sum in \eqref{TTy}, with all mixed ``off-diagonal'' contributions eliminated
and only the $N$th powers of the individual terms retained.
We call this the quantum Frobenius property, in analogy with the quantum
Frobenius morphisms for quantum groups and quantized function algebras at
roots of unity, through which classical structures emerge from suitable
$N$th-power elements \cite{L90,DCL94,G07}.
A distinctive feature of the present result is that this Frobenius phenomenon
incorporates the spectral parameter, through the product over its full
$\ve$-orbit $y,\ve y,\ldots,\ve^{N-1}y$.

Our proof consists of two main steps, carried out in Sections \ref{ss:yf} and \ref{ss:tf}.
They establish the quantum Frobenius property
with respect to the expansions in \eqref{TTai} and \eqref{TTgv}, respectively.
The proof of Theorem \ref{th:main} is completed at the end of
Section \ref{ss:tf}, where the two steps are combined.

\begin{remark}\label{re:uwrep}
The representation \eqref{repW} is only one particular choice.
More generally, since $e^{N\uu_v}$ and $e^{N\ww_v}$ are central at
$q=\ve$, their images may be prescribed as arbitrary nonzero scalar
multiples of $\mathbb{I}$.  Further families of representations are
obtained by pullback along nontrivial automorphisms of the $q$-Weyl
algebra.
Theorem~\ref{th:main} remains valid for these more general
representations, provided that the RHS of \eqref{pdef} is
replaced by
\[
\mathscr{P}_N(y)
=
\sum_{\mathbf{b}\in\{0,1\}^n,\gamma}
\Gamma_\gamma(\mathbf{a},\mathbf{b},\mathbf{i})^N
y^{||\mathbf{b}||}e^{N\vv_\gamma}.
\]
Here $e^{N\vv_\gamma}$ denotes the scalar by which this central element
acts in the chosen representation.  Indeed, the proof uses only the
scalarity of the $N$th powers of the $q$-Weyl generators, rather than
the normalization $e^{N\uu_v}=e^{N\ww_v}=1$ imposed in \eqref{repW}.
\end{remark}

\subsection{Frobenius property for the expansion \eqref{TTai}}\label{ss:yf}
 
The following lemma is a spectral-parameter-dependent version of the 
quantum Frobenius property 
with respect to the two-term decomposition \eqref{TTd}.
The functional relation \eqref{ss01} plays a key role.
 
\begin{lem}\label{le:tttt}
The transfer matrices $\mathbb{T}(y), \mathbb{T}_0(y)$ and 
$\mathbb{T}_1(y)$ satisfy the functional relation
\begin{equation}\label{tttt}
\begin{split}
&\mathbb{T}(y)\mathbb{T}(\ve y) \cdots \mathbb{T}(\ve^{N-1}y)
\\
& =  \mathbb{T}_0(y)\mathbb{T}_0(\ve y) \cdots \mathbb{T}_0(\ve^{N-1}y)
+ y^{Ns_y(n)}
\mathbb{T}_1(y)\mathbb{T}_1(\ve y) \cdots \mathbb{T}_1(\ve^{N-1}y),
\end{split}
\end{equation}
where $s_y(n)$ is defined in \eqref{sdy}.
\end{lem}
 
\begin{proof}
By the commutativity of $\mathbb{T}(y)$ and \eqref{qwn},
the LHS is equal to
\begin{equation}\label{cal}
\begin{split}
& \mathbb{T}(\omega^N y)\mathbb{T}(\omega^{N-1} y) 
 \cdots \mathbb{T}(\omega y)
 \\
 &=
 (\mathbb{T}_0(\omega^N y) + \omega^{Ns_y(n)} y^{s_y(n)}\mathbb{T}_1(\omega^N y))
 \cdots 
 (\mathbb{T}_0(\omega y) + \omega^{s_y(n)} y^{s_y(n)}\mathbb{T}_1(\omega y))
 \\
 &=
\sum_{r=0}^N y^{s_y(n)r}
\sum_{\substack{i_1,\ldots, i_N=0,1 \\ i_1+\cdots + i_N=r}}
\omega^{(i_1+2i_2+\cdots Ni_N)s_y(n)}
\mathbb{T}_{i_N}(\omega^N y) \cdots  \mathbb{T}_{i_1}(\omega y)
\\
&=\sum_{r=0}^N y^{s_y(n)r}
\mathbb{T}_{1}(\omega^N y) \cdots  \mathbb{T}_{1}(\omega^{N-r+1} y)
\mathbb{T}_{0}(\omega^{N-r} y) \cdots  \mathbb{T}_{0}(\omega y)
\sum_{\substack{i_1,\ldots, i_N=0,1 \\ i_1+\cdots + i_N=r}}
\omega^{(i_1+2i_2+\cdots Ni_N)s_y(n)}.
\end{split}
\end{equation}
Here we have used \eqref{TTd}, and in the last step the sequence of indices
$i_N,\ldots, i_1$ has been rearranged into
$\overbrace{1,\ldots,1}^r\overbrace{0,\ldots,0}^{N-r}$
by repeated use of \eqref{ss01}.
The multiple sum appearing here is equal to
$\delta_{r,0}+\delta_{r,N}$, since it is the coefficient of
$\chi^r$ in the expansion of
\begin{equation*}
\begin{split}
&\sum_{i_1,\ldots, i_N=0,1}
(\chi \omega^{s_y(n)})^{i_1}\cdots (\chi \omega^{s_y(n)N})^{i_N}
\\
&=(1+\chi\omega^{s_y(n)})(1+\chi\omega^{2s_y(n)})\cdots 
(1+\chi\omega^{Ns_y(n)}) = 1+\chi^N.
\end{split}
\end{equation*}
Thus only the terms with $r=0$ and $r=N$ survive in \eqref{cal}, and the
resulting expression coincides with the RHS of \eqref{tttt} by
\eqref{qwn} and \eqref{TTc}.
\end{proof}
 
The main result of this subsection is the following, where
$||\mathbf{b}||$ is defined in \eqref{ztc}:
\begin{prop}
\begin{equation}\label{FTy}
\mathbb{T}(y)\mathbb{T}(\ve y) \cdots \mathbb{T}(\ve^{N-1}y)
= \sum_{\mathbf{b} \in \{0,1\}^n}
\bigl(\mathbb{T}^{\mathbf{a}, \mathbf{b}}_{\mathbf{i}, \mathbf{b}}\bigr)^N
y^{N ||\mathbf{b}||}
\end{equation}
\end{prop}

\begin{proof}
In view of \eqref{TTai} and \eqref{TTs},
the decomposition \eqref{TTd} amounts to separating the summands in
\eqref{TTai} according to the last boundary variable $b_n=\beta=0,1$.
Lemma \ref{le:tttt} shows that, upon taking the product over the
$\ve$-orbit $y,\ve y,\ldots,\ve^{N-1}y$, the mixed contributions between
these two sectors disappear, while the $\ve$-orbit product structure is preserved
within each sector.

The same argument can be repeated after fixing $b_n$.
Indeed, the commutativity and the functional relation used in the proof
remain valid when additional north-south boundary variables are fixed.
We may therefore successively separate the two possibilities
$b_{n-1}=0,1$, then $b_{n-2}=0,1$, and so on, until all the north and south
boundary conditions are fixed by a single
$\mathbf{b}\in \{0,1\}^n$.

At the end of this iteration, each sector contains only
$\mathbb{T}^{\mathbf{a}, \mathbf{b}}_{\mathbf{i}, \mathbf{b}}
y^{||\mathbf{b}||}$.
The corresponding $\ve$-orbit product gives
$\bigl(\mathbb{T}^{\mathbf{a}, \mathbf{b}}_{\mathbf{i}, \mathbf{b}}\bigr)^N
y^{N||\mathbf{b}||}$.
Summing over all $\mathbf{b}\in\{0,1\}^n$ yields \eqref{FTy}.
\end{proof}
 
\subsection{Frobenius property for the expansion \eqref{TTgv}}\label{ss:tf}

For clarity, we formulate the second step of the proof intrinsically in terms of
simple tangles.  Let $\mathscr{T}$ be a simple tangle with $L$ oriented wires,
and let $A^{\bm{\mu}}_{\bm{\nu}}\in\mathcal{W}_K(q)$ be its fixed-boundary
amplitude with $\bm{\mu},\bm{\nu}\in\{0,1\}^L$, as in
Section~\ref{ss:ta}.  Throughout this subsection, we regard
$\mathscr{T}$, $\bm{\mu}$, $\bm{\nu}$, and the signs
$s(1),\ldots,s(L)$ as the basic data, without keeping track of their origin
from an admissible diagram $G$ and the boundary data
$\mathbf{a},\mathbf{b},\mathbf{i},\mathbf{j}$.
The correspondence with the original formulation will be restored only at the
end, through \eqref{TZ}.

Write the monomial expansion of the amplitude as
\begin{equation}\label{Aexp}
A^{\bm{\mu}}_{\bm{\nu}}
=\sum_\gamma \Gamma_\gamma(\bm{\mu},\bm{\nu})e^{\vv_\gamma},
\end{equation}
where each $\mathsf{v}_\gamma$ is a $\Z$-linear form 
in the canonical variables $\uu_v, \ww_v$, and 
$\Gamma_\gamma(\bm{\mu},\bm{\nu})$ is a monomial in the parameters
$r_v,s_v,f_v,g_v$ attached to the vertices of $\mathscr{T}$.
As in \eqref{tgv}, these coefficients have no explicit dependence on $q$.
At $q=\ve$, define the corresponding amplitude in $\mathrm{End}(\mathscr{V})$ by
\begin{equation}\label{Aroot}
\mathbb{A}^{\bm{\mu}}_{\bm{\nu}}
:=\varrho\bigl(A^{\bm{\mu}}_{\bm{\nu}}|_{q=\ve}\bigr)
=\sum_\gamma \Gamma_\gamma(\bm{\mu},\bm{\nu})e^{\vv_\gamma}.
\end{equation}
Here and below, following the convention introduced after \eqref{repW},
we suppress $\varrho$ from the notation for the specialized $q$-Weyl
generators.  Thus, $e^{\vv_\gamma}$ in \eqref{Aroot} is understood as
$\varrho\bigl(e^{\vv_\gamma}|_{q=\ve}\bigr)$.

Since $\varrho$ is an algebra homomorphism, the commutation relations in
Lemma~\ref{le:TT} specialize directly to the corresponding relations for
$\mathbb{A}$ with $q$ replaced by $\ve$.
In particular, the commutation factors that occur there are
$\ve^{\pm1}$ and $\ve^2$; all of them are primitive $N$th roots of unity
because $N$ is odd.  We shall repeatedly use the following elementary fact.

\begin{lem}\label{le:qfrob}
Let $X$ and $Y$ be elements of a $\C$-algebra, and let $\xi$ be a primitive
$N$th root of unity.  If
\begin{equation*}
XY=\xi YX,
\end{equation*}
then
\begin{equation*}
(X+Y)^N=X^N+Y^N.
\end{equation*}
\end{lem}

\begin{proof}
This follows immediately from the quantum binomial theorem: all intermediate
quantum binomial coefficients vanish when the deformation parameter is a
primitive $N$th root of unity.
\end{proof}

As a simple example, consider the two-term local weight
$\LL^{01}_{10}=rs e^{\ww}+fg e^{2\uu+\ww}$ in \eqref{L3}.
At $q=\ve$, its two summands $X=fg e^{2\uu+\ww}$ and
$Y=rs e^{\ww}$ satisfy $XY=\ve^2YX$, so Lemma~\ref{le:qfrob} gives
$(X+Y)^N=X^N+Y^N$.  In the representation \eqref{repW}, this reads
$(\varrho(\LL^{01}_{10}|_{q=\ve}))^N=
((fg)^N+(rs)^N)\mathbb{I}$.

We now prove the Frobenius property for fixed-boundary amplitudes.
For the induction argument, it is convenient to relax the global
requirement that the tangle arise from an actual admissible transformation
of an admissible diagram $G$.
The definitions of the fixed-boundary amplitude
$A^{\bm{\mu}}_{\bm{\nu}}$, its expansion \eqref{Aexp}, and its specialization
$\mathbb{A}^{\bm{\mu}}_{\bm{\nu}}$ in \eqref{Aroot} apply without change
to the resulting finite vertical concatenations.

\begin{prop}\label{prop:ZF}
Let $\mathscr{T}$ be a finite simple tangle as in Definition \ref{def:stan}.
Then, for every fixed boundary condition, the expansion \eqref{Aroot}
satisfies
\begin{equation}\label{FZ}
\bigl(\mathbb{A}^{\bm{\mu}}_{\bm{\nu}}\bigr)^N
=
\sum_\gamma
\Gamma_\gamma(\bm{\mu},\bm{\nu})^N\,\mathbb{I}.
\end{equation}
\end{prop}
\begin{proof}
We proceed by induction on the number of elementary pieces of $\mathscr{T}$.
For the empty tangle the amplitude is either $0$ or $1$, according to the
boundary condition, and the assertion is immediate.

Consider first a tangle built from
${\rm(o)},{\rm(t)},{\rm(I)},{\rm(iI)}$.
Write an $(\ell+1)$-piece tangle as
$\mathscr{T}_{\ell+1}=\mathscr{T}_1\mathscr{T}_{\ell}$, where
$\mathscr{T}_1$ is the uppermost elementary piece, and assume the assertion
for $\mathscr{T}_{\ell}$ with arbitrary fixed boundary conditions.
We suppress all unchanged boundary entries and write
$\mathbb{A}_{\ell}^{ij}$ for the amplitude of $\mathscr{T}_{\ell}$ with
the two boundary states joined to $\mathscr{T}_1$ fixed to $(i,j)$. 
See Figure~\ref{fig:conc}.

\begin{figure}[H]
\centering
\includegraphics[page=2,trim=150bp 80bp 150bp 610bp,clip,width=.70\textwidth]{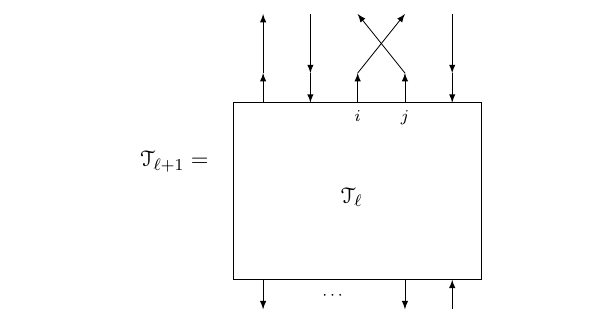}
\caption{Concatenation of tangle diagrams: the case of
$\mathscr{T}_{\ell+1} = \mathscr{T}_1 \mathscr{T}_{\ell}$ with
$\mathscr{T}_1 =$ (o) as the uppermost piece.
The signs specifying the
orientations of the two consecutive wires under consideration are
$(+1,+1)$.}
\label{fig:conc}
\end{figure}

For $\mathscr{T}_1={\rm(o)}$, the four possible states on the two upper
ends of $\mathscr{T}_1$ give, respectively,
\begin{equation}\label{oZ}
\begin{aligned}
(0,0):\quad & r\,\mathbb{A}_{\ell}^{00}, \\
(1,1):\quad & s\,\mathbb{A}_{\ell}^{11}, \\
(0,1):\quad & f e^{\uu}\mathbb{A}_{\ell}^{10}
          +e^{-\ww}\mathbb{A}_{\ell}^{01}, \\
(1,0):\quad & g e^{\uu}\mathbb{A}_{\ell}^{01}
          +(rs e^{\ww}+fg e^{2\uu+\ww})\mathbb{A}_{\ell}^{10}.
\end{aligned}
\end{equation}
The first two cases follow directly from the induction hypothesis.
For the third case, set
\begin{equation*}
X=f e^{\uu}\mathbb{A}_{\ell}^{10},\qquad
Y=e^{-\ww}\mathbb{A}_{\ell}^{01}.
\end{equation*}
The specialized form of \eqref{ZZ1}, together with the Weyl relation
\eqref{qwv}, gives
\begin{equation*}
XY=\ve^{-2}YX.
\end{equation*}
Since $\ve^{-2}$ is a primitive $N$th root of unity,
Lemma~\ref{le:qfrob} and the induction hypothesis give \eqref{FZ}.

For the last case in \eqref{oZ}, set
\begin{equation*}
X'=g e^{\uu}\mathbb{A}_{\ell}^{01},\qquad
Y'=(rs e^{\ww}+fg e^{2\uu+\ww})\mathbb{A}_{\ell}^{10}.
\end{equation*}
Again by \eqref{ZZ1} and \eqref{qwv},
\begin{equation*}
X'Y'=\ve^2Y'X'.
\end{equation*}
Moreover, the two summands in the local factor occurring in $Y'$ satisfy
\begin{equation*}
(rs e^{\ww})(fg e^{2\uu+\ww})
=\ve^{-2}(fg e^{2\uu+\ww})(rs e^{\ww}).
\end{equation*}
Thus Lemma~\ref{le:qfrob}, applied first to $X'+Y'$ and then to the two
summands within the local factor in $Y'$, together with the induction
hypothesis, gives \eqref{FZ}.

For $\mathscr{T}_1={\rm(t)}$, the signs specifying the orientations of
the relevant two consecutive wires are $(-1,-1)$. 
The corresponding nontrivial cases are
\begin{equation}\label{tZ}
\begin{split}
(0,1):\quad
&f e^{\uu}\mathbb{A}_{\ell}^{10}
 +(rs e^{\ww}+fg e^{2\uu+\ww})\mathbb{A}_{\ell}^{01},
\\
(1,0):\quad
&g e^{\uu}\mathbb{A}_{\ell}^{01}
 +e^{-\ww}\mathbb{A}_{\ell}^{10}.
\end{split}
\end{equation}
Here the relevant relation in \eqref{ZZ1} is
\begin{equation}
\mathbb{A}_{\ell}^{01}\mathbb{A}_{\ell}^{10}
=
\ve^{-1}\mathbb{A}_{\ell}^{10}\mathbb{A}_{\ell}^{01}.
\end{equation}
The same argument as above, using Lemma~\ref{le:qfrob}, proves
\eqref{FZ}.  The $(0,0)$ and $(1,1)$ cases are immediate.

If $\mathscr{T}_1={\rm(I)}$, then
$\mathbb{A}_{\ell+1}=\mathbb{A}_{\ell}$, so there is nothing to prove.
If $\mathscr{T}_1={\rm(iI)}$, then
\begin{equation}
\mathbb{A}_{\ell+1}
=
\mathbb{A}_{\ell}^{00}+\mathbb{A}_{\ell}^{11}.
\end{equation}
The specialized form of \eqref{ZZ2} gives
\begin{equation}
\mathbb{A}_{\ell}^{00}\mathbb{A}_{\ell}^{11}
=
\ve^2\mathbb{A}_{\ell}^{11}\mathbb{A}_{\ell}^{00},
\end{equation}
and hence \eqref{FZ} follows again from Lemma~\ref{le:qfrob} and the
induction hypothesis.  This completes the induction for the first class
in \eqref{tcls}.

For a tangle built from
${\rm(o)},{\rm(t)},\text{\rm(I')},\text{\rm(iI')}$, write instead
$\mathscr{T}_{\ell+1}=\mathscr{T}_{\ell}\mathscr{T}_1$, with the
elementary piece $\mathscr{T}_1$ at the bottom.
Thus the newly appended bottom piece again appears as the rightmost factor,
in agreement with the convention following Figure~\ref{fig:equiv}.
The same argument applies, using the relations in Lemma~\ref{le:TT}
corresponding to variations of the lower indices.
This proves \eqref{FZ} for the second class in
\eqref{tcls} as well.
\end{proof}

We now return to the fixed-boundary partition functions of the original
Q6V model.

\begin{cor}\label{cor:FTgv}
For the fixed-boundary matrix element in \eqref{TTgv},
\begin{equation}\label{FTgv}
\bigl(\mathbb{T}^{\mathbf{a},\mathbf{b}}_{\mathbf{i},\mathbf{b}}\bigr)^N
=
\sum_\gamma
\Gamma_\gamma(\mathbf{a},\mathbf{b},\mathbf{i})^N\mathbb{I}.
\end{equation}
\end{cor}

\begin{proof}
Set $\mathbf{j}=\mathbf{b}$ in the correspondence \eqref{TZab}.
The corresponding tangle is simple, so
Proposition~\ref{prop:ZF} applies.
The amplitude expansion \eqref{Aexp} then becomes \eqref{tgv}, while its root-of-unity
specialization \eqref{Aroot} is precisely
$\mathbb{T}^{\mathbf{a},\mathbf{b}}_{\mathbf{i},\mathbf{b}}$ in
\eqref{TTgv}.  This gives \eqref{FTgv}.
\end{proof}

Having established the two requisite Frobenius relations, we are finally
in a position to complete the proof of our main theorem.
\begin{proof}[Proof of Theorem~\ref{th:main}]
Substituting \eqref{FTgv} in Corollary~\ref{cor:FTgv} into \eqref{FTy}, we obtain
\begin{equation*}
\begin{split}
\mathbb{T}(y)\mathbb{T}(\ve y)\cdots\mathbb{T}(\ve^{N-1}y)
&=
\sum_{\mathbf{b}\in\{0,1\}^n,\gamma}
\Gamma_\gamma(\mathbf{a},\mathbf{b},\mathbf{i})^N
y^{N||\mathbf{b}||}\mathbb{I}
=\mathscr{P}_N(y^N)\mathbb{I},
\end{split}
\end{equation*}
where the last equality follows from \eqref{pdef}, as required.
\end{proof}

\begin{example}\label{ex:main}
We conclude this section by spelling out Theorem~\ref{th:main} for the
configuration of Example~\ref{ex:tabij3}.
After specialization to $q=\ve$, the decomposition \eqref{Tdec}, together
with $s_y(3)=1$, becomes
\[
\mathbb{T}(y)=\mathbb{T}_0(y)+y\mathbb{T}_1(y).
\]
Reading off the monomial coefficients from \eqref{T01e} and using
\eqref{pdef}, we obtain
\[
\begin{aligned}
\mathscr{P}_N(y)={}&\bigl((f_3g_4r_1)^N+(r_1r_2)^N\bigr)y
+(s_2s_3s_4)^N y^3
\\[-2pt]
&+\bigl((f_3g_4s_2)^N+(r_1s_3s_4)^N
+(f_2g_2)^N+(r_2s_2)^N\bigr)y^2.
\end{aligned}
\]
In particular, the two monomials in
$T^{11,101}_{00,101}$ displayed in \eqref{Tex} contribute separately as
$(f_2g_2)^N$ and $(r_2s_2)^N$ to the coefficient of $y^2$.
The elimination of their mixed powers is due to the fixed-boundary
Frobenius relation \eqref{FTgv}.

A closer examination of the two sectors gives the $\ve$-orbit products
\[
\begin{aligned}
\prod_{j=0}^{N-1}\mathbb{T}_0(\ve^j y)
&=\bigl((f_3g_4r_1)^Ny^N+(f_3g_4s_2)^Ny^{2N}\bigr)\mathbb{I},
\\
\prod_{j=0}^{N-1}\mathbb{T}_1(\ve^j y)
&=\bigl((r_1r_2)^N
+\bigl((r_1s_3s_4)^N+(f_2g_2)^N+(r_2s_2)^N\bigr)y^N
+(s_2s_3s_4)^Ny^{2N}\bigr)\mathbb{I}.
\end{aligned}
\]
Therefore, \eqref{tttt} gives
\[
\prod_{j=0}^{N-1}\mathbb{T}(\ve^j y)
=
\prod_{j=0}^{N-1}\mathbb{T}_0(\ve^j y)
+y^N\prod_{j=0}^{N-1}\mathbb{T}_1(\ve^j y)
=\mathscr{P}_N(y^N)\mathbb{I}.
\]
\end{example}

\subsection{Free-parafermion spectrum from the Frobenius property}\label{ss:fps}

The significance of the functional equation \eqref{main} is that
it determines the spectrum of $\mathbb{T}(y)$ up to multiplicities, as we shall now explain.
This fact has been recognized in several earlier works, for example,
\cite[Sec.~9]{B04}, \cite[Sec.~3]{B14}, and \cite[Sec.~V]{AP20}.

Recall that $\mathbb{T}(y)$ \eqref{TTai} is a
Laurent polynomial in $y$ obtained by specializing
$T(\mathbf{i},\mathbf{a}|y)$ in \eqref{tai} to $q=\ve$.
Using the definition of $||\mathbf{b}||$ in \eqref{ztc} and \eqref{sdy},
we set
\begin{align}
\kappa 
=\mathrm{min}\{||\mathbf{b}|| \mid \mathbf{b} \in \{0,1\}^n, 
T^{\mathbf{a},\mathbf{b}}_{\mathbf{i}, \mathbf{b}}\neq 0\},
\quad
\kappa+d
=\mathrm{max}\{||\mathbf{b}|| \mid \mathbf{b} \in \{0,1\}^n, 
T^{\mathbf{a},\mathbf{b}}_{\mathbf{i}, \mathbf{b}}\neq 0\}
\end{align}
and assume $d\ge 1$.
Then $\mathbb{T}(y)$ has  the expansion
\begin{equation}\label{tyexp}
\mathbb{T}(y) = \mathbb{H}_0 y^\kappa - \mathbb{H}_1 y^{\kappa+1}
+ \cdots + (-1)^d \mathbb{H}_d y^{\kappa+d},
\end{equation}
where we have introduced the commuting ``Hamiltonians'' 
$\mathbb{H}_0, \ldots, \mathbb{H}_d 
\in \mathcal{W}_K(q)|_{q=\ve}$.

According to \eqref{TTy} and \eqref{pdef}, the Laurent polynomial
$\mathscr{P}_N(y)$ has the parallel expansion
\begin{equation}\label{pex}
\mathscr{P}_N(y)
= C_0 y^\kappa - C_1 y^{\kappa+1}
+ \cdots + (-1)^d C_d y^{\kappa+d},
\end{equation}
where the scalars $C_0,\ldots,C_d$ are obtained directly from
the monomial expansion \eqref{TTy}.
In particular, $\mathbb{H}_0^N= C_0\mathbb{I}$ follows from \eqref{main}.
We write the factorized form of \eqref{pex} as
\begin{equation}\label{pconc}
\mathscr{P}_N(y)
= C_0 y^\kappa
\prod_{i=1}^d\Bigl(1-\frac{y}{\zeta_i}\Bigr),
\end{equation}
where $\zeta_1,\ldots,\zeta_d$ are nonzero scalars.
Let $\Lambda(y)$ be an eigenvalue of $\mathbb{T}(y)$.
The commutativity of $\mathbb{T}(y)$ implies that $\Lambda(y)$ is also
a Laurent polynomial in $y$ involving the powers
$y^\kappa,\ldots,y^{\kappa+d}$.

Matching its leading coefficient with \eqref{pex}, we may write
\begin{equation}\label{Lac}
\Lambda(y)
=
\ve^{\ell_0}C_0^{1/N} y^\kappa
\prod_{i=1}^d\Bigl(1-\frac{y}{\lambda_i}\Bigr),
\qquad
\ell_0\in\Z_N,
\end{equation}
for some nonzero scalars $\lambda_1,\ldots,\lambda_d$.
Substituting \eqref{pconc} and \eqref{Lac} into the quantum Frobenius relation
for the eigenvalues,
\begin{equation}
\Lambda(y) \cdots \Lambda(\ve^{N-1}y) = \mathscr{P}_N(y^N),
\end{equation}
we find that the $\lambda_i$ are determined by
\begin{equation}
\lambda_i = \ve^{-\ell_i}\zeta_i^{1/N}
\qquad (\ell_i \in \Z_N),
\qquad
\mathscr{P}_N(\zeta_i) = 0
\qquad (i = 1,\ldots,d),
\end{equation}
up to a permutation of the indices.
The latter relation just states that $\zeta_1,\ldots,\zeta_d$ are the
nonzero roots of the degree-$d$ polynomial
$y^{-\kappa}\mathscr{P}_N(y)$ in \eqref{pconc}.
(It may be viewed as playing the role of a Bethe equation.)
Throughout, $C_0^{1/N}$ and $\zeta_i^{1/N}$ denote fixed choices of
the $N$th roots.  The integers
$\ell_0,\ell_1,\ldots,\ell_d$ account for the corresponding
root-of-unity ambiguities.

Comparing the expansions \eqref{tyexp} and \eqref{Lac},
we find that the spectra of the Hamiltonians satisfy
\begin{align}\label{Hpfk}
\mathrm{Spec}(\mathbb{H}_k) \subseteq
\left\{
\ve^{\ell_0}C_0^{1/N}
\mathsf{e}_k(\ve^{\ell_1}\zeta^{-1/N}_1,\ldots,
\ve^{\ell_d}\zeta^{-1/N}_d)
\;\middle|\;
\ell_0,\ell_1,\ldots,\ell_d\in\Z_N
\right\}
\quad (0 \le k \le d),
\end{align}
where $\mathsf{e}_k$ denotes the $k$th elementary symmetric polynomial.
In particular, for the first Hamiltonian $\mathbb{H}_1$,
the shift $\ell_0$ can be absorbed into
$\ell_k \mapsto \ell_k+\ell_0$ $(1\le k\le d)$.
Therefore we obtain
\begin{align}\label{Hpf}
\mathrm{Spec}(\mathbb{H}_1) \subseteq
\left\{
\ve^{\ell_1}\mathcal{E}_1+\cdots+\ve^{\ell_d}\mathcal{E}_d
\;\middle|\;
\ell_1,\ldots,\ell_d\in\Z_N
\right\},
\qquad
\mathcal{E}_i=(C_0/\zeta_i)^{1/N}.
\end{align}

Note that $\dim \mathscr{V}=N^K$ in \eqref{Vcyc}, whereas
the RHS of \eqref{Hpf} is parametrized by $N^d$ choices.
This naturally raises interesting questions concerning the structure
and origin of spectral degeneracies.

Finally, since $\mathbb{H}_0, \ldots, \mathbb{H}_d$ are generally non-Hermitian,
the associated model should be understood 
as a quantum spin system in a generalized sense.

\section{Local gauge freedom}\label{s:t2}

In this section, we study local gauge changes of the 3D
$L$-operator and their effect on the associated transfer matrices.
We show that the commutativity and quantum Frobenius
properties persist when the two gauges are chosen independently
at each vertex.  

\subsection{3D $L$-operator in another gauge}\label{ss:Lag}

\begin{figure}[H]
\centering
{\unitlength 0.011in
\begin{picture}(490,75)(-15,30)
\put(6,80){
\put(-11,0){\vector(1,0){23}}\put(0,-10){\vector(0,1){22}}
}
\multiput(81,80.5)(75,0){6}{
\put(-11,0){\vector(1,0){23}}\put(0,-10){\vector(0,1){22}}
}
\put(-74,0){
\put(60.5,77){$i$}\put(77.5,60){$j$}
\put(96,77){$a$}\put(77.5,96.5){$b$}
}
\put(61,77){0}\put(78,60){0}\put(96,77){0}\put(78,96.5){0}
\put(75,0){
\put(61,77){1}\put(78,60){1}\put(96,77){1}\put(78,96.5){1}
}
\put(150,0){
\put(61,77){1}\put(78,60){0}\put(96,77){1}\put(78,96.5){0}
}
\put(225,0){
\put(61,77){0}\put(78,60){1}\put(96,77){0}\put(78,96.5){1}
}
\put(300,0){
\put(61,77){0}\put(78,60){1}\put(96,77){1}\put(78,96.5){0}
}
\put(375,0){
\put(61,77){1}\put(78,60){0}\put(96,77){0}\put(78,96.5){1}
}
\put(78,40){
\put(-77,0){$\bar{\mathscr{L}}^{ab}_{ij}$}
\put(0,0){$r$} \put(75,0){$s$} \put(144,0){$f e^\uu$}
\put(219,0){$g e^\uu$} 
\put(260,0){$rs e^{-\ww}\!+\! fg e^{2\uu-\ww}$}
\put(375,0){$e^\ww$}
}
\end{picture}
}
\caption{The operator $\bar{\mathscr{L}}^{ab}_{ij}(r,s,f,g;q)$.
Compared with Figure \ref{fig:6v}, the fifth and sixth weights are different.}
\label{fig:6vv}
\end{figure}
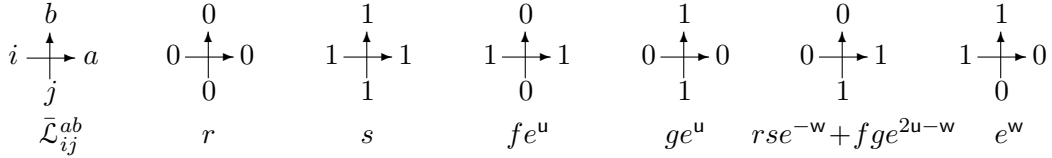

In this subsection, we assume that $q$ is generic.
For the 3D $L$-operator $\mathcal{L}=\mathcal{L}(r,s,f,g;q)$ in \eqref{L1},
we introduce an alternative gauge 
$\bar{\mathcal{L}} = \bar{\mathcal{L}}(r,s,f,g;q)$ by
\begin{subequations}\label{adL}
\begin{align}
\bar{\mathcal{L}}
&= (1 \otimes 1 \otimes \mathrm{Ad}_{\Psi(\lambda e^{2\uu})  e^{\kappa \uu}})(\mathcal{L}),
\label{Lad}\\
\Psi(z) &= \prod_{n=0}^\infty(1+q^{2n+1}z),
\quad
\lambda = \frac{fg}{rs},\quad 
q^{-\kappa} = rs,
\label{Psi}
\end{align}
\end{subequations}
where the adjoint,  defined by 
$\mathrm{Ad}_X(Y)=X Y X^{-1}$, acts on the 
third $q$-Weyl component.
From \eqref{qh}, \eqref{qcom} and the Baker-Campbell-Hausdorff formula, we have
\begin{equation*}
\begin{split}
&\Psi(\lambda e^{2\uu})e^{\kappa \uu} e^{-\ww} e^{-\kappa \uu} \Psi(\lambda e^{2\uu})^{-1}
= q^{-\kappa} \Psi(\lambda e^{2\uu}) e^{-\ww} \Psi(\lambda e^{2\uu})^{-1} 
= rs e^{-\ww} \Psi(\lambda q^{-2} e^{2\uu})\Psi(\lambda e^{2\uu})^{-1}
\\
& = rs e^{-\ww}(1+q^{-1}\lambda e^{2\uu})
= rs (e^{-\ww}+ \lambda e^{-\ww+ 2\uu}) = rs e^{-\ww} + fg e^{-\ww+2\uu}.
\end{split}
\end{equation*}
A similar calculation leads to 
\begin{subequations}\label{Lbdef}
\begin{align}
&\bar{\mathcal{L}}(r,s,f,g;q) 
= \sum_{a,b,i,j=0,1} E_{ai}\otimes E_{bj} \otimes \bar{\mathcal{L}}^{ab}_{ij}
\in \mathrm{End}(V \otimes V)  \otimes \mathcal{W}(q),
\label{Lb1}
\end{align}
where 
$\bar{\mathcal{L}}^{ab}_{ij} =
\bar{\mathcal{L}}^{ab}_{ij}(r,s,f,g;q)
= \mathrm{Ad}_{\Psi(\lambda e^{2\uu}) e^{\kappa \uu}}(\LL^{ab}_{ij})$
is given by 
\begin{align}
&\bar{\mathcal{L}}^{ab}_{ij}=0\; \text{unless}\; a+b=i+j,
\label{Lb2}
\\
&\bar{\mathcal{L}}^{00}_{00} = r,\;\;  \bar{\mathcal{L}}^{11}_{11} = s,\;\;
\bar{\mathcal{L}}^{10}_{10} = f e^\uu,\;\;
\bar{\mathcal{L}}^{01}_{01} = g e^\uu,
\;\;  \bar{\mathcal{L}}^{10}_{01} =  rs e^{-\ww} + fg e^{-\ww+2\uu}, \;\;
\bar{\mathcal{L}}^{01}_{10} = e^\ww,
\label{Lb3}
\end{align}
\end{subequations}
A graphical representation is given in Figure~\ref{fig:6vv}.

\subsection{Transfer matrices with independently chosen local gauges}\label{ss:Tag}

We continue to assume that $q$ is generic.
In Section \ref{ss:ct}, we defined the $\mathcal{W}_K(q)$-valued 
partition function $T^{\mathbf{a}, \mathbf{b}}_{\mathbf{i}, \mathbf{j}}$
with fixed boundary conditions
as the sum of products of local Boltzmann weights $\LL^{ab}_{ij}(r_v,s_v,f_v,g_v;q)$ 
assigned to the vertices $v=1,\ldots, K$ of an admissible diagram $G$.
For each vertex, set
\begin{equation}\label{Lpm}
\mathcal{L}^{(+)}_v=\mathcal{L}(r_v,s_v,f_v,g_v;q),
\qquad
\mathcal{L}^{(-)}_v=\bar{\mathcal{L}}(r_v,s_v,f_v,g_v;q).
\end{equation}
Choose
$\boldsymbol{\sigma}=(\sigma_v)_{v=1}^K\in\{+,-\}^K$, independently at
each vertex, and define
$T^{\mathbf{a}, \mathbf{b}}_{\mathbf{i}, \mathbf{j}}(\boldsymbol{\sigma})$
by using the local Boltzmann weights
$(\mathcal{L}^{(\sigma_v)}_v)^{ab}_{ij}$ at the vertex $v$.
This gives a family of $2^K$ gauge-equivalent fixed-boundary partition functions.
The partition function considered in the previous sections corresponds to the
all-plus choice $\boldsymbol{\sigma}=(+,\ldots,+)$.
By the construction \eqref{adL}, we have
\begin{subequations}\label{tep}
\begin{align}
T^{\mathbf{a}, \mathbf{b}}_{\mathbf{i}, \mathbf{j}}(\boldsymbol{\sigma})
&= \mathrm{Ad}_{X(\boldsymbol{\sigma})}
\bigl(T^{\mathbf{a}, \mathbf{b}}_{\mathbf{i}, \mathbf{j}}\bigr) \in \mathcal{W}_K(q),
\\
X(\boldsymbol{\sigma})
&= \prod_{1\le v\le K; \,\sigma_v=-}
\Psi(\lambda_v e^{2\uu_v})e^{\kappa_v \uu_v},
\quad 
\lambda_v = \frac{f_vg_v}{r_vs_v},\quad 
q^{-\kappa_v} = r_vs_v,
\end{align}
\end{subequations}
where the ordering in the product does not matter due to 
the commutativity \eqref{qwv}.

In view of \eqref{tep}, all the relations among the transfer matrices
that are adjoint invariant persist for each fixed $\boldsymbol{\sigma}$.
In particular, Theorem \ref{thm:TT2}, Proposition \ref{pr:tai}, Proposition \ref{pr:T01},
Proposition \ref{pr:t01}, Corollary \ref{co:t01} and Lemma \ref{le:TT} remain valid.

Let $T_{\boldsymbol{\sigma}}(y)$ be the transfer matrix obtained from
$T^{\mathbf{a},\mathbf{b}}_{\mathbf{i},\mathbf{j}}(\boldsymbol{\sigma})$.
Then Proposition \ref{pr:tai} extends naturally to
$[T_{\boldsymbol{\sigma}}(y),T_{\boldsymbol{\sigma}}(z)]=0$.

\subsection{Root-of-unity specialization.}
Let $q=\ve$ be as in \eqref{qwn}.
For each choice $\boldsymbol{\sigma}$, 
we define the partition functions at $q=\ve$ by specializing the matrix elements
directly:
\begin{equation}\label{TTsg}
\mathbb{T}_{\boldsymbol{\sigma}}(y)
=\varrho\bigl(T_{\boldsymbol{\sigma}}(y)|_{q=\ve}\bigr)
\in\mathrm{End}(\mathscr{V}).
\end{equation}
Since the commutation relations above are algebraic identities in the
explicit $q$-Weyl algebra matrix elements, they admit this specialization.
In particular, for every fixed $\boldsymbol{\sigma}$,
\begin{equation}\label{TTcm}
[\mathbb{T}_{\boldsymbol{\sigma}}(y),
 \mathbb{T}_{\boldsymbol{\sigma}}(z)]=0.
\end{equation}

For later reference, write the fixed-boundary matrix elements and the
transfer matrix in monomial form as
\begin{align}
T^{\mathbf{a},\mathbf{b}}_{\mathbf{i},\mathbf{b}}
 (\boldsymbol{\sigma})
&=
\sum_\gamma
\Gamma_{\boldsymbol{\sigma},\gamma}
 (\mathbf{a},\mathbf{b},\mathbf{i})
e^{\vv_{\boldsymbol{\sigma},\gamma}},
\label{tgs}
\\
T_{\boldsymbol{\sigma}}(y)
&=
\sum_{\mathbf{b}\in\{0,1\}^n,\gamma}
\Gamma_{\boldsymbol{\sigma},\gamma}
 (\mathbf{a},\mathbf{b},\mathbf{i})
y^{\|\mathbf{b}\|}e^{\vv_{\boldsymbol{\sigma},\gamma}}.
\label{tys}
\end{align}
As in \eqref{tgv},  $\vv_{\boldsymbol{\sigma},\gamma}$ 
ranges over distinct nonzero $\Z$-linear
combinations of the canonical variables $\uu_v$ and $\ww_v$.
Define
\begin{equation}\label{pds}
\mathscr{P}_{N,\boldsymbol{\sigma}}(y)
=
\sum_{\mathbf{b}\in\{0,1\}^n,\gamma}
\Gamma_{\boldsymbol{\sigma},\gamma}
 (\mathbf{a},\mathbf{b},\mathbf{i})^N
y^{\|\mathbf{b}\|}.
\end{equation}
Notice that this polynomial generally depends on $\boldsymbol{\sigma}$.

\begin{thm}[Quantum Frobenius property under independent local gauge changes]
\label{th:mixedF}
For every fixed $\boldsymbol{\sigma}\in\{+,-\}^K$, the directly specialized transfer
matrix \eqref{TTsg} satisfies
\begin{equation}\label{mixF}
\mathbb{T}_{\boldsymbol{\sigma}}(y)
\mathbb{T}_{\boldsymbol{\sigma}}(\ve y)\cdots
\mathbb{T}_{\boldsymbol{\sigma}}(\ve^{N-1}y)
=\mathscr{P}_{N,\boldsymbol{\sigma}}(y^N)\mathbb{I}.
\end{equation}
For the all-plus choice this is Theorem~\ref{th:main}.
\end{thm}

\begin{proof}
The proof of Theorem~\ref{th:main} has two ingredients, and both are
stable under an independent change of gauge at each vertex.  First, the
proof of the spectral-parameter Frobenius relation in
Lemma~\ref{le:tttt} uses only the two-term decomposition \eqref{TTd}, the
commutativity relations \eqref{TTc}, and the exchange relation
\eqref{ss01}.  The corresponding transfer-matrix relations persist for
each fixed $\boldsymbol{\sigma}$ by \eqref{tep}; being algebraic
identities in the explicit matrix elements, they remain valid after the
direct specialization $q=\ve$.

Second, the fixed-boundary Frobenius relation in
Corollary~\ref{cor:FTgv} is local in the $q$-Weyl algebras.  Replacing
$\mathcal{L}^{(+)}_v$ by $\mathcal{L}^{(-)}_v$ amounts, before
specialization, to pulling back the $v$th local representation by the
automorphism in \eqref{adL}.  The tangle proof is unchanged under such
independent local pullbacks; this is precisely the representation
independence explained in Remark~\ref{re:uwrep}.  Its final identities
involve only the explicit Laurent-polynomial matrix elements, so they can
again be specialized directly even though the conjugating infinite product
$\Psi$ itself is not specialized.  Consequently,
\[
\bigl(\mathbb{T}^{\mathbf{a},\mathbf{b}}_{\mathbf{i},\mathbf{b}}
(\boldsymbol{\sigma})\bigr)^N
=\sum_\gamma
\Gamma_{\boldsymbol{\sigma},\gamma}
(\mathbf{a},\mathbf{b},\mathbf{i})^N\mathbb{I}.
\]
Combining the two ingredients exactly as in the proof of
Theorem~\ref{th:main} gives \eqref{mixF}.
\end{proof}

This completes the extension to independent local gauge choices.

\section{The $\tau_2$ model and its free-parafermion specialization}
\label{s:tau2}

In this section, we exploit the local gauge freedom introduced in
Section~\ref{s:t2} for a two-row Q6V model at a root of unity and
identify its reduced transfer matrix with that of the $\tau_2$ model.
Under this identification, we also recover the $\tau_2$ Hamiltonian
and its free-parafermion specialization.

The $\tau_2$ model is known as an intermediate vertex model connecting
the six-vertex and chiral Potts models \cite{BS90,BBP90,B04}.
It includes, as a special case, the $\Z_N$-clock model \cite{B89}, in
which a free-parafermion spectrum was first observed.
After a period of relative inactivity, the subject was revived by
\cite{F14}, where the underlying integrable structure was studied under
the name {\em free parafermions}, leading to further developments in
\cite{AYP14,B14}.
More recent extensions and related perspectives can be found in
\cite{AP20,BHL23,MEWC25}.

\subsection{Q6V model on two-row admissible diagram}
We now show that a Q6V model on a simple two-row diagram
with appropriate gauges gives rise to the $\tau_2$ model at $q=\ve$.
We focus mainly on fixed boundary states, which are relevant to the
free-parafermion spectrum.  The identification also holds under periodic
boundary conditions upon taking the auxiliary trace, as noted after
\eqref{t2id}.  The argument below generalizes \cite[Sec.~5.2]{IKTY25}
without imposing the five-vertex parameter specialization
\cite[(5.7)]{IKTY25}.

\begin{figure}[H]
\[
\begin{tikzpicture}
\begin{scope}[>=latex,xshift=0pt]
\draw [<-] (-0.5,1)--(4.5,1);
\draw [->] (-0.5,0)--(4.5,0);
\draw [->] (0,-0.5)--(0,1.5);
\draw [->] (1,-0.5)--(1,1.5);
\draw [->] (2,-0.5)--(2,1.5);
\draw [->] (4,-0.5)--(4,1.5);
\node[left] at (-0.5,1) {$i_1$};
\node[left] at (-0.5,0) {$i_2$};
\node[right] at (4.5,1) {$a_1$};
\node[right] at (4.5,0) {$a_2$};
\node at (-2.0,1) {$\mathcal{L}^{(+)}$};
\node at (-2.0,0) {$\mathcal{L}^{(-)}$};
\draw [-] (-0.5,1.5)--(4.5,1.5);
\draw [-] (-0.5,-0.5)--(4.5,-0.5);
\draw [-] (-0.5,-0.5)--(-0.5,1.5);
\draw [-] (4.5,-0.5)--(4.5,1.5);
\draw(0,1) node[below right]{$\scriptstyle{1'}$};
\draw(0,0) node[below right]{$\scriptstyle{1}$};
\draw(1,1) node[below right]{$\scriptstyle{2'}$};
\draw(1,0) node[below right]{$\scriptstyle{2}$};
\draw(2,1) node[below right]{$\scriptstyle{3'}$};
\draw(2,0) node[below right]{$\scriptstyle{3}$};
\draw(4,1) node[below right]{$\scriptstyle{L'}$};
\draw(4,0) node[below right]{$\scriptstyle{L}$};
\end{scope}
\end{tikzpicture}
\]
\caption{The admissible diagram $G$ underlying the $\tau_2$ model.
The upper row carries $\mathcal{L}^{(+)}$, whereas the lower row carries
$\mathcal{L}^{(-)}$.}
\label{fig:t2G}
\end{figure}
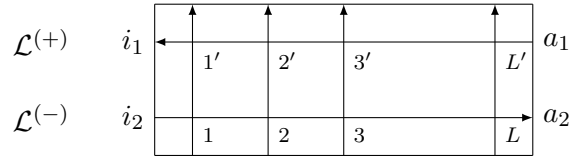

For an integer $L\geq 2$, let $G$ be the $2\times L$ square lattice
whose arrows and vertices are specified in Figure~\ref{fig:t2G}.
The graph $G$ is admissible.  We label the vertices in the lower row by
$1,\ldots,L$ and those in the upper row by $1',\ldots,L'$.  In the
notation \eqref{Lpm}, we assign
\begin{equation}\label{t2gs}
\mathcal{L}^{(+)}_{i'}
=\mathcal{L}(r_{i'},s_{i'},f_{i'},g_{i'};q),
\qquad
\mathcal{L}^{(-)}_i
=\bar{\mathcal{L}}(r_i,s_i,f_i,g_i;q)
\qquad (1\leq i\leq L)
\end{equation}
to the upper and lower rows, respectively.  At $q=\ve$, these local
operators are understood by the direct specialization described in the
preceding subsection.  Thus the two rows carry opposite gauges, as
indicated in the figure.  The four parameters $r_v,s_v,f_v,g_v$ at every
vertex are kept unrestricted.

\begin{figure}[H]
\centering
\def\acbase{%
  \draw[<-] (-0.4,0.6)--(1.6,0.6);
  \draw[->] (-0.4,0)--(1.6,0);
  \draw[->] (0,-0.4)--(0,1);
  \draw[->] (0.6,-0.4)--(0.6,1);
  \draw[->] (1.2,-0.4)--(1.2,1);
  \draw (-0.4,1)--(1.6,1);
  \draw (-0.4,-0.4)--(1.6,-0.4);
  \draw (-0.4,-0.4)--(-0.4,1);
  \draw (1.6,-0.4)--(1.6,1);
}
\begin{tabular}{@{}c@{\qquad}c@{\qquad}c@{}}
\begin{tikzpicture}[>=latex]
\acbase
\draw[very thick,->] (0,-0.4)--(0,1);
\node at (0.6,-0.75) {$(j,k)=(1,1)$};
\end{tikzpicture}
&
\begin{tikzpicture}[>=latex]
\acbase
\draw[very thick,->] (0.6,-0.4)--(0.6,1);
\node at (0.6,-0.75) {$(j,k)=(2,2)$};
\end{tikzpicture}
&
\begin{tikzpicture}[>=latex]
\acbase
\draw[very thick,->] (1.2,-0.4)--(1.2,1);
\node at (0.6,-0.75) {$(j,k)=(3,3)$};
\end{tikzpicture}
\\[5mm]
\begin{tikzpicture}[>=latex]
\acbase
\draw[very thick] (0,-0.4)--(0,0)--(0.6,0)--(0.6,0.6)--(0,0.6);
\draw[very thick,->] (0,0.6)--(0,1);
\node at (0.6,-0.75) {$(j,k)=(1,2)$};
\end{tikzpicture}
&
\begin{tikzpicture}[>=latex]
\acbase
\draw[very thick] (0,-0.4)--(0,0)--(1.2,0)--(1.2,0.6)--(0,0.6);
\draw[very thick,->] (0,0.6)--(0,1);
\node at (0.6,-0.75) {$(j,k)=(1,3)$};
\end{tikzpicture}
&
\begin{tikzpicture}[>=latex]
\acbase
\draw[very thick] (0.6,-0.4)--(0.6,0)--(1.2,0)--(1.2,0.6)--(0.6,0.6);
\draw[very thick,->] (0.6,0.6)--(0.6,1);
\node at (0.6,-0.75) {$(j,k)=(2,3)$};
\end{tikzpicture}
\end{tabular}
\caption{The one-path configurations contributing to the coefficient of $y^1$ in
$T(y)$ for $L=3$.  Thick and thin edges carry the local states $1$ and
$0$, respectively.  The label $(j,k)$, with $1\leq j\leq k\leq L$,
records the columns through which the thick path enters from the lower
boundary and exits at the upper boundary.  The boundary condition
$\mathbf{a}=\mathbf{i}=(0,0)$ requires the leftmost and rightmost
horizontal edges to be thin.}
\label{fig:ac}
\end{figure}
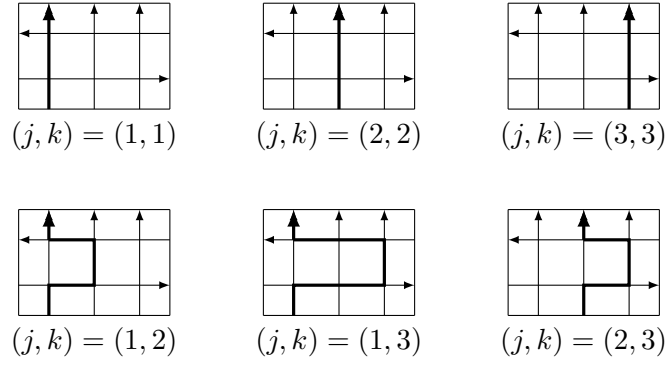

Let $T(y)=T(\mathbf{i},\mathbf{a}|y)$ be the commuting transfer matrix
defined in \eqref{ryak} and \eqref{tai}, with
$\mathbf{a}=(a_1,a_2)$ and $\mathbf{i}=(i_1,i_2)$ as indicated in
Figure~\ref{fig:t2G}.  We henceforth restrict to
$\mathbf{a}=\mathbf{i}=(0,0)$.  Then $T(y)$ is a polynomial of degree
$L$ in $y$.  The coefficient of $y^1$ has a particularly transparent
path interpretation: it receives contributions from $L(L+1)/2$
configurations labeled by $(j,k)$, where $1\leq j\leq k\leq L$.
The six configurations for $L=3$ are displayed in
Figure~\ref{fig:ac}.  We postpone the evaluation of their Boltzmann
weights until after establishing the model identification below.

\subsection{Reduction to $L$ cyclic degrees of freedom}
Set
\begin{equation}\label{Dred}
D_j=e^{\uu_j-\uu_{j'}} \qquad (1\leq j\leq L).
\end{equation}
The conservation law \eqref{L2}, applied to the two vertices in each column
and combined with the boundary condition
$\mathbf{a}=\mathbf{i}=(0,0)$, implies
\begin{equation}\label{TDred}
[T(y),D_j]=0 \qquad (1\leq j\leq L).
\end{equation}

We now specialize to $q=\ve$ as in \eqref{qwn} and use the cyclic
representation \eqref{repW}.  For the present $2L$-vertex graph, the space
$\mathscr{V}$ in \eqref{Vcyc} has dimension $N^{2L}$.  In accordance with
the two-row notation of Figure~\ref{fig:t2G}, we write the upper-row labels $m_{j'}$
above the lower-row labels $m_j$.  
By \eqref{TDred}, $\mathbb{T}(y)$ preserves
each common eigenspace of $D_1,\ldots,D_L$.  In general, these eigenspaces
are characterized by fixed shifts $m_{j'}=m_j+c_j$ with
$c_j\in\mathbb{Z}_N$.  We consider the zero-shift sector, in which
$c_j=0$ for all $j$:
\begin{equation}\label{Vred}
\begin{aligned}
\mathscr{V}_{\mathrm{red}}
&=\bigcap_{j=1}^L\ker(D_j-\mathbb{I})
=\bigoplus_{\mathbf{m}\in\mathbb{Z}_N^L}
  \mathbb{C}|\mathbf{m}\rangle_{\mathrm{red}},
\\[-1mm]
|\mathbf{m}\rangle_{\mathrm{red}}
&=\left.
\left|
\begin{matrix}
m_{1'}&\cdots&m_{L'}\\
m_1&\cdots&m_L
\end{matrix}
\right\rangle
\right|_{m_{j'}=m_j\; (\forall j)},
\end{aligned}
\end{equation}
The subspace $\mathscr{V}_{\mathrm{red}} \subset 
\mathscr{V}$ has dimension $N^L$.  Thus the two
cyclic degrees of freedom in each column can be reduced to one.
Since this subspace is invariant under $\mathbb{T}(y)$, we set
\begin{equation}\label{that}
\widehat{\mathbb{T}}(y)
:=\left.\mathbb{T}(y)\right|_{\mathscr{V}_{\mathrm{red}}} \in 
\mathrm{End}(\mathscr{V}_{\mathrm{red}}).
\end{equation}

On $\mathscr{V}_{\mathrm{red}}$, define
\begin{equation}\label{XZred}
X_j=\left.e^{\uu_j+\uu_{j'}}\right|_{\mathscr{V}_{\mathrm{red}}},
\qquad
Z_j=\left.e^{-\ww_j-\ww_{j'}}\right|_{\mathscr{V}_{\mathrm{red}}}.
\end{equation}
Using the unit vector $\mathbf{e}_j=(\delta_{j1},\ldots, \delta_{jL})$ 
in $\mathbb{Z}_N^L$, one has
\begin{equation}\label{XZac}
X_j|\mathbf{m}\rangle_{\mathrm{red}}
=\omega^{m_j}|\mathbf{m}\rangle_{\mathrm{red}},
\qquad
Z_j|\mathbf{m}\rangle_{\mathrm{red}}
=|\mathbf{m}-\mathbf{e}_j\rangle_{\mathrm{red}},
\qquad
Z_jX_k=\omega^{\delta_{jk}}X_kZ_j,
\end{equation}
where $\omega=\ve^2$ as in \eqref{qwn}.  In particular,
\begin{equation}\label{u2red}
\left.e^{2\uu_j}\right|_{\mathscr{V}_{\mathrm{red}}}
=\left.e^{2\uu_{j'}}\right|_{\mathscr{V}_{\mathrm{red}}}
=X_j.
\end{equation}

\subsection{Generalized $\tau_2$ local matrix}
Although the horizontal states are fixed to $(0,0)$ at the boundaries, the
intermediate ones can be either $(0,0)$ or $(1,1)$.  Thus each
column is naturally represented by a $2\times2$ matrix on this two-state space.

\begin{samepage}
The corresponding local matrices can be displayed directly. 
Take the $2\times1$ specialization of Figure~\ref{fig:t2G}
consisting of the lower vertex $j$ and the upper vertex $j'$.
In the following single-column formulas, we abbreviate
$(r_j,s_j,f_j,g_j)$ and $(r_{j'},s_{j'},f_{j'},g_{j'})$ as
$(r,s,f,g)$ and $(r',s',f',g')$, respectively. Ordering the rows by
$\mathbf{i}=(0,0),(1,1)$ and the columns by
$\mathbf{a}=(0,0),(1,1)$, set
\def\tblock#1#2#3#4{%
\begin{tikzpicture}[baseline={(current bounding box.center)},>=latex,
  x=0.52cm,y=0.39cm,font=\tiny]
\draw[<-] (-0.5,0.85)--(0.5,0.85);
\draw[->] (-0.5,0.15)--(0.5,0.15);
\draw[->] (0,-0.42)--(0,1.42);
\draw (-0.5,-0.42) rectangle (0.5,1.42);
\node[left] at (-0.5,0.85) {$#1$};
\node[left] at (-0.5,0.15) {$#2$};
\node[right] at (0.5,0.85) {$#3$};
\node[right] at (0.5,0.15) {$#4$};
\node[above=-1pt] at (0,1.42) {$b$};
\node[below=-1pt] at (0,-0.42) {$b$};
\end{tikzpicture}%
}
\begin{equation}
\mathsf{L}^{(b)}=
\begin{pmatrix}
T^{00,b}_{00,b} & T^{11,b}_{00,b}\\[1mm]
T^{00,b}_{11,b} & T^{11,b}_{11,b}
\end{pmatrix}
\quad =\quad
\begin{pmatrix}
\tblock{0}{0}{0}{0} & \tblock{0}{0}{1}{1}\\[2mm]
\tblock{1}{1}{0}{0} & \tblock{1}{1}{1}{1}
\end{pmatrix}
\quad (b=0,1).
\end{equation}
Set
$\mathcal{L}^{(-)}_{\mathrm{low}}=
\bar{\mathcal{L}}(r,s,f,g;q)$ and
$\mathcal{L}^{(+)}_{\mathrm{up}}=
\mathcal{L}(r',s',f',g';q)$.
If $c\in\{0,1\}$ denotes the internal vertical state in a column, then,
more explicitly,
\begin{equation}\label{t2ct}
(\mathsf{L}^{(b)})_{\mathbf{i},\mathbf{a}}
=\sum_{c=0,1}
\bigl(\mathcal{L}^{(-)}_{\mathrm{low}}\bigr)^{a_2\, c}_{i_2\, b}
\bigl(\mathcal{L}^{(+)}_{\mathrm{up}}\bigr)^{b\, i_1}_{c \, a_1}.
\end{equation}
\end{samepage}
At $q=\ve$, the two choices of the vertical boundary state give
\begin{subequations}\label{t2Lb}
\begin{align}
\mathsf{L}^{(0)}
&=
\begin{pmatrix}
rr' & 0\\[1mm]
e^{\ww+\ww'}
  \bigl(r's'+\ve f'g'e^{2\uu'}\bigr)
& fg'e^{\uu+\uu'}
\end{pmatrix},
\label{t2L0}
\\[2mm]
\mathsf{L}^{(1)}
&=
\begin{pmatrix}
gf'e^{\uu+\uu'}
& e^{-\ww-\ww'}
  \bigl(rs+\ve^{-1}fge^{2\uu}\bigr)\\[1mm]
0 & ss'
\end{pmatrix}.
\label{t2L1}
\end{align}
\end{subequations}
Thus, their weighted sum on
$\mathscr{V}_{\mathrm{red}}$ is
\begin{equation}\label{t2L}
\mathsf{L}(y)
=\left.(\mathsf{L}^{(0)}+y\mathsf{L}^{(1)})\right|_{\mathscr{V}_{\mathrm{red}}}
=
\begin{pmatrix}
rr'+ygf'X
& yZ\bigl(rs+\ve^{-1}fgX\bigr)\\[1mm]
Z^{-1}\bigl(r's'+\ve f'g'X\bigr)
& fg'X+yss'
\end{pmatrix}.
\end{equation}
This agrees, up to an overall factor, with the local matrix in
\cite[(6)]{AYP14}.  More precisely, set $\omega=\ve^2$ according to 
\eqref{qwn}, and define
$\mathsf{L}_j(y)$ from $\mathsf{L}(y)$ by the substitutions
\begin{equation}\label{AYPd}
\begin{gathered}
X\mapsto X_j,\qquad Z\mapsto Z_j,\\
r=b_{2j-2},\qquad s=c_{2j-1},\qquad
f=-\ve a_{2j-1},\qquad g=d_{2j-2},\\
r'=-\omega^{-1}b_{2j-1},\qquad s'=c_{2j-2},\qquad
f'=d_{2j-1},\qquad g'=\ve^{-1}a_{2j-2}.
\end{gathered}
\end{equation}
Then the $2\times2$ matrix formed by the four entries in \cite[(6)]{AYP14}
is related to $\mathsf{L}_j(y)$ as
\begin{equation}\label{AYPm}
\mathcal{L}^{\mathrm{AYP}}_j(t)
:=
\begin{pmatrix}
\mathcal{L}_j(0,0) & \mathcal{L}_j(0,1)\\
\mathcal{L}_j(1,0) & \mathcal{L}_j(1,1)
\end{pmatrix}
=-\omega\,\mathsf{L}_j(t).
\end{equation}
The concrete cyclic bases used in \cite{AYP14} and in \eqref{XZac} are
related by a discrete Fourier transform.

Multiplying the local matrices in column order gives the following identity:
\begin{subequations}\label{t2id}
\begin{align}
\mathsf{L}_1(t)\cdots\mathsf{L}_L(t)
&=(-\omega^{-1})^L
  \mathcal{L}^{\mathrm{AYP}}_1(t)\cdots
  \mathcal{L}^{\mathrm{AYP}}_L(t).
\label{t2M}
\end{align}
Our transfer matrix with the
fixed horizontal boundary states is given by
\begin{align}
\widehat{\mathbb{T}}(t)
&=\bigl[\mathsf{L}_1(t)\cdots\mathsf{L}_L(t)\bigr]_{00}.
\label{t2T}
\end{align}
\end{subequations}
Here $[\,\cdot\,]_{00}$ denotes the $(0,0)$ matrix element in the
two-state space.  Apart from the overall factor in \eqref{t2M},
equation~\eqref{t2T} is precisely the open-boundary transfer matrix of
\cite{AYP14}.  Indeed, the boundary matrix
$\mathcal{L}_0=\operatorname{diag}(1,0)$ in \cite[(8)]{AYP14} is a
projector, so the trace in \cite[(4)]{AYP14} selects this matrix element.
Thus, we have
\begin{equation}\label{tht2}
\widehat{\mathbb{T}}(t) = (-\omega^{-1})^L
\boldsymbol{\tau}_2(t),
\end{equation}
where the RHS is given by \cite[(10), (13), (14)]{AYP14}.
Replacing
$[\,\cdot\,]_{00}$ in \eqref{t2T} by the trace gives the
corresponding periodic-boundary transfer matrix.

\subsection{Hamiltonians}
We now return to the one-path configurations in Figure~\ref{fig:ac}.
For $1\leq j\leq k\leq L$, let $h_{jk}$ denote the Boltzmann weight,
evaluated on $\mathscr{V}_{\mathrm{red}}$, of the configuration labeled
by $(j,k)$.  Direct evaluation at $q=\ve$ gives
\begin{subequations}\label{hjkr}
\begin{align}
 h_{jj}
&=\Bigl(\prod_{\ell \in [1,j) \cup (j,L]}r_\ell r_{\ell'}\Bigr)
  g_jf_{j'}X_j,
\label{hrd}
\\
 h_{jk}
&=\Bigl(\prod_{\ell \in [1,j) \cup (k,L]}r_\ell r_{\ell'}\Bigr)
  Z_j\bigl(r_js_j+\ve^{-1}f_jg_jX_j\bigr)
\Bigl(\prod_{\ell=j+1}^{k-1}f_\ell g_{\ell'}X_\ell\Bigr)
  Z_k^{-1}\bigl(r_{k'}s_{k'}+\ve f_{k'}g_{k'}X_k\bigr).
\label{hro}
\end{align}
\end{subequations}
Here and below, an empty product is understood to be $1$.
The expansion of the transfer matrix in the spectral parameter $y$ takes the form 
(the convention in \eqref{THexp} differs slightly from that in \eqref{tyexp}):
\begin{subequations}\label{Tyred}
\begin{align}
\widehat{\mathbb{T}}(y)
&=\mathsf{c}_0\mathbb{I}
  +y\widehat{\mathbb{H}}+\sum_{j=2}^Ly^j\widehat{\mathbb{H}}_j,
\label{THexp}\\
\mathsf{c}_0
&=\prod_{j=1}^{L}r_jr_{j'}
=(-\omega^{-1})^L\prod_{j=0}^{2L-1}b_j,
\label{Tc0}\\
\widehat{\mathbb{H}}
&=\sum_{1\leq j\leq k\leq L} h_{jk},
\qquad
\widehat{\mathbb{H}}_L = \prod_{j=1}^Lh_{jj}.
\label{HH}
\end{align}
\end{subequations}
Equation~\eqref{Tc0} follows from the substitutions \eqref{AYPd}.
On the other hand, according to \cite[(10), (13), (14)]{AYP14}, the
fixed-boundary transfer matrix of the $\tau_2$ model has the expansion
$\boldsymbol{\tau}_2(t)=A_0(\mathbb{I}+\omega t\mathcal{H}
+\mathcal{O}(t^2))$.
It then follows from \eqref{tht2} that
\begin{equation}\label{HAYP}
\widehat{\mathbb{H}}=\omega\mathsf{c}_0\mathcal{H}.
\end{equation}

Let us verify the consistency of \eqref{HAYP}.
In $\widehat{\mathbb{H}}$, defined by \eqref{HH} and \eqref{hjkr}, we
apply the substitutions \eqref{AYPd} column by column, interpreting
$(r,s,f,g)$ and $(r',s',f',g')$ as
$(r_j,s_j,f_j,g_j)$ and
$(r_{j'},s_{j'},f_{j'},g_{j'})$, respectively.  We further set
$b_j=1$ for all $j$, and hence
$r_\ell=-\omega r_{\ell'}=1$ for all $\ell$, in accordance with the
normalization in \cite[(3.6)]{B14}.  Then
$(\omega\mathsf{c}_0)^{-1}\widehat{\mathbb{H}}$ reproduces precisely the
Hamiltonian $\mathcal{H}$ in \cite[(3.22)]{B14}.  Indeed, the part with
$k=j$ in the first double sum of \cite[(3.22)]{B14} gives
$(\omega\mathsf{c}_0)^{-1}\sum_j h_{jj}$ from \eqref{hrd}.  The part
with $j<k$ in the same sum, together with the other three double sums in
\cite[(3.22)]{B14}, coincides with
$(\omega\mathsf{c}_0)^{-1}\sum_{j<k}h_{jk}$ from \eqref{hro}.

Note that $\widehat{\mathbb{H}}_L$ measures the ``total charge'':
$\widehat{\mathbb{H}}_L|\mathbf{m}\rangle_{\mathrm{red}}
=
\mathrm{const}\cdot \omega^{m_1+\cdots+m_L}
|\mathbf{m}\rangle_{\mathrm{red}}$.
Hence the commutativity
$[\widehat{\mathbb{T}}(y),\widehat{\mathbb{H}}_L]=0$
implies that $\mathscr{V}_{\mathrm{red}}$ decomposes into a direct sum of charge sectors
labeled by $m_1+\cdots+m_L\in\mathbb{Z}_N$.

The free-parafermion Hamiltonian is obtained from the $\tau_2$ Hamiltonian
$\widehat{\mathbb{H}}$ in \eqref{HAYP} by imposing
$f_j=0$ for $1\leq j<L$ and $g_{j'}=0$ for $2\leq j\leq L$,
which is equivalent to the specialization
$a_1=\cdots=a_{2L-2}=0$ in \cite[(3.25)]{B14}.
Under this
specialization, $h_{jk}$ vanishes unless $k=j$ or $k=j+1$.  With the
normalization specified above, one obtains
\begin{equation}\label{hfp}
\widehat{\mathbb{H}} = (-\omega^{-1})^{L-1}
\bigl(\sum_{j=1}^L g_jf_{j'}X_j 
+ \sum_{j=1}^{L-1}s_js_{(j+1)'}Z_jZ^{-1}_{j+1}\bigr).
\end{equation}
Up to a change of notation for the parameters entering the coefficients,
this reproduces the original free-parafermion Hamiltonian in
\cite[(1)]{B89}.

\begin{remark}
In \cite[Sec.~5.2]{IKTY25}, the free-parafermion Hamiltonian
\eqref{hfp} was obtained without employing both gauges
$\LL^{(+)}$ and $\LL^{(-)}$.  This simplification relies on the
specialization $f_j=0$ for $1\leq j<L$ and $g_{j'}=0$ for $2\leq j\leq L$.
For the general $\tau_2$ model considered here, however, both gauges are
essential.
\end{remark}

\section{Relativistic quantum Toda chain}\label{s:td}

In this section, we apply the quantum Frobenius property to the
relativistic quantum Toda chain.  We first derive a $2$ by $2$
matrix-product representation for the four nontrivial Q6V transfer
matrices with fixed boundary states.
For the homogeneous chain at an odd root of unity, we then compute the
associated scalar polynomials explicitly in terms of Chebyshev
polynomials and use their zeros to describe the free-parafermion
spectra associated with these transfer matrices.
We conclude by comparing our results with those of \cite{PS02}.

\subsection{Q6V realization and boundary conditions}\label{ss:tb}

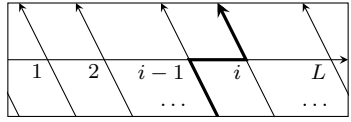
\begin{figure}[H]
\centering
\begin{tikzpicture}[>=stealth,scale=0.75,font=\scriptsize]
\draw[->] (0.3,1) -- (6.3,1);

\begin{scope}
\clip (0.3,0) rectangle (6.3,2);
\draw[->] (0.5,0) -- (-0.5,2);
\draw[->] (6.5,0) -- (5.5,2);
\end{scope}

\draw[->] (1.5,0) -- (0.5,2);
\draw[->] (2.5,0) -- (1.5,2);
\draw[->] (4,0) -- (3,2);
\draw[->] (5,0) -- (4,2);

\node at (3.25,0.2) {$\dots$};
\node at (5.75,0.2) {$\dots$};

\node[below left=-2pt] at (1,1) {$1$};
\node[below left=-2pt] at (2,1) {$2$};
\node[below left=-2pt] at (3.5,1) {$i-1$};
\node[below left=-2pt] at (4.5,1) {$i$};
\node[below left=-2pt] at (6,1) {$L$};

\draw[very thick,->]
(4,0) -- (3.5,1) -- (4.5,1) -- (4,2);

\draw (0.3,0) rectangle (6.3,2);
\end{tikzpicture}
\caption{The admissible diagram for the relativistic quantum
Toda chain, adapted from \cite[Fig.~11]{IKTY25}.
The chain length is denoted by $N$ in \cite{IKTY25} and by $L$ here.
The configuration shown contributes to the Toda Hamiltonian.}
\label{fig:ftod}
\end{figure}

Recall from \cite[Sec.~5.3]{IKTY25} that the relativistic quantum
Toda chain is realized as the Q6V model on the admissible diagram
shown in Figure~\ref{fig:ftod}.

We adopt the normalization $r_i=f_i=1$ and introduce the $2$ by $2$
matrix
\begin{equation}\label{tdai}
A_i(y)=
\begin{pmatrix}
1&yv_i\\
z_i&ys_i
\end{pmatrix},
\qquad
z_i=e^{-\ww_i},
\qquad
v_i=s_i e^{\ww_i}+g_i e^{\ww_i+2\uu_i}.
\end{equation}

There are two wire segments crossing each of the east and west
boundaries of the fundamental domain.  The boundary arrays carry
signed weights determined by the orientations of these segments.
By \eqref{ztc} and \eqref{sdy}, we have
$\lVert\mathbf{a}\rVert=a_1-a_2$ and
$\lVert\mathbf{i}\rVert=i_1-i_2$.
The constraint
$\lVert\mathbf{a}\rVert=\lVert\mathbf{i}\rVert$ therefore leaves
precisely the six boundary conditions
\begin{equation}\label{tdbc}
(\mathbf{i},\mathbf{a})
=
(10,10),\ (01,01),\ (00,00),\
(11,00),\ (00,11),\ (11,11).
\end{equation}
Here the first and second entries in each pair specify the west and
east boundary states $\mathbf{i}$ and $\mathbf{a}$, respectively.
Recall that $T(\mathbf{i},\mathbf{a}|y)$, defined in \eqref{tai},
denotes the transfer matrix with boundary states
$\mathbf{i}$ and $\mathbf{a}$.  The first two cases are trivial:
\begin{equation}\label{tdtr}
T(10,10|y)=\prod_{i=1}^L e^{\uu_i},
\qquad
T(01,01|y)=y^L\prod_{i=1}^L g_i e^{\uu_i}.
\end{equation}
We henceforth focus on the remaining four sectors.  Summing over the
internal Q6V states yields
\begin{equation}\label{tdmo}
\begin{pmatrix}
T(00,00|y)&T(11,00|y)\\
T(00,11|y)&T(11,11|y)
\end{pmatrix}
=
A_L(y)A_{L-1}(y)\cdots A_1(y).
\end{equation}
Here $\mathbf{i}$ and $\mathbf{a}$ label the columns and rows,
respectively.  Formula \eqref{tdmo} can also be obtained by solving
the corresponding recursion relations in the chain length $L$.

For later reference, we record the leading and next-to-leading
terms in the expansions in $y$ of the four matrix elements in
\eqref{tdmo}.  For $L\geq3$, direct expansion gives
\begin{subequations}\label{tdha}
\begin{align}
T(00,00|y)
&=
1+y\sum_{i=2}^L v_i z_{i-1}
+O(y^2),
\label{tdh1}\\
T(11,00|y)
&=
yv_1
+y^2\left(
s_1v_2
+v_1\sum_{i=3}^L v_i z_{i-1}
\right)
+O(y^3),
\label{tdh2}\\
T(00,11|y)
&=
z_L
+y\left(
z_L\sum_{i=2}^{L-1}v_i z_{i-1}
+s_Lz_{L-1}
\right)
+O(y^2),
\label{tdh3}\\
T(11,11|y)
&=
yv_1z_L
+y^2\left[
z_L\left(
s_1v_2
+v_1\sum_{i=3}^{L-1}v_i z_{i-1}
\right)
+s_Lz_{L-1}v_1
\right]
+O(y^3).
\label{tdh4}
\end{align}
\end{subequations}
In particular,
\begin{subequations}
\begin{align}
&T(00,00|y)+T(11,11|y)
=
1+y\mathcal{H}+O(y^2),
\\
&\mathcal{H}
=
\sum_{i=1}^L
\left(
s_i e^{\ww_i-\ww_{i-1}}
+g_i e^{2\uu_i+\ww_i-\ww_{i-1}}
\right),
\quad
(\ww_0=\ww_L).
\end{align}
\end{subequations}
Under the identification in \cite[(5.27)]{IKTY25},
$\mathcal{H}$ is the relativistic Toda Hamiltonian
\cite[(5.18)]{IKTY25} with \emph{periodic} boundary conditions.
At odd roots of unity, our general result instead determines the
spectra associated with the individual fixed-boundary-state transfer
matrices $T(\mathbf{i},\mathbf{a}|y)$.
As seen from \eqref{tdha}, the coefficients in their expansions do not
in general admit an interpretation as Toda Hamiltonians with boundary
terms or fixed boundary conditions.

\subsection{Homogeneous specialization}\label{ss:tdf}

We now specialize to $q=\ve$, where $\ve$ is a primitive odd $N$th
root of unity, and use the cyclic representation \eqref{repW}.
We assume that the parameters are homogeneous:
\[
s_i=s,\qquad g_i=g\qquad (i=1,\ldots,L),
\]
and set
\begin{equation}\label{tdsg}
Y=y^N,\qquad S=s^N,\qquad G=g^N,\qquad V=S+G.
\end{equation}
Since $N$ is odd, $\ve^2$ is again a primitive $N$th root of unity.
The elementary quantum Frobenius identity, recalled in
Lemma~\ref{le:qfrob}, therefore yields
\[
\left(s e^{\ww}+g e^{\ww+2\uu}\right)^N=S+G=V,
\qquad
\left(e^{-\ww}\right)^N=1.
\]
It follows from Theorem~\ref{th:main} that the quantum Frobenius
relations
\begin{equation}\label{tdac0}
\begin{pmatrix}
\prod_{j=0}^{N-1}\mathbb{T}(00,00|\ve^j y)&
\prod_{j=0}^{N-1}\mathbb{T}(11,00|\ve^j y)\\
\prod_{j=0}^{N-1}\mathbb{T}(00,11|\ve^j y)&
\prod_{j=0}^{N-1}\mathbb{T}(11,11|\ve^j y)
\end{pmatrix}
=
\begin{pmatrix}
\mathscr P_N^{00,00}(y^N)&\mathscr P_N^{11,00}(y^N)\\
\mathscr P_N^{00,11}(y^N)&\mathscr P_N^{11,11}(y^N)
\end{pmatrix}
\end{equation}
hold.  
Applying to \eqref{tdmo} the same prescription that produces
\eqref{pdef} from \eqref{TTy}, we obtain the matrix on the RHS
explicitly as
\begin{equation}\label{tdac}
\begin{pmatrix}
\mathscr P_N^{00,00}(Y)&\mathscr P_N^{11,00}(Y)\\
\mathscr P_N^{00,11}(Y)&\mathscr P_N^{11,11}(Y)
\end{pmatrix}
=
\mathcal A(Y)^L,
\qquad
\mathcal A(Y)=
\begin{pmatrix}
1&YV\\
1&YS
\end{pmatrix}.
\end{equation}

For later use, set
\begin{equation}\label{tdtd}
\tau(Y)=\operatorname{tr}\mathcal A(Y)=1+SY,
\qquad
\Delta(Y)=\det\mathcal A(Y)=-GY,
\qquad
X(Y)=\frac{\tau(Y)}{2\sqrt{\Delta(Y)}}.
\end{equation}
We denote the Chebyshev polynomial of the second kind by
$\mathsf U_n$.  It is defined by
\begin{equation}\label{tdcd}
\mathsf U_n(\cos\theta)
=
\frac{\sin((n+1)\theta)}{\sin\theta},
\end{equation}
or, equivalently, by
\[
\mathsf U_0(x)=1,\qquad
\mathsf U_1(x)=2x,\qquad
\mathsf U_{n+1}(x)
=
2x\mathsf U_n(x)-\mathsf U_{n-1}(x).
\]
Define the polynomials $F_n(Y)$ by
\begin{equation}\label{tdfr}
F_{-1}(Y)=0,\qquad F_0(Y)=1,\qquad
F_n(Y)=\tau(Y)F_{n-1}(Y)-\Delta(Y)F_{n-2}(Y).
\end{equation}
For $n\geq1$, they may equivalently be written as
\begin{align}\label{tdfd}
F_n(Y)
&=
\det
\begin{pmatrix}
\tau(Y)&\Delta(Y)&&&\\
1&\tau(Y)&\Delta(Y)&&\\
&1&\ddots&\ddots&\\
&&\ddots&\tau(Y)&\Delta(Y)\\
&&&1&\tau(Y)
\end{pmatrix}_{n\times n}
=
\Delta(Y)^{n/2}
\mathsf U_n\left(X(Y)\right).
\end{align}
Although the last expression contains a square root, it is the
polynomial specified by the recurrence \eqref{tdfr}.
The Cayley--Hamilton identity gives
$\mathcal A(Y)^2=\tau(Y)\mathcal A(Y)-\Delta(Y)I$,
where $I$ denotes the $2$ by $2$ identity matrix.
Multiplying this identity successively by $\mathcal A(Y)$ and using
the recurrence \eqref{tdfr}, we obtain
\begin{equation}\label{tdch}
\mathcal A(Y)^L
=
F_{L-1}(Y)\mathcal A(Y)
-\Delta(Y)F_{L-2}(Y)I.
\end{equation}
Consequently,
\begin{subequations}\label{tdfp}
\begin{align}
\mathscr P_N^{00,00}(Y)
&=
F_{L-1}(Y)-\Delta(Y)F_{L-2}(Y),
\label{tdp1}\\
\mathscr P_N^{11,00}(Y)
&=
YV F_{L-1}(Y),
\label{tdp2}\\
\mathscr P_N^{00,11}(Y)
&=
F_{L-1}(Y),
\label{tdp3}\\
\mathscr P_N^{11,11}(Y)
&=
YS F_{L-1}(Y)-\Delta(Y)F_{L-2}(Y).
\label{tdp4}
\end{align}
\end{subequations}

\subsection{Free-parafermion spectra}\label{ss:ts}

Following \eqref{tyexp}, we write the expansion of the transfer
matrix as
\begin{align}
\mathbb{T}(\mathbf{i},\mathbf{a}|y)
=
\mathbb{H}^{\mathbf{i},\mathbf{a}}_0y^\kappa
-\mathbb{H}^{\mathbf{i},\mathbf{a}}_1y^{\kappa+1}
+O(y^{\kappa+2}).
\end{align}
Here we explicitly indicate the dependence of the Hamiltonians
on the boundary states $\mathbf{i}$ and $\mathbf{a}$.  After the
root-of-unity specialization, their explicit forms can be read
off from \eqref{tdha}.
Let
\begin{equation}\label{pian}
\mathscr P_N^{\mathbf{i},\mathbf{a}}(Y)
=
C_0Y^\kappa
\prod_{k=1}^{d}
\left(1-\frac{Y}{\zeta_k}\right)
\end{equation}
be any of the four nontrivial polynomials in \eqref{tdfp}.
For generic parameters, $d=L-1$, and the leading data are
\begin{subequations}
\begin{align}
&(\kappa,C_0)=(0,1)
\quad\,\text{for }(00,00)\text{ and }(00,11),\\
&(\kappa,C_0)=(1,V)
\quad\text{for }(11,00)\text{ and }(11,11).
\end{align}
\end{subequations}
These statements follow from $F_n(0)=1$ for $n\geq0$.
Section~\ref{ss:fps} then gives
\begin{equation}\label{tdsp}
\mathrm{Spec}\bigl(\mathbb{H}_1^{\mathbf{i},\mathbf{a}}\bigr)
\subseteq
\Bigl\{
\ve^{\ell_1}\mathcal{E}_1+\cdots+
\ve^{\ell_{L-1}}\mathcal{E}_{L-1}
\;\Bigm|\;
\ell_1,\ldots,\ell_{L-1}\in\Z_N
\Bigr\},
\quad
\mathcal{E}_k
=
\left(\frac{C_0}{\zeta_k}\right)^{1/N}.
\end{equation}

The two polynomials corresponding to the mixed boundary conditions,
given in \eqref{tdp2} and \eqref{tdp3}, have the same nonzero roots:
they are precisely the zeros of $F_{L-1}(Y)$.  By \eqref{tdfd} and
\eqref{tdcd}, these roots are characterized by
\begin{equation}\label{tdoz}
\mathsf U_{L-1}\left(X(\zeta_k)\right)=0,
\qquad
X(\zeta_k)=\cos\frac{k\pi}{L},
\qquad
k=1,\ldots,L-1.
\end{equation}
Changing the choice of the square root in $X(Y)$, defined in
\eqref{tdtd}, only interchanges the indices $k$ and $L-k$ and
hence leaves the set of roots unchanged.

For the two polynomials corresponding to the diagonal boundary
conditions, given in \eqref{tdp1} and \eqref{tdp4}, the nonzero
roots are determined by
\begin{subequations}\label{tddz}
\begin{align}
\mathsf U_{L-1}\left(X(\zeta)\right)
&=
\sqrt{\Delta(\zeta)}\,
\mathsf U_{L-2}\left(X(\zeta)\right)
\quad\text{for }(\mathbf{i},\mathbf{a})=(00,00),
\label{tdza}\\
\zeta S\,
\mathsf U_{L-1}\left(X(\zeta)\right)
&=
\sqrt{\Delta(\zeta)}\,
\mathsf U_{L-2}\left(X(\zeta)\right)
\quad\text{for }(\mathbf{i},\mathbf{a})=(11,11).
\label{tdzb}
\end{align}
\end{subequations}
Despite their appearance, these equations are independent of the
choice of the square root.  Equivalently, by \eqref{tdp1} and
\eqref{tdp4}, they are polynomial equations in $\zeta$.
Their roots determine the corresponding free-parafermion energies
through \eqref{tdsp}.

We finally compare these results with the relativistic Toda chain
at a root of unity studied in \cite{PS02}.  In the notation of that
paper, the four entries $a(\lambda),b(\lambda),c(\lambda),d(\lambda)$
of the quantum monodromy matrix satisfy cyclic product relations
whose RHSs are the corresponding entries
$A(\lambda^N),B(\lambda^N),C(\lambda^N),D(\lambda^N)$ of the
classical monodromy matrix; see \cite[Prop.~2.1]{PS02}.  After
matching the gauge and spectral-parameter conventions, these
relations correspond to the specialization of the quantum
Frobenius relation \eqref{main} 
to the four Q6V transfer matrices with fixed boundary states in \eqref{tdac0}.

For the homogeneous chain, \cite{PS02} parametrizes the zeros of
the off-diagonal monodromy element by $\phi_k=k\pi/L$ for
$k=1,\ldots,L-1$; see \cite[(2.29)--(2.33)]{PS02}, with their chain
length $M$ replaced by $L$.  The angles $\phi_k$ are precisely
those appearing in the zeros of the Chebyshev polynomial
$\mathsf U_{L-1}$, so this parametrization is equivalent to the
Chebyshev condition \eqref{tdoz}.
The emphasis in \cite{PS02}, however, is on the periodic transfer
matrix $a(\lambda)+d(\lambda)$ and on the use of the zeros of
$b(\lambda)$ as separation variables.  In the present formulation,
the four monodromy entries are interpreted as Q6V transfer matrices
with fixed boundary states.  The quantum Frobenius
property, together with the argument of Section~\ref{ss:fps}, then
determines the free-parafermion energies directly from the zeros
of the corresponding scalar polynomials
$\mathscr P_N^{\mathbf{i},\mathbf{a}}(Y)$, as in \eqref{tdsp}.
    
\section{Summary and outlook}\label{s:ol}

We have established the quantum Frobenius property of
the layer transfer matrix $\mathbb{T}(y)$ \eqref{TTai}
of the Q6V model at odd roots of unity
(Theorems \ref{th:main} and \ref{th:mixedF}).
The result holds for any admissible diagram $G$ (Definition \ref{def:adm})
and for arbitrary choices of the boundary conditions $\mathbf{a}, \mathbf{i}$;
see, for example, \eqref{Tabij}.
Our proof exploits the underlying three-dimensional integrability,
whose essential structures are most transparently revealed through the
tetrahedron equations.
See, for example, Figures \ref{fig:mtt}, \ref{fig:emtt}, \ref{fig:rtttt}, and \ref{fig:mzz}.
The tetrahedron equations already imply the standard commutativity
of the transfer matrices, as in Proposition \ref{pr:tai} and
\cite[Thm.~3.4]{IKTY25}.  In the present work, we push their use further to derive
nontrivial functional relations for the transfer matrix
(Proposition \ref{pr:T01}) and the $q$-commutativity
of the quantum amplitudes (Lemma \ref{le:TT}).

As explained in Section \ref{ss:fps},
the quantum Frobenius property implies the free-parafermion spectrum
for the associated Hamiltonians of generalized quantum spin systems.
This yields a broad family of models: besides the freedom to choose
a genuinely two-dimensional diagram $G$, one may choose the boundary
data $\mathbf{a},\mathbf{i}$, as well as the parameters
$r,s,f,g$ and the gauge \eqref{Lpm} independently at each site.
For suitable choices of $G$, the Q6V model
admits a specialization to the one-dimensional $\tau_2$ model, as shown in \eqref{tht2}.
From this viewpoint, our main formulas \eqref{main} and \eqref{mixF}
may be regarded as far-reaching generalizations of the spectral relations
in \cite[(76)]{B04} and \cite[(3.13)]{B14}, and ultimately of the original
observation going back to \cite{B89}.

For the relativistic quantum Toda chain, realized by the admissible
diagram in Figure~\ref{fig:ftod}, we identify the four nontrivial
transfer matrices with fixed, possibly different, boundary
states as the four entries of the matrix product \eqref{tdmo}.  For
the homogeneous chain at an odd root of unity, the corresponding
scalar polynomials in the quantum Frobenius relations are obtained as
the entries of $\mathcal A(Y)^L$ and are expressed explicitly in terms
of Chebyshev polynomials.  Their zeros then determine the corresponding
free-parafermion spectra through \eqref{tdsp}.  These formulas also
recover the homogeneous parametrization in
\cite[(2.29)--(2.33)]{PS02} and clarify its relation to the cyclic
product relations of \cite[Prop.~2.1]{PS02}.  

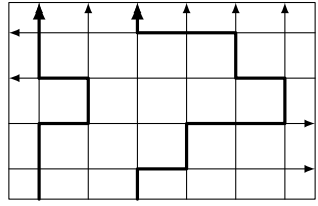
\begin{figure}[H]
\centering
\begin{tikzpicture}[>=latex,line cap=round,line join=round]
\draw[<-] (-0.4,1.8)--(3.65,1.8);
\draw[<-] (-0.4,1.2)--(3.65,1.2);
\draw[->] (-0.4,0.6)--(3.65,0.6);
\draw[->] (-0.4,0)--(3.65,0);
\foreach \x in {0,0.65,1.3,1.95,2.6,3.25}
  \draw[->] (\x,-0.4)--(\x,2.2);
\draw (-0.4,-0.4) rectangle (3.65,2.2);
\draw[very thick]
  (0,-0.4)--(0,0.6)--(0.65,0.6)--(0.65,1.2)--(0,1.2);
\draw[very thick,->] (0,1.2)--(0,2.2);
\draw[very thick]
  (1.3,-0.4)--(1.3,0)--(1.95,0)--(1.95,0.6)--
  (3.25,0.6)--(3.25,1.2)--(2.6,1.2)--(2.6,1.8)--(1.3,1.8);
\draw[very thick,->] (1.3,1.8)--(1.3,2.2);
\end{tikzpicture}
\caption{A two-path configuration on a $4 \times 6$ admissible diagram.  Thick and
thin edges carry the local states $1$ and $0$, respectively.}
\label{fig:2p}
\end{figure}

The results of Sections~\ref{s:tau2} and \ref{s:td} place the
$\tau_2$ model and the relativistic quantum Toda chain on the same
footing within the Q6V framework: in both cases, free-parafermion
spectra at odd roots of unity emerge from the quantum Frobenius
property.
Moreover, a comparison of the admissible diagrams in
Figures~\ref{fig:t2G} and \ref{fig:ftod} suggests that, even within
the class of regular lattices, the $\tau_2$ and Toda cases may be
viewed as the first two members of a natural sequence.  From this
perspective, it would be interesting to investigate the models
associated with more general choices of $G$.  Figure~\ref{fig:2p}
provides an example of such an admissible diagram $G$, together with
a two-path configuration contributing to the $y^2$ term in the
expansion of $\mathbb{T}(y)$.

For generic choices of the local parameters, the resulting Hamiltonians
may involve interactions extending along the paths; 
it would be interesting to seek suitable model settings
for which genuinely \emph{local} two-dimensional Hamiltonians emerge.

The present work also points to several outstanding problems, including
(i) the model at even roots of unity,
(ii) the construction of ``shift'' operators as in \cite[Sec.5]{F14}, 
(iii) the structure and origin of spectral degeneracies, and
(iv) connections with other models possessing known free-parafermion spectra;
see, for example, \cite{AP20}.

\section*{Acknowledgments}
The authors would like to thank
Murray Batchelor, Souta Oikawa, and Yuan Miao for their kind interest in this work.
RI is supported by JSPS KAKENHI Grant Number 23K03048. 
AK is supported by JSPS KAKENHI Grant Number 24K06882.
YT is supported by JSPS KAKENHI Grant Numbers 21K03240, 22H01117, and 25K06969.
JY is supported by NSFC Grant Number 12375064.

Generative AI tools were used on a limited basis for language editing,
literature searches, and preliminary mathematical discussions.
All mathematical content and references were independently verified by the authors.

\appendix

\section{Graphical representations of tetrahedron and inversion relations}\label{s:ap}

We provide graphical representations of the
tetrahedron equations and the inversion relations in Proposition \ref{pr:te4}.
The figures below are taken from \cite{IKTY25} for the reader's convenience.

\begin{itemize}
\item[(o)] Ordinary type: 
$\MM_{126} \MM_{346} \LL_{135} \LL_{245} 
= \LL_{245} \LL_{135} \MM_{346} \MM_{126}$. 
\begin{align}\label{g:Lo}
\begin{tikzpicture}
\begin{scope}[>=latex,xshift=0pt]
\draw [-] (0,0) coordinate(A) to [out = 0, in = -135] (2,0.5) coordinate(B);
\draw [-] (1,1) coordinate(C) to [out = 0, in = 135] (B);
\draw [-] (C) to [out = 90, in = -45] (0.5,2) coordinate(D);
\draw [-] (A) to [out = 90, in = -135] (D);
\draw [-] (-0.5,0) node[left]{$1$} --(A); \draw [-] (0,-0.5) node[below]{$3$}--(A);
{\color{blue} 
\draw [<-] (-0.5,-0.5)--(A)--(C)--(1.5,1.5) node[above right]{$5$}[thick]; %
}
{\color{green}
\draw [<-] (2,0)--(B) to [out = 90, in = 0] (D) -- (0,2) node[left]{$6$}[thick];}
\draw [<-] (2.5,0)--(B); \draw [<-] (2.5,1)--(B);
\draw [-] (1,0.5) node[below]{$4$}--(C); \draw [-] (0.5,1) node[left]{$2$}--(C);
\draw [<-] (1,2.5)--(D); \draw [<-] (0,2.5)--(D);
\draw (3.3,1) node {$=$}; 
\end{scope}
\begin{scope}[>=latex,xshift=135pt]
\draw [-] (2,2) coordinate(A) to [out = 180, in = 45] (0,1.5) coordinate(B);
\draw [-] (1,1) coordinate(C) to [out = 180, in = -45] (B);
\draw [-] (C) to [out = -90, in = 135] (1.5,0) coordinate(D);
\draw [-] (A) to [out = -90, in = 45] (D);
\draw [<-] (2.5,2)--(A); \draw [<-] (2,2.5)--(A);
{\color{blue} 
\draw [->] (2.5,2.5) node[above right]{$5$}--(A)--(C)--(0.5,0.5) [thick]; %
}
{\color{green}
\draw [->] (0,2) node[above]{$6$}--(B) to [out = -90, in = 180] (D)--(2,0)[thick]; 
}
\draw [-] (-0.5,2) node[left]{$2$}--(B); \draw [-] (-0.5,1) node[left]{$1$}--(B);
\draw [<-] (1.5,1)--(C); \draw [<-] (1,1.5)--(C);
\draw [-] (1,-0.5) node[below left]{$3$}--(D); \draw [-] (2,-0.5) node[below right]{$4$}--(D); 
\end{scope}
\end{tikzpicture}
\end{align}

\item[(h)] Horizontally reversed type:
$[\MM_{346} \LL_{315} \LL_{425} \MM_{126}]_{\ast 6} 
= [\MM_{126} \LL_{425} \LL_{315} \MM_{346}]_{\ast 6}$ for $r'=s'$ and $g' = qf'$.
\begin{align}\label{g:Lh}
\begin{tikzpicture}
\begin{scope}[>=latex,xshift=0pt]
\draw [-] (0,0) coordinate(A) to [out = 0, in = -135] (2,0.5) coordinate(B);
\draw [-] (1,1) coordinate(C) to [out = 0, in = 135] (B);
\draw [-] (C) to [out = 90, in = -45] (0.5,2) coordinate(D);
\draw [-] (A) to [out = 90, in = -135] (D);
\draw [<-] (-0.5,0) node[left]{$1$} --(A); \draw [-] (0,-0.5) node[below]{$3$}--(A);
{\color{blue} 
\draw [-] (A)--(C) [thick]; %
\draw [<-] (-0.5,-0.5)--(A)[thick];%
\draw [-] (1.5,1.5) node[above right]{$5$}--(C)[thick]; %
}
{\color{green}
\draw [-] (B) to [out = 90, in = 0] (D)[thick]; 
\draw [<-] (2,0)--(B) [thick]; 
\draw [-] (0,2) node[left]{$6$}--(D)[thick]; 
}
\draw [-] (2.5,0)--(B); \draw [-] (2.5,1)--(B);
\draw [-] (1,0.5) node[below]{$4$}--(C); \draw [<-] (0.5,1) node[left]{$2$}--(C);
\draw [<-] (1,2.5)--(D); \draw [<-] (0,2.5)--(D);
\draw (3.3,1) node {$=$}; 
\end{scope}
\begin{scope}[>=latex,xshift=135pt]
\draw [-] (2,2) coordinate(A) to [out = 180, in = 45] (0,1.5) coordinate(B);
\draw [-] (1,1) coordinate(C) to [out = 180, in = -45] (B);
\draw [-] (C) to [out = -90, in = 135] (1.5,0) coordinate(D);
\draw [-] (A) to [out = -90, in = 45] (D);
\draw [-] (2.5,2)--(A); \draw [<-] (2,2.5)--(A);
{\color{blue} 
\draw [-] (A)--(C) [thick]; %
\draw [-] (2.5,2.5) node[above right]{$5$} --(A) [thick];%
\draw [<-] (0.5,0.5)--(C)[thick];%
}
{\color{green}
\draw [-] (B) to [out = -90, in = 180] (D)[thick]; 
\draw [-] (0,2) node[above]{$6$}--(B) [thick]; 
\draw [<-] (2,0)--(D)[thick]; 
}
\draw [<-] (-0.5,2) node[left]{$2$}--(B); \draw [<-] (-0.5,1) node[left]{$1$}--(B);
\draw [-] (1.5,1)--(C); \draw [<-] (1,1.5)--(C);
\draw [-] (1,-0.5) node[below left]{$3$}--(D); \draw [-] (2,-0.5) node[below right]{$4$}--(D); 
\end{scope}
\end{tikzpicture}
\end{align}
\item[(v)] Vertically reversed type: 
$\MM_{126} \LL_{315} \LL_{425} \MM_{346} 
= \MM_{346} \LL_{425} \LL_{315} \MM_{126}$ for $r'=s'$ and $g'=q^{-1}f'$.
\begin{align}\label{g:Lv}
\begin{tikzpicture}
\begin{scope}[>=latex,xshift=0pt]
\draw [-] (0,0) coordinate(A) to [out = 0, in = -135] (2,0.5) coordinate(B);
\draw [-] (1,1) coordinate(C) to [out = 0, in = 135] (B);
\draw [-] (C) to [out = 90, in = -45] (0.5,2) coordinate(D);
\draw [-] (A) to [out = 90, in = -135] (D);
\draw [-] (-0.5,0) node[left]{$1$} --(A); \draw [<-] (0,-0.5) node[below]{$3$}--(A);
{\color{blue} 
\draw [-] (A)--(C) [thick]; %
\draw [<-] (-0.5,-0.5)--(A)[thick];%
\draw [-] (1.5,1.5) node[above right]{$5$}--(C)[thick]; %
}
{\color{green}
\draw [-] (B) to [out = 90, in = 0] (D)[thick]; 
\draw [<-] (2,0)--(B) [thick]; 
\draw [-] (0,2) node[left]{$6$}--(D)[thick]; 
}
\draw [<-] (2.5,0)--(B); \draw [<-] (2.5,1)--(B);
\draw [<-] (1,0.5) node[below]{$4$}--(C); \draw [-] (0.5,1) node[left]{$2$}--(C);
\draw [-] (1,2.5)--(D); \draw [-] (0,2.5)--(D);
\draw (3.3,1) node {$=$}; 
\end{scope}
\begin{scope}[>=latex,xshift=135pt]
\draw [-] (2,2) coordinate(A) to [out = 180, in = 45] (0,1.5) coordinate(B);
\draw [-] (1,1) coordinate(C) to [out = 180, in = -45] (B);
\draw [-] (C) to [out = -90, in = 135] (1.5,0) coordinate(D);
\draw [-] (A) to [out = -90, in = 45] (D);
\draw [<-] (2.5,2)--(A); \draw [-] (2,2.5)--(A);
{\color{blue} 
\draw [-] (A)--(C) [thick]; %
\draw [-] (2.5,2.5) node[above right]{$5$} --(A) [thick];%
\draw [<-] (0.5,0.5)--(C)[thick];%
}
{\color{green}
\draw [-] (B) to [out = -90, in = 180] (D)[thick]; 
\draw [-] (0,2) node[above]{$6$}--(B) [thick]; 
\draw [<-] (2,0)--(D)[thick]; 
}
\draw [-] (-0.5,2) node[left]{$2$}--(B); \draw [-] (-0.5,1) node[left]{$1$}--(B);
\draw [<-] (1.5,1)--(C); \draw [-] (1,1.5)--(C);
\draw [<-] (1,-0.5) node[below left]{$3$}--(D); \draw [<-] (2,-0.5) node[below right]{$4$}--(D); 
\end{scope}
\end{tikzpicture}
\end{align}
\item[(t)] Totally reversed type: 
$\LL_{135} \LL_{245} \MM_{126} \MM_{346} 
= \MM_{346} \MM_{126} \LL_{245} \LL_{135}$. 
\begin{align}\label{g:Lt}
\begin{tikzpicture}
\begin{scope}[>=latex,xshift=0pt]
\draw [-] (0,0) coordinate(A) to [out = 0, in = -135] (2,0.5) coordinate(B);
\draw [-] (1,1) coordinate(C) to [out = 0, in = 135] (B);
\draw [-] (C) to [out = 90, in = -45] (0.5,2) coordinate(D);
\draw [-] (A) to [out = 90, in = -135] (D);
\draw [<-] (-0.5,0) node[left]{$1$} --(A); \draw [<-] (0,-0.5) node[below]{$3$}--(A);
{\color{blue} 
\draw [-] (A)--(C) [thick]; %
\draw [<-] (-0.5,-0.5)--(A)[thick];%
\draw [-] (1.5,1.5) node[above right]{$5$}--(C)[thick]; %
}
{\color{green}
\draw [-] (B) to [out = 90, in = 0] (D)[thick]; 
\draw [<-] (2,0)--(B) [thick]; 
\draw [-] (0,2) node[left]{$6$}--(D)[thick]; 
}
\draw [-] (2.5,0)--(B); \draw [-] (2.5,1)--(B);
\draw [<-] (1,0.5) node[below]{$4$}--(C); \draw [<-] (0.5,1) node[left]{$2$}--(C);
\draw [-] (1,2.5)--(D); \draw [-] (0,2.5)--(D);
\draw (3.3,1) node {$=$}; 
\end{scope}
\begin{scope}[>=latex,xshift=135pt]
\draw [-] (2,2) coordinate(A) to [out = 180, in = 45] (0,1.5) coordinate(B);
\draw [-] (1,1) coordinate(C) to [out = 180, in = -45] (B);
\draw [-] (C) to [out = -90, in = 135] (1.5,0) coordinate(D);
\draw [-] (A) to [out = -90, in = 45] (D);
\draw [-] (2.5,2)--(A); \draw [-] (2,2.5)--(A);
{\color{blue} 
\draw [-] (A)--(C) [thick]; %
\draw [-] (2.5,2.5) node[above right]{$5$} --(A) [thick];%
\draw [<-] (0.5,0.5)--(C)[thick];%
}
{\color{green}
\draw [-] (B) to [out = -90, in = 180] (D)[thick]; 
\draw [-] (0,2) node[above]{$6$}--(B) [thick]; 
\draw [<-] (2,0)--(D)[thick]; 
}
\draw [<-] (-0.5,2) node[left]{$2$}--(B); \draw [<-] (-0.5,1) node[left]{$1$}--(B);
\draw [-] (1.5,1)--(C); \draw [-] (1,1.5)--(C);
\draw [<-] (1,-0.5) node[below left]{$3$}--(D); \draw [<-] (2,-0.5) node[below right]{$4$}--(D); 
\end{scope}
\end{tikzpicture}
\end{align}
\end{itemize} 

\begin{itemize}
\item[(I)] Ordinary type: $\MM(s',r',qg',q^{-1}f';q)\MM(r',s',f',g';q) = r's' \mathrm{Id}$.
\begin{align}\label{g:M-I}
\begin{tikzpicture}
\begin{scope}[>=latex,xshift=0pt]
{\color{green} 
\draw [->] (-0.5,1.5) node[left]{$3$}-- (0,1.5) coordinate(A) to [out = 0, in = 90] (1.5,0) coordinate(B) -- (1.5,-0.5)[thick];
}
\draw [->] (-0.5,2) node[above left]{$2$} -- (2,-0.5);
\draw [->] (0,2) node[above]{$1$}-- (A) to [out=-90,in=180] (B) --(2,0); 
\draw (3,0.5) node {$=$}; 
\end{scope}
\begin{scope}[>=latex,xshift=135pt]
\draw [->] (-0.5,2) node[above left]{$2$}-- (2,-0.5);
\draw [->] (-0.3,2.2) node[above]{$1$}-- (2.2,-0.3);
{\color{green} 
\draw [->] (-0.7,1.8) node[left]{$3$}-- (1.8,-0.7)[thick];
}
\draw (2.3,0.5) node {$\times \,r's'$}; 
\end{scope}
\end{tikzpicture}
\end{align}

\item[(I')] Reversed type: $[\MM(r',s',f',g';q)\MM(s',r',q^{-1}g',qf';q)]_{\ast 3} = r's' \mathrm{Id}$.
\begin{align}\label{g:Ip}
\begin{tikzpicture}
\begin{scope}[>=latex,xshift=0pt]
{\color{green} 
\draw [->] (-0.5,1.5) node[left]{$3$}-- (0,1.5) coordinate(A) to [out = 0, in = 90] (1.5,0) coordinate(B) -- (1.5,-0.5)[thick];
}
\draw [<-] (-0.5,2) node[above left]{$1$}-- (2,-0.5);
\draw [<-] (0,2) node[above]{$2$}-- (A) to [out=-90,in=180] (B) --(2,0); 
\draw (3,0.5) node {$=$}; 
\end{scope}
\begin{scope}[>=latex,xshift=135pt]
\draw [<-] (-0.5,2) node[above left]{$1$}-- (2,-0.5);
\draw [<-] (-0.3,2.2) node[above]{$2$}-- (2.2,-0.3);
{\color{green} 
\draw [->] (-0.7,1.8)node[left]{$3$} -- (1.8,-0.7)[thick];
}
\draw (2.3,0.5) node {$\times \,r's'$}; 
\end{scope}
\end{tikzpicture}
\end{align}
\end{itemize}


\end{document}